\documentclass{aastex62}

\usepackage[version=4]{mhchem} 
\usepackage{amsmath}
\usepackage{amssymb}
\usepackage{graphics}
\usepackage{rotating}
\usepackage{color}
\usepackage{epstopdf}
\definecolor{alizarin}{rgb}{0.82, 0.1, 0.26}
\definecolor{red}{rgb}{1.0, 0.0, 0.0}
\definecolor{green}{rgb}{0.3, 0.73, 0.09}
\definecolor{blue}{rgb}{0.0, 0.0, 1.0}

\newcommand{\dme}{CH$_{3}$OCH$_{3}$}
\newcommand{\mf}{CH$_{3}$OCHO}
\newcommand{\actd}{CH$_{3}$CHO}

\graphicspath{{./}{figures/}}

\submitjournal{ApJS}

\shorttitle{Formation of Complex Organic Molecules in Cold Environments}
\shortauthors{Jin \& Garrod}

\begin{document}

\title{Formation of Complex Organic Molecules in Cold Interstellar Environments through non-diffusive grain-surface and ice-mantle chemistry}

\correspondingauthor{Mihwa Jin}
\email{mj2ua@virginia.edu}

\author{Mihwa Jin}
\affil{Department of Chemistry, University of Virginia, Charlottesville, VA, 22904, USA}
\nocollaboration

\author{Robin T. Garrod}
\affiliation{Department of Chemistry, University of Virginia, Charlottesville, VA, 22904, USA}
\affiliation{Department of Astronomy, University of Virginia, Charlottesville, VA, 22904, USA}
\nocollaboration

\begin{abstract}

A prevailing theory for the interstellar production of complex organic molecules (COMs) involves formation on warm dust-grain surfaces, via the diffusion and reaction of radicals produced through grain-surface photodissociation of stable molecules. However, some gas-phase O-bearing COMs, notably acetaldehyde (\actd), methyl formate (\mf), and dimethyl ether (\dme), are now observed at very low temperatures, challenging the warm scenario. Here, we introduce a selection of new non-diffusive mechanisms into an astrochemical model, to account for the failure of the standard diffusive picture and to provide a more generalized scenario of COM formation on interstellar grains. New generic rate formulations are provided for cases where: (i) radicals are formed by reactions occurring close to another reactant, producing an immediate follow-on reaction; (ii) radicals are formed in an excited state, allowing them to overcome activation barriers to react with nearby stable molecules; (iii) radicals are formed through photo-dissociation close to a reaction partner, followed by immediate reaction. Each process occurs without the diffusion of large radicals. The new mechanisms significantly enhance cold COM abundances, successfully reproducing key observational results for prestellar core \object{L1544}. H-abstraction from grain-surface COMs, followed by recombination, plays a crucial role in amplifying chemical desorption into the gas phase. The UV-induced chemistry produces significant COM abundances in the bulk ices, which are retained on the grains and may persist to later stages. O$_2$ is also formed strongly in the mantle though photolysis, suggesting cometary O$_2$ could indeed be interstellar.

\end{abstract}


\section{Introduction} \label{sec:intro}

Complex organic molecules (COMs), usually defined as carbon-bearing molecules with six or more atoms, have been detected within the interstellar medium and in various protoplanetary environments \citep{blake87, fayolle15, bottinelli10, arce08, oberg10, bacmann12}. COMs synthesized at the early stages of star formation are suggested to have been a starting point for the organic materials that went on to seed the nascent solar system \citep{herbstvandishoeck09}. While the degree to which the interstellar synthesis of COMs may contribute to pre-biotic/biotic chemistry on Earth is still a matter of debate, many recent studies have shed light on the possible interstellar/protostellar origins of chemical complexity. For instance, the sugar-like molecule glycolaldehyde (CH$_2$(OH)CHO) has been detected toward the class 0 protostellar binary source IRAS16293-2422 \citep{jorgensen12}, as well as the Galactic Centre source Sgr B2(N) and other hot cores \citep{hollis00, beltran09}. A related molecule, ethylene glycol ($\ce{HOCH$_2$CH$_2$OH}$), has also been found toward low-mass protostars \citep{maury14,jorgensen16}, having first been detected in SgrB2(N)~\citep{hollis02}. The simplest amino acid, glycine ($\ce{NH2CH2COOH}$), has not been detected in the interstellar medium, but numerous amino acids have been found in meteorites \citep{kvenvolden70}, and glycine has been detected in two comets: Wild 2 \citep{elsila09} and Churyumov-Gerasimenko~\citep{altwegg16}. In the laboratory, \citep{meinert16} observed the generation of numerous sugar molecules, including the aldopentose ribose, through photochemical and thermal processing of interstellar ice analogs initially composed of $\ce{H2O, CH3OH, and NH3}$.

One of the prevailing theories explaining the formation of interstellar COMs is based on radical chemistry occurring on dust-grain particles, following the formation of ice mantles on their surfaces. In dark clouds and prestellar cores, where temperatures may be as low as around 10~K, hydrogen atoms are the most mobile species on the grain. Their high mobility allows them to find reaction partners on the surface easily. Thus, the accretion of other atoms and simple molecules, such as O, C, N and CO, onto the grain surfaces often leads to reactions with atomic H, producing an ice mantle composed of commonly-observed stable hydrides like water, methane, and ammonia, through repetitive hydrogenation. In dense regions, almost all of the CO produced in the gas phase may ultimately freeze out onto the dust grains \citep[e.g.][]{caselli99}, producing an outer ice surface that is rich in CO. Atomic H may also react with CO to produce methanol (CH$_3$OH); although there are activation energy barriers that prevent all of this CO being hydrogenated, a substantial fraction of the ice -- up to as much as around 30\% with respect to water toward some protostars, but generally of the order of 5\% \citep{boogert15} -- is found observationally to be composed of $\ce{CH3OH}$.

Grain-surface methanol, and indeed other molecules, may be broken down into radicals through UV-induced photolysis, caused either by external photons or by the secondary UV field induced by cosmic-ray collisions with gas-phase H$_2$; however, in the standard picture used in most astrochemical models, these radicals are still immobile on the grain surfaces at low temperatures, hindering their ability to react with anything other than mobile H atoms. As a star-forming core evolves to be heated to temperatures greater than $\sim$20 K, radicals such as CH$_3$ and HCO become more mobile, allowing them to meet via diffusion and quickly react to form larger COMs \citep{garrodandherbst06, garrod08a}. At this stage, most of the COMs produced in this way are unable to desorb effectively into the gas phase where they may be directly detected; only non-thermal mechanisms such as chemical desorption \citep{garrod07} and photo-desorption \citep{oberg09a,oberg09b} are active, with the former expected to yield only around 1\% of reaction products to the gas phase, while the latter would be weaker still in such regions where visual extinction is high. Only when the dust-grains in the core reach a high enough temperature (typically on the order of $\sim$100~K), through protostellar heating, can large COMs sublimate efficiently from the grains. This process is reviewed in more detail by \citet{oberg10}.

According to this scenario, it would be unlikely that COMs other than methanol would be effectively synthesized under the low-temperature conditions of dark clouds and prestellar cores, due to the immobility of heavy species on the grains, while methanol would still be formed as the result of the diffusion and reaction of H alone. The detection of COMs in these cold environments therefore presents a significant challenge to current astrochemical models based on diffusive grain-surface chemistry. In particular, three O-bearing COMs, acetaldehyde (AA; \actd), methyl formate (MF; \mf), and dimethyl ether (DME; \dme), have been detected in prestellar cores such as L1689B, \object{L1544} and B1-b~\citep{bacmann12, vastel14, cernicharo12} as well as in the cold outer envelopes of protostars ~\citep{oberg10, bergner17a}. 

Many chemical mechanisms have been proposed to explain the presence of COMs in cold cores: The Eley-Rideal process combined with complex-induced reactions on the grains \citep{ruaud15}; non-diffusive reactions \citep{changandherbst16}; oxygen insertion reactions \citep{bergner17b}; cosmic-ray induced chemistry \citep{shingledecker18}; and gas-phase formation through radical reactions \citep{balucani15}. In particular, the latter gas-phase formation scheme is noteworthy in that it suggests a chemical link between gas-phase methyl formate and dimethyl ether via the radical $\ce{CH3OCH2}$, which would be formed through the abstraction of a hydrogen atom from DME by atoms such as F and Cl. The dimethyl ether itself would form through the efficient radiative association of the radicals CH$_3$ and CH$_3$O in the gas phase.

Although many of the above-mentioned studies provide plausible routes to one or more complex molecules, some suffer from poorly constrained parameters due to the lack of experimental data or detailed computational chemistry studies (e.g. the radiative association reaction to produce DME), or require somewhat extreme chemical conditions. For example, the fractional abundance of methoxy in the model of \citet{balucani15} is more than one order of magnitude larger than the observed value, making it unclear whether dimethyl ether can be formed solely in the gas phase in cold cores~\citep{vasyuninandherbst13,changandherbst16}. Also, the chemical networks for F and Cl are relatively sparse, introducing further uncertainties; the models of \citet{vasyunin17} suggest that OH is instead a more important agent of H-abstraction from dimethyl ether. 

Cosmic ray-induced radiolysis models by \citet{shingledecker18} appear promising, although it is unclear whether they can account for large abundances of COMs on their own, or whether they require radicals also to be produced by UV photons. For example, those authors suggest an average rate of solid-phase water dissociation by cosmic rays of around $6\times10^{-16}$ s$^{-1}$ at the canonical cosmic-ray ionization rate. However, the typically-assumed rate used in astrochemical models for the separate process of UV photo-dissociation of {\em gas-phase} water by cosmic ray-induced photons is around $1.3\times10^{-14}$ s$^{-1}$. Even if the rates of UV photo-dissociation in the solid phase are around three times lower than those of gas-phase molecules \citep{kalvans18}, the influence of cosmic ray-induced photons in producing radicals in ice mantles would, on average, appear to be greater than that of direct cosmic-ray impingement by close to an order of magnitude. Other calculations for CR-induced water dissociation \citep{garrod19} would suggest an even greater discrepancy. The formation of COMs through direct cosmic-ray impacts into icy grains may therefore rely on the presence of pre-existing radicals produced by CR-induced UV.

In order for chemical modeling studies to be directly comparable with observations, they should ideally seek to reproduce the spatio-physical conditions of the target source. In pre-stellar cores such as \object{L1544}, strong gradients in physical parameters such as dust temperature, density, and visual extinction may sample a range of chemical regimes. The recent detection of a chemically-active outer shell around the core center of \object{L1544} sheds light on the complexity of its chemical structure~\citep{bizzocchi14}. Methanol column density in this source appears to peak at a radius around 4000~au distant from the source center. Abundances of other COMs seem to be enhanced at this position also \citep{jimenez-serra16}.

The modeling study by \citet{vasyunin17} is notable in its use of an explicit spatio-physical model to perform chemical modeling of \object{L1544}, successfully reproducing a similar feature to the observed methanol peak. Those authors used a surface-specific treatment of chemical desorption from grains in which a pure water-ice surface would allow only minimal desorption of newly-formed molecules, while a more CO-rich surface would allow molecules including methanol to desorb more easily. However, the peak fractional abundances of methanol produced in this model appear high ($\sim$$10^{-7}$ with respect to total hydrogen) compared with the usual values observed in cold sources ($\sim$$10^{-9}$). The efficiency of the chemical desorption of methanol (which is responsible for its gas-phase abundance) and many other species on water ice is poorly constrained by experiment, with an upper limit of 8\% for methanol formed through H addition \citep{minissale16c}; thus, there is room for the efficiency of this mechanism to be varied in models to produce the correct observed quantity. However, the \citet{vasyunin17} models rely on gas-phase methanol as the feedstock for the production of other, larger COMs. Lower peak abundances of gas-phase methanol would almost certainly render the gas-phase routes investigated by those authors too weak to account for the abundances of DME, MF and AA. Furthermore, using gas-phase reactions between CH and CH$_3$OH as a mechanism to form CH$_3$CHO, they erroneously extrapolated experimental reaction rates determined at room temperature~\citep{johnson00} down to 10~K; this produces impossibly large rate values that are around 3 orders of magnitude greater than the collisional rates for neutrals at such temperatures. It is unclear how much influence their alternative route for AA production, the radiative association of CH$_3$ and HCO radicals, would have on that molecule's peak gas-phase abundance; however, in their network they assume a rate that is on the order of the collisional rate (i.e. high efficiency), based on high-pressure experiments that would in fact be more representative of collisional (rather than radiative) de-excitation, and which therefore say nothing certain about radiative association. Until detailed calculations for this, or indeed for the radiative association of CH$_3$O with CH$_3$, are available, those rates may be considered to be somewhat optimistic.

Recent laboratory results have added new impetus to understanding the behavior of grain-surface chemistry at very low temperatures. \citet{fedoseev15, fedoseev17} considered the co-deposition of CO and H at low temperatures, resulting in significant production of complex organics such as glycolaldehyde, ethylene glycol, and glycerol. They postulate that reactions between adjacent HCO radicals, and/or HCO and CO, may produce such species with minimal thermal diffusion required -- a result not predicted by standard gas-grain chemical models, due to the lack of diffusion involved. So-called microscopic Monte Carlo models, unlike the more usual rate-equation models, are capable of simulating the relative positions of all atoms and molecules on a grain surface at each moment, allowing them automatically to account for this non-diffusive process. Such models seem to indicate that COM production through non-diffusive mechanisms is plausible~\citep[][Ioppolo et al. 2020, {\em subm.}, Garrod et al. 2020, {\em in prep.}]{fedoseev15,changandherbst16,dulieu19}.

It would appear, therefore, that -- regardless of other mechanisms such as gas-phase processes or radiolysis -- the standard rate description of grain-surface reactions is insufficient to treat all temperature regimes; there are situations in which reactants may be produced and rapidly meet (and react), either without diffusion of radicals or with some minimal amount of diffusion that does not obey the more general rate treatment.

In this study, we present a relatively simple formulation for non-diffusive chemistry, for use in standard gas-grain chemical models, that allows a newly-formed reaction product to react further with some other chemical species that happens to be in close proximity to the product(s) of the first reaction. Due to the instantaneous nature of this process, we refer to it here as a ``three-body reaction mechanism'' (3-B). We also consider a similar, related mechanism in which the new product has sufficient excitation energy to allow it to overcome the activation-energy barrier to its reaction with some nearby species; here, specifically with CO or H$_2$CO. This mechanism is referred to here as the ``three-body excited formation mechanism'' (3-BEF). 

A functionally-similar mechanism was included in the recent model of the solid-phase chemistry of cometary nuclei by \citet{garrod19}, but rather with the initiating process being the production of radicals by UV-induced photo-dissociation. In preliminary versions of that model that did not include that mechanism, it was found that the photo-dissociation of bulk ice molecules at temperatures 5--10~K was capable of producing implausibly high abundances of reactive radicals, which were unable to react due to the lack of bulk thermal diffusion at those temperatures. The \citet{garrod19} model includes a new reaction process whereby a newly-formed photo-product may react immediately with a nearby reaction partner in the ice. This mechanism is included in the present model also, which we label as the ``photodissociation-induced reaction mechanism'' (PDI).

\citet{garrodandpauly11} used a similar mechanism to the 3-body mechanisms to explain the formation of $\ce{CO2}$ ice at low temperatures. In their treatment, the production of an OH radical via the reaction of H and O atoms in proximity to a CO molecule could allow the immediate formation of CO$_2$ (overcoming a modest activation energy barrier). Their models successfully reproduced the observed behavior of CO, $\ce{CO2}$, and water ice in dense clouds, and showed that such non-diffusive processes could be handled within a standard gas-grain chemical model. More recently, \citet{changandherbst16} implemented a similar process, which they called a ``chain reaction mechanism" in their microscopic Monte Carlo simulation, achieving abundances of gas-phase COMs high enough to reproduce the observational values toward cold cores (at a temperature of $\sim$10 K), using a chemical desorption efficiency of 10$\%$ per reaction. \citet{dulieu19}, seeking to explain the surface production of NH$_2$CHO in laboratory experiments involving H$_2$CO, NO and H deposition, introduced a non-diffusive reaction treatment for a single reaction (see also Sec. \ref{subsec:three-body}).

Finally, a simple treatment for the Eley-Rideal process is included in the present model, in which an atom or molecule from the gas-phase is accreted directly onto a grain-surface reaction partner, resulting in immediate reaction (mediated by an activation-energy barrier, where appropriate). Such processes have been included in similar models before \citep[e.g.][]{ruaud15}, but are included here for completeness in the consideration of all mechanisms by which reactants may instantly be brought together without a mediating diffusion mechanism.

The formulations presented here for the above processes also allow, through a repetitive application of the main non-diffusive reaction process (i.e. the three-body mechanism), for the products of each of those processes themselves to be involved in further non-diffusive reaction events (in cases where such processes are allowed by the reaction network). Thus, for example, an Eley-Rideal reaction may be followed by an immediate secondary non-diffusive reaction. The importance of such repetitive processes will diminish with each iteration.

All of the above non-diffusive reaction mechanisms are considered in the model, with a particular emphasis on the production of the O-bearing COMs that are now detected in the gas-phase in cold prestellar cores. The formulations corresponding to each of the new mechanisms presented here are functionally similar to each other, but quite different from the standard diffusive reaction formula used in typical astrochemical models. However, they are fully compatible with the usual treatment and may be used in tandem with it.

With the introduction of the new mechanisms, we run multi-point chemical models of prestellar core \object{L1544} to test their effectiveness in an appropriate environment. A spectroscopic radiative-transfer model is implemented here as a means to evaluate the observable column densities of molecules of interest, allowing the direct comparison of the model results with observations.

This paper is structured as follows. The chemical model and the newly-implemented mechanisms are described in \S~2. The results of the models are explored in \S~3, with discussion in \S~4. Conclusions are summarized in \S~5.

\section{Chemical model} \label{sec:model}

We use the astrochemical kinetic code {\em MAGICKAL} to study new grain-surface/ice-mantle mechanisms that may effectively form complex organic molecules in cold environments. The model uses a three-phase approach, based on that described by \cite{garrodandpauly11} and \cite{garrod13}, in which the coupled gas-phase, grain/ice-surface and bulk-ice chemistry are simulated through the solution of a set of rate equations. Initial chemical abundances used in the model are shown in Table~\ref{init_element}, with elemental values based on those used by \citet{garrod13}. The initial H/H$_2$ abundances are chosen to agree approximately with the steady-state values appropriate to our initial physical conditions, as determined by the chemical model. Bulk diffusion is treated as described by \citet{garrod17}. Although the bulk ice is technically chemically active in this model, at the low temperatures employed in this work, diffusive reactions in general are negligibly slow, excluding processes involving H or H$_2$ diffusion. However, the addition of non-diffusive reactions to the model increases significantly the degree of chemical activity within the bulk. 

\begin{deluxetable}{ccccc}
\tablewidth{0pt}
\tabletypesize{\footnotesize}
\tablecolumns{5}
\tablecaption{Initial elemental and chemical abundances \label{init_element}}
\scriptsize
\tablehead{
\colhead{Species, \textit{i}} & \colhead{$n(i)/n_\textrm{H}^{a}$ \tablenotemark{a}
}}
\startdata
$\ce{H}$ & 5.0(-4) \\
$\ce{H2}$ & 0.49975\\
$\ce{He}$ & 0.09\\
$\ce{C}$ & 1.4(-4)\\ 
$\ce{N}$ & 7.5(-5)\\
$\ce{O}$ & 3.2(-4)\\
$\ce{S}$ & 8.0(-8)\\
$\ce{Na}$ & 2.0(-8)\\
$\ce{Mg}$ & 7.0(-9)\\
$\ce{Si}$ & 8.0(-9)\\
$\ce{P}$ & 3.0(-9)\\
$\ce{Cl}$ & 4.0(-9)\\
$\ce{Fe}$ & 3.0(-9)\\
\enddata
\tablecomments{$^{a}A(B)=A^{B}$}
\end{deluxetable}

The model uses the modified-rate treatment for grain-surface chemistry presented by \citep{garrod08b}, which allows the stochastic behavior of the surface chemistry to be approximated; the back-diffusion treatment of \citet{willisandgarrod17} is also used. Surface diffusion barriers ($E_{\mathrm{dif}}$) are related to desorption (i.e. binding) energies ($E_{\mathrm{des}}$) such that $E_{\mathrm{dif}}$=0.35$E_{\mathrm{des}}$ for all molecular species, with bulk diffusion barriers taking values twice as high \citep{garrod13}. However, the recent study by \citet{minissale16b} estimated surface diffusion barriers for atomic species such as N and O to be $E_{\mathrm{dif}}$=0.55$E_{\mathrm{des}}$. We adopt a similar value $E_{\mathrm{dif}}/E_{\mathrm{des}}$=0.6 for all atomic species; the impact of this parameter is discussed in \S \ref{bindingenergy}. These basic surface diffusion and desorption parameters are also adjusted according to the time-dependent abundance of H$_2$ in the surface layer, following the method of \cite{garrodandpauly11}. All diffusion is assumed to be thermal, with no tunneling component (see also Sec. \ref{bindingenergy}).

Chemical desorption, whereby grain-surface reactions allow some fraction of their products to desorb into the gas phase, is treated using the RRK formulation of \cite{garrod07} with an efficiency factor $a_{\mathrm{RRK}}=0.01$.

The grain-surface/ice-mantle photodissociation rates used in {\em MAGICKAL} are based on the equivalent gas-phase rates, in the absence of other evidence \citep[e.g.][]{garrod08a}, and they likewise assume the same product branching ratios. Following the work of \citet{kalvans18}, we adopt photodissociation rates on the grain-surfaces and in the ice mantles that are a factor of 3 smaller than those used for the gas phase. Photodissociation may be caused either by external UV, or by the induced UV field caused by cosmic-ray collisions with H$_2$ molecules, and both sources of dissociation are included in the model.

Methanol in cold clouds is mainly formed on the grain surfaces through ongoing hydrogenation of CO~\citep{fuchs09}. The methanol production network used in the present network follows from that implemented by \citet{garrod13}, and includes not only forward conversion of CO to methanol but also the backward reactions of each intermediate species with H atoms. 

The overall chemical network used here is based on that of \citet{garrod17}, with a few exceptions. In particular, a new chemical species, CH$_3$OCH$_2$, has been added along with a set of associated gas-phase and grain-surface reactions/processes listed in Tables \ref{gas_rxn} (gas-phase) and \ref{solid_rxn} (solid-phase). This radical is a key precursor of DME in our new treatments (see \S \ref{subsec:3-bef}). Its inclusion also allows the addition of a grain-surface H-abstraction reaction from DME, making this species consistent with methyl formate and acetaldehyde.

An additional reaction was included in the surface network, corresponding to H-abstraction from methane (CH$_4$) by an O atom~\citep[$E_A$=4380~K; ][]{herronandhuie73}, as a means to ensure the fullest treatment for the CH$_3$ radical in the network. 

The final change to the network is the adjustment of the products of $\ce{CH3OH}$ photo-desorption to be $\ce{CH3}$+$\ce{OH}$ rather than $\ce{CH3OH}$, roughly in line with the recent experimental study by \citet{bertin16}.

Each of the generic non-diffusive mechanisms that we include in the model (as described below) is allowed to operate on the full network of grain-surface and ice-mantle reactions that are already included for the regular diffusive mechanism; reactive desorption, where appropriate, is also allowed to follow on from each of these, in the case of surface reactions. The full model therefore includes around 1600 surface and 1100 bulk-ice reaction processes for each new generic mechanism included in the model (excluding 3-BEF, see Section 2.5). All the grain-surface and bulk-ice reactions allowed in the network are presented in machine-readable format in Table~\ref{solid_rxn}. The rate formulations for diffusive and non-diffusive reaction mechanisms used in the model are described below.

\subsection{New chemical mechanisms} \label{sec:mechanisms}

The standard formulation, as per e.g. \citet{hasegawa92}, for the treatment of a diffusive grain-surface chemical reaction (also known as the Langmuir-Hinshelwood or L-H mechanism) between species $A$ and $B$ is based on: the hopping rates of the two reactants, $k_{\mathrm{hop}}(A)$ and $k_{\mathrm{hop}}(B)$; the abundances of both species on the grains, $N(A)$ and $N(B)$, which here are expressed as the average number of atoms or molecules of that species present on an individual grain; the total number of binding sites on the grain surface, $N_S$, often assumed to be on the order of 1 million for canonically-sized grains; and an efficiency related to the activation energy barrier (if any), $f_{\mathrm{act}}(AB)$, which takes a value between zero and unity. Thus, the total rate of production (s$^{-1}$) may be expressed in the following form (which is arranged in such a way as to demonstrate its correspondence with the non-diffusive mechanisms discussed later):

\begin{equation}
R_{AB} = f_{\mathrm{act}}(AB) \, \left[ k_{\mathrm{hop}}(A) \, N(A) \right] \frac{N(B)}{N_S} \, 
       + f_{\mathrm{act}}(AB) \, \left[ k_{\mathrm{hop}}(B) \, N(B) \right] \frac{N(A)}{N_S} 
\end{equation}

In the first term, the expression within square brackets corresponds to the total rate at which particles of species $A$ may hop to an adjacent surface binding site. The ratio $N(B)/N_S$ gives the probability for each such hop to result in a meeting with a particle of species $B$. Multiplying these by the reaction efficiency gives the reaction rate associated solely with the diffusion of species $A$. The reaction rate associated with diffusion solely of species $B$ is given by the second term. The total reaction rate is commonly expressed more succinctly thus:

\begin{gather}
R_{AB} = k_{AB} \, N(A) \, N(B) \\
k_{AB} = f_{\mathrm{act}}(AB) \, (k_{\mathrm{hop}}(A) + k_{\mathrm{hop}}(B)) / N_S
\end{gather}

\noindent which provides a more standard-looking second-order reaction rate. The rate coefficient, $k_{AB}$, may be further adjusted to take account of random walk, in which a reactant that has yet to meet a reaction partner may re-visit previous, unsuccessful sites. This effect typically reduces the overall reaction rate by no more than a factor of a few \citep[e.g.][]{charnley05, lohmarandkrug06, lohmar09, willisandgarrod17}.

The individual hopping rate for some species $i$ is assumed in this model to be a purely thermal mechanism, given by
\begin{equation}
k_{\mathrm{hop}}(i) = \nu(i) \, \exp{\left( \frac{-E_{\mathrm{dif}}(i)}{T} \right)}
\end{equation}

\noindent where $\nu(i)$ is the characteristic vibrational frequency of species $i$ and $E_{\mathrm{dif}}(i)$ is the barrier against diffusion in units of K.

The reaction efficiency factor for a reaction between species $A$ and $B$ considers the case where, if there is an activation energy barrier, the diffusion of either species {\em away} from the other may compete with the reaction process itself \citep[see e.g.][]{garrodandpauly11}, thus:

\begin{equation}
f_{\mathrm{act}}(AB) = \frac{ \nu_{AB} \, \kappa_{AB} }{  \nu_{AB} \, \kappa_{AB} + k_{\mathrm{hop}}(A) + k_{\mathrm{hop}}(B) }
\end{equation}

\noindent where $\nu_{AB}$ is taken as the faster of either $\nu(A)$ or $\nu(B)$, and $\kappa_{AB}$ is a Boltzmann factor or tunneling efficiency for the reaction \citep[see][]{hasegawa92}. The denominator represents the total rate at which an event may occur when species A and B are in a position to react.

In order to formulate rates for {\em non-diffusive} reaction processes of whatever kind, the total rate must again be decomposed into its constituent parts, which, unlike in Eqs. (2) and (3), cannot generally be re-combined.

The generic form that we adopt for such processes is:

\begin{equation}
R_{AB} = f_{\mathrm{act}}(AB) \, R_{\mathrm{comp}}(A) \, \frac{N(B)}{N_S} \, + f_{\mathrm{act}}(AB) \, R_{\mathrm{comp}}(B) \, \frac{N(A)}{N_S} 
\end{equation}

\noindent where $R_{\mathrm{comp}}(i)$ is labeled the ``completion rate'' for the reaction, corresponding specifically to the ``appearance'' of species $i$. The determination of $R_{\mathrm{comp}}(i)$ values is explained in more detail in the following subsections for each of the specific reaction mechanisms considered. The above form is essentially the same as that given by \citet{garrod19} for photodissociation-induced reactions. The correspondence of Eq. (1) (for diffusive reactions) with Eq. (6) is clear; the latter may be considered a more general description of a surface reaction rate, which can be applied to both diffusive and non-diffusive processes, according to the chosen form of $R_{\mathrm{comp}}(i)$. The regular diffusive mechanism would use $R_{\mathrm{comp}}(i) = k_{\mathrm{hop}}(i) \, N(i)$.

While the general form given in Eq. (6) is set up to describe grain-surface processes, it may easily be adapted for bulk-ice processes by substituting $N_S$ for $N_M$, the total number of particles in the ice mantle, with $N(i)$ now representing the number of atoms/molecules of species $i$ present in the mantle. In this case, mantle-specific diffusion rates should be used.

The several non-diffusive processes incorporated into the chemical model, based on Eq. (6), are described below. Table~\ref{modeltickbox} indicates which specific new mechanisms are included in each of the model setups tested in the present study.

\subsection{Eley-Rideal reactions}

The Eley-Rideal (E-R) reaction process occurs when some atom or molecule that is adsorbing/accreting from the gas phase onto the grain surface immediately encounters its grain-surface reaction partner as it adsorbs. \cite{ruaud15} considered a more intricate treatment than we use here, in which an adsorbing carbon atom could enter into a bound complex with a surface molecule. Here we adopt a more generalized treatment in which we do not differentiate between the binding properties of local surface species.

The rates for the Eley-Rideal process can easily be represented using Eq. (6). For reactions that have no activation-energy barrier (and for which $f_{\mathrm{act}}(AB)$ is therefore close to unity), this is achieved by setting $R_{\mathrm{comp}}(i) = R_{\mathrm{acc}}(i)$, the total accretion rate of species $i$ from the gas phase.

In the interests of completeness, it is necessary also to consider how to treat the kinetics of E-R reactions that have at least some modest activation-energy barrier. To this end, one may consider a hypothetical case where oxygen atoms are slowly accreting onto an otherwise pure CO surface. For purposes of illustration, it is initially assumed here that the surface-diffusion rates of both O and CO are negligible, with the result that $f_{\mathrm{act}}(\mathrm{O+CO})$=1.

The reaction O + CO $\rightarrow$ CO$_2$ has an activation energy on the order of 1000~K; although the reaction will not be instant, in the absence of all other competing processes it should nevertheless occur on some finite timescale. Thus, the total timescale for the complete E-R process for an individual accreting O atom encountering and reacting with a surface CO molecule would be the sum of: (i) the accretion timescale of the oxygen atom onto the surface and (ii) the lifetime against its subsequent reaction with CO, i.e. $1 / R_{\mathrm{acc}}(\mathrm{O}) + 1 / ( \nu_{\mathrm{O+CO}} \, \kappa_{\mathrm{O+CO}} )$. This would provide a total completion rate associated with O accretion (to be employed in Eq. 6) of:

\begin{equation}
R_{\mathrm{comp}}(\mathrm{O}) = \frac{1}{ 1 / R_{\mathrm{acc}}(\mathrm{O}) + 1 / ( \nu_{\mathrm{O+CO}} \, \kappa_{\mathrm{O+CO}} )  } \nonumber
\end{equation}\

The completion rate $R_{\mathrm{comp}}(\mathrm{O})$ should be viewed as the rate at which the reaction process occurs {\em successfully} from the point of view of an individual accreting O atom, taking into account all sequential steps in the completion of the reaction process. Note that, in the full description, the probability of encountering a CO molecule on the surface, $N(\mathrm{CO}) / N_S$, and the reaction efficiency, $f_{\mathrm{act}}(\mathrm{O+CO})$, should both remain outside of the formula for $R_{\mathrm{comp}}(\mathrm{O})$, as per Eq. (6). Neither of these values affects the actual {\em timescale} over which an individual O atom successfully accretes and reacts with a surface CO molecule; rather, they affect the {\em probability} that a single such event is successful.

This expression for $R_{\mathrm{comp}}(\mathrm{O})$ could result in one of two important outcomes, depending on the relative rates of accretion and reaction. If accretion of O is very slow, and therefore reaction is comparatively fast, then $R_{\mathrm{comp}}(\mathrm{O}) \simeq R_{\mathrm{acc}}(\mathrm{O})$. Since $N($CO$) / N_S \simeq 1$, this means that the total Eley-Rideal production rate would initially be $R_{\mathrm{O+CO}} \simeq R_{\mathrm{acc}}(\mathrm{O})$. In other words, the overall production rate of CO$_2$ is only limited by the rate of O accretion onto the surface, which is as one would expect for this case.

However, if reaction is slower than or comparable to the initiating accretion process, each accretion of O would be followed by some significant lag-time between accretion and reaction, which must be accounted for in the overall rate; the incorporation of the above expression for $R_{\mathrm{comp}}(\mathrm{O})$ into Eq. (6) indeed does this. Without this expression, and instead using the value $R_{\mathrm{comp}}(\mathrm{O}) = R_{\mathrm{acc}}(\mathrm{O})$, the rate of conversion of O and CO into CO$_2$ would incorrectly be set to the accretion rate of O. The correct formulation gives a total E-R reaction rate that is less than the total accretion rate, allowing the build-up of O on the surface.

The final adjustment to the barrier-mediated E-R treatment comes into play when one or other surface diffusion rate is non-negligible. If diffusion of (say) O is indeed fast compared to reaction, then the reaction efficiency, $f_{\mathrm{act}}(\mathrm{O+CO})$, becomes small, which reduces the total rate of the reaction as per Eq. (6). However, the completion rate $R_{\mathrm{comp}}(\mathrm{O})$ must also be adjusted, to correspond only to the instances in which the O+CO reaction is actually successful. Successful reactions would have to occur before the diffusive separation of the two reactants could render the process unsuccessful, so the reaction timescale would become shorter, even though the reaction {\em probability} (i.e. $f_{\mathrm{act}}$) were reduced. For this reason, diffusion rates must also be considered when formulating $R_{\mathrm{comp}}(\mathrm{O})$. Using a more general description for reactants $A$ and $B$, the average lifetime against some event occurring (including reaction itself), once the reactants are in a position to react, may be described more fully by the expression:

\begin{equation}
t_{\mathrm{AB}} = 1 / ( \nu_{AB} \, \kappa_{AB} + k_{\mathrm{hop}}(A) + k_{\mathrm{hop}}(B) )
\end{equation}

\noindent which can then be used in the general definitions:
\begin{eqnarray}
R_{\mathrm{comp}}(A) = \frac{1}{ 1 / R_{\mathrm{app}}(A) + t_{AB} }\\
R_{\mathrm{comp}}(B) = \frac{1}{ 1 / R_{\mathrm{app}}(B) + t_{AB} }
\end{eqnarray}

\noindent where $R_{\mathrm{app}}(i)$ is the ``appearance rate'' of species $i$, which in the case of the E-R mechanism is simply $R_{\mathrm{acc}}(i)$.

It should be noted that once diffusion becomes significant, a model even as simple as the one used above to describe pure Eley-Rideal reaction processes would be incomplete; the standard diffusive reactions described by Eq. (1) must also be considered (as an entirely separate process) in such a model, to handle the occasions where accreting atoms (e.g. O) do not immediately react with their reaction partners (e.g. CO) before they diffuse away to another binding site, where they may also have the ability to react. In this case, the Eley-Rideal expressions would depend much less strongly on the time-lag effect described above, meaning that $R_{\mathrm{comp}}$ and $R_{\mathrm{acc}}$ would be similar in cases where diffusion of either reactant were relatively fast. In practical application to astrochemical models, for non-diffusive reactions whose reactants have slow or negligible diffusion rates, other processes could also act to interfere with the reaction; for example, the UV-induced dissociation of one or other reactant might occur on a shorter timescale than a very slow reaction, or a hydrogen atom might arrive to react with one of other reactant, before the reaction in question could occur. Competition from processes such as these would prevent very slow reactions (i.e. those with large activation barriers) from becoming important, even where diffusion of the reactants were negligible. A yet more complete treatment of reaction competition would include rates for these processes in Eqs. (5) and (7).

In our chemical model {\em MAGICKAL}, Eqs. (6)--(9) are used to set up Eley-Rideal versions of all allowed grain-surface reactions in the network. Because the E-R process is exclusively a surface process, no such processes in the ice mantle are included. Note that, when incorporating the Eley-Rideal mechanism into a model, no modification of the accretion (adsorption) rates themselves is required, since the Eley-Rideal mechanism does not {\em replace} any part of the adsorption rate. Rather, the E-R mechanism occurs immediately after adsorption, and therefore acts as a sink on the surface populations of the reactants, even though its rate is driven by the rate of arrival from the gas phase of one or other reactant.

Equations with the general form of Eqs. (6)--(9) are used also to formularize the remaining non-diffusive reaction mechanisms described below, where $R_{\mathrm{app}}(i)$ is the only quantity to vary between processes. These formulations can be used equally well for processes with or without activation-energy barriers.

While Eq.~(6) is still valid for the regular diffusive reaction mechanism with completion rates of $R_{\mathrm{comp}}(i) = k_{\mathrm{hop}}(i) \, N(i)$, no adjustment following Eqs. (7)--(9) should be used, nor would be needed. The formulation required to model any lag-time for diffusive reactions is different from that of non-diffusive processes (because $N(i) / N_S$ and $f_{\mathrm{act}}(AB)$ cannot remain outside the $R_{\mathrm{comp}}(i)$ expression), but there are no circumstances in which such a lag-time would be significant.

\subsection{Photodissociation-induced reactions}

\citet{garrod19} suggested that the omission of non-diffusive, photodissociation-induced reactions from models of interstellar ice chemistry may result in the photolytic production of COMs being severely underestimated. Past models of chemistry in star-forming regions \citep[e.g.][]{garrod08a,garrod17} have allowed photodissociation to contribute to the production of COMs in the surface and bulk-ice phases in only an indirect way, mediated by thermal diffusion. That is, photodissociation of various molecules produces radicals, which are separately allowed to react through the standard diffusive mechanism. Thus, at very low temperatures, no significant COM production is seen via radical-radical recombination, as diffusion of radicals is minimal. However, the presence of radicals in or upon the ice means that in some fraction of photodissociation events, the products may sometimes be formed with other reactive radicals already present nearby. In this case, the immediate products of photodissociation could react with the pre-existing radicals either without diffusion, or following some short-ranged, non-thermal diffusion process (possibly enabled by the excitation of the dissociation products).

Eqs. (6)--(9) can again be used to describe this process, with an appropriate choice for $R_{\mathrm{app}}(i)$, which is simply the total rate of production of photo-product $i$ caused by all possible photodissociation processes:

\begin{equation}
R_{\mathrm{app}}(i) = \sum_{\mathrm{all} \, j} \, R_{j}(i) \\
\end{equation}

\noindent where $R_{j}(i)$ is the production rate of $i$ via an individual photodissociation process $j$. For the radical CH$_3$, for example, this would include the photodissociation of CH$_3$OH, CH$_4$, and various larger molecules containing a methyl group. 

If one were to consider, for example, the production of dimethyl ether through this mechanism, an important reaction would be CH$_3$ + CH$_3$O $\rightarrow$ CH$_3$OCH$_3$, which is usually assumed to be barrierless. For this reaction, in Eq. (6), species $A$ = CH$_3$ and species $B$ = CH$_3$O; the appearance rate of CH$_3$ would be as described above. The main contribution to the appearance rate of CH$_3$O would likely be the photodissociation process CH$_3$OH + h$\nu$ $\rightarrow$ CH$_3$O + H. The formulation used for the dimethyl ether-producing reaction simply states that some fraction of CH$_3$ produced by photodissociation of various molecules in the ice will immediately meet a CH$_3$O radical that it can react with, and vice versa. 

Reactions affected by this PD-induced mechanism need not only be radical-radical recombination reactions; the production, via photodissociation, of atomic H in close proximity to CO, for example, could enhance the rate of the reaction H + CO $\rightarrow$ HCO, which has an activation-energy barrier. The treatment of barrier-mediated reactions in the generic Eqs. (7)--(9) is used again for this purpose. 

This treatment does not take into account any explicit consideration of excitation of the photo-products, which could also enhance reaction rates \citep[as per e.g.][in the case of cosmic-ray induced dissociation]{shingledecker18}. It is also implicitly assumed that the rates of photodissociation used in the network represent the rates at which dissociation occurs without immediate recombination of those same photo-products.

It is trivial to adapt the equations used for surface reactions to deal instead with ice-mantle related processes, and this is indeed implemented in the simulations presented here.

\subsection{Three-body reactions} \label{subsec:three-body}

The laboratory results of authors such as \citet{fedoseev15, fedoseev17}, in which H, CO and/or other species are deposited onto a cold surface, indicate that surface reactions between radicals of low mobility may produce COMs, even at low temperatures and without any energetic processing. The suggested explanation is that pairs of radical species such as HCO may, on occasion, be formed in close proximity to each other, allowing them to react either immediately or after a very small number of surface hops. The HCO radicals themselves would initially be formed through a more typical diffusive (Langmuir-Hinshelwood) process or through an Eley-Rideal process, via the barrier-mediated reaction of H and CO. \citet{fedoseev15} suggest that reactions of HCO with CO may also be active, which would require no diffusion of HCO at all, if the HCO itself is formed through the reaction of atomic H and CO on top of a CO-rich surface.

In a similar vein, \cite{garrodandpauly11} found, using chemical kinetics modeling, that the interstellar abundance of solid-phase CO$_2$ could be explained by the reaction H + O $\rightarrow$ OH occurring on a CO-rich dust-grain ice surface. This allows the newly-produced OH to react rapidly with the CO without any intervening thermal diffusion. They introduced into their models a new reaction rate specifically for this process that was functionally similar to Eq. (6).

Here, we use Eqs. (6)--(9) to calculate rates for what may be termed three-body reactions, which include the above examples. This approach is extended to all grain-surface reactions in the network, with a similar treatment for bulk-ice processes. To do this, another dedicated expression for the appearance rate $R_{\mathrm{app}}(i)$ to be used in Eqs. (8) and (9) must be constructed specifically for three-body reactions. Eq. (10) can again be used, this time where $R_{j}(i)$ is the production rate of $i$ (as determined using Eq. 1) resulting from any diffusive (Langmuir-Hinshelwood) reaction, or for any non-diffusive Eley-Rideal or photodissociation-induced reaction, whose rates are described above. Thus, $R_{\mathrm{app}}(i)$ includes the production rates of $i$ for all reactive mechanisms $j$ that could lead to a subsequent reaction. From a technical point of view, the rates of all such reaction processes must therefore be calculated in advance of the calculations for any three-body reactions.

Using the example of the process considered by \cite{garrodandpauly11}, the reaction under consideration as a three-body process would be OH + CO $\rightarrow$ CO$_2$; thus, in Eq. (6), $A$ = OH and $B$ = CO. There are several reactions in our network that could produce OH, but the main one is indeed H + O $\rightarrow$ OH. The sum of the production rates of OH from all of these reactions would comprise $R_{\mathrm{app}}($OH$)$. The appearance rate for CO would also be constructed from the CO production rates of all reactions leading to its formation.

In this way, CO$_2$ could be formed via a three-body reaction process in which, for example, an H and an O atom diffuse on a surface until they happen to meet in a binding site where CO is in close proximity, they react to form OH in the presence of the CO, and the OH and CO then subsequently react with no further diffusion required. Alternatively, an oxygen atom could be situated in contact with a CO molecule when an incoming H atom from the gas phase initiates an Eley-Rideal process, leading to OH formation, followed by reaction with the CO. Or, a CH$_4$ molecule in close proximity to a CO molecule and an O atom could be dissociated to H and CH$_3$, with the H quickly reacting with the O atom to produce OH, which would then react with CO. The prescription above would allow many such scenarios to be included in the overall production rate of CO$_2$, including others relating to the formation of a CO molecule in close proximity to an OH radical. The adoption of this generalized process means that the special-case prescription for the OH + CO reaction introduced by \cite{garrodandpauly11} is no longer required.

In the kind of chemical system considered by \citet{fedoseev15,fedoseev17}, in which H and CO are deposited onto a surface, complex molecules could be built up via three-body reactions between HCO radicals, initiated either by E-R or L-H production of HCO. 

Note that the new treatment does not explicitly differentiate between the case where the newly-formed reactant is immediately in contact with the next reaction partner, and the case where it has sufficient excess energy to allow it to undergo a thermal hop in order to find its next reaction partner. It is in fact highly probable that the products of exothermic reactions (which includes virtually every surface reaction included in the network) would have sufficient energy to allow some degree of non-thermal hopping immediately following formation. The possibility of such energy also allowing barrier-mediated three-body reactions to occur more rapidly is considered in the next subsection.

To go yet a stage further, one may imagine a scenario in which the products of three-body reactions themselves could also be involved in subsequent non-diffusive three-body reactions. This possibility is also included in our model, using the same equations as before, with appearance rates defined by:

\begin{equation}
R_{\mathrm{app}}(i) = \sum_{\mathrm{all} \, j} \, R_{j,\mathrm{3B}}(i) \\
\end{equation}

\noindent where $R_{j,\mathrm{3B}}(i)$ is the production rate of $i$ caused by the three-body reaction labelled $j$. Although these appearance rates will usually be lower than those used in the first round of three-body reactions, the second three-body reaction could be the most important for certain species if they have no more dominant production mechanism. In the present models, we allow a total of three rounds of three-body reactions to take place. Although this could in theory be increased to any arbitrary number of rounds, the influence of those processes rapidly diminishes beyond the second round.

As with the photodissociation-induced reactions, a similar method is employed also for reactions in the bulk ice. In this case, the appearance rates of reactants in the first round of three-body reactions would generally all be products of photodissociation-induced reactions, as the Eley-Rideal process is exclusively a surface mechanism, while the thermal diffusion of all species in the bulk -- excluding arguably H and H$_2$ -- would be very slow at the temperatures considered in the simulations presented here.

Finally, we note that a method for treating what we label three-body reactions was recently employed by \citet{dulieu19}, for a single reaction between H$_2$CO and newly-formed H$_2$NO. Those authors constructed a separate chemical species to represent H$_2$NO that is formed in contact with H$_2$CO on a surface, then included a special reaction in their network that occurs at a rate equal to the vibrational collision frequency of the two contiguous species. Although apparently different from our approach, such a treatment should also provide the correct result; this is because, assuming that there are no competing processes with the reaction in question, the abundance of the special H$_2$NO species is determined solely by its formation rate and its one destruction rate. In that case, both the specific abundance of the special H$_2$NO species and its reaction rate coefficient are cancelled out in the overall rate calculation, giving a total production rate for the reaction that is equal to the rate that we employ in the present work. In such a case, therefore, the chosen rate coefficient becomes immaterial to the result. It is presumably possible to set up a large network of such reactions for newly-formed species; however, the requirement to include a new chemical species for each reactant pair would likely make this method prohibitive for large networks.

\subsubsection{Specific reactions}

Although the full model includes a range of three-body (3-B) processes capable of producing acetaldehyde (CH$_3$CHO), methyl formate (CH$_3$OCHO) and dimethyl ether (CH$_3$OCH$_3$), the dominant mechanisms for each (based on model results) are presented below.

For acetaldehyde, the most important three-body mechanism is made up of a pair of sequential two-body processes as follows:
\begin{eqnarray}
\rm H~+~CH_{2} &\rightarrow & \rm CH_{3} \\
\rm CH_{3}~+~HCO &\rightarrow & \rm CH_{3}CHO
\end{eqnarray}

The most important sequential mechanisms for the other two COMs are:
\begin{eqnarray}
\rm H~+~CH_{2} &\rightarrow & \rm CH_{3}\\
\rm CH_{3}~+~CH_{3}O &\rightarrow & \rm CH_{3}OCH_{3}
\end{eqnarray}
\begin{eqnarray}
\rm H~+~CO &\rightarrow & \rm HCO\\
\rm HCO~+~CH_{3}O &\rightarrow & \rm CH_{3}OCHO
\end{eqnarray}

Each of these reaction pairs involves the addition of radicals in the second step, and two of them involve the addition of atomic H to a radical in the first step. The production of the COMs through these mechanisms should therefore have a strong dependence on the instantaneous abundances of short-lived reactive radicals.

The full network used in the models includes three-body versions of all the reactions used for regular diffusive chemistry, for all surface and mantle species.

\subsection{Excited three-body reactions} \label{subsec:3-bef}

\begin{figure}
\centering
\includegraphics[width=0.7\textwidth]{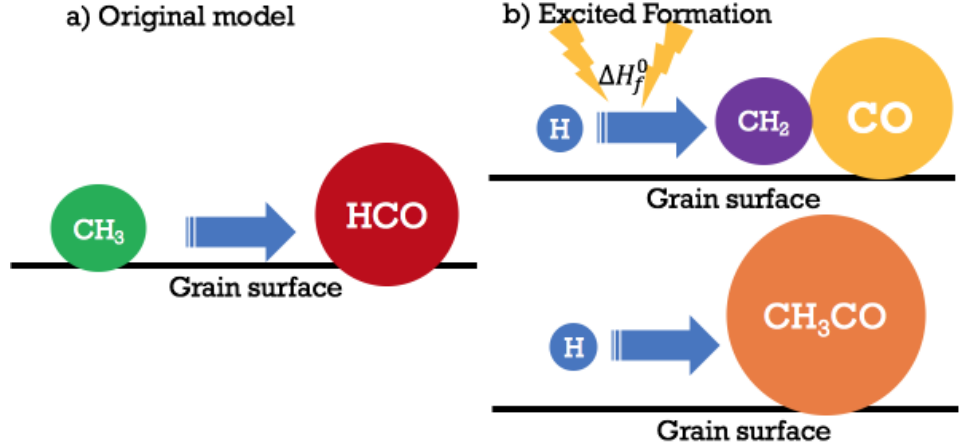}
\caption{Illustration of alternative mechanisms for acetaldehyde formation: (a) the regular diffusive grain-surface reaction between radicals CH$_3$ and HCO, and (b) a postulated three-body excited-formation mechanism involving H, CH$_2$ and CO, followed by a regular diffusive reaction between the radical product, CH$_3$CO, and another H atom. In case (a), reaction is slow at low temperatures. In case (b), a rapid initiating reaction between H and CH$_2$, with exothermicity $\Delta H_{f}^{0}$ = 4.80 eV, provides enough energy to the product CH$_3$ that it may immediately overcome the barrier to reaction with a neighboring CO molecule. This produces a precursor to acetaldehyde, CH$_3$CO, that can easily be hydrogenated by a mobile H atom to form \actd.}
\label{EF_Diagram}
\end{figure}

Besides the three-body reaction process described in \S 2.4, we also consider a mechanism whereby the initiating reaction produces a product that is sufficiently excited that it is able to overcome the activation energy barrier to a subsequent reaction. This is of particular interest if it may allow a reaction with either CO or H$_2$CO -- both abundant surface species -- that would result in the production of a precursor to an important O-bearing COM. In this picture, the energy of formation released by a reaction is held in the vibrational excitation of the product species. That excited species can then immediately react with a contiguous reaction partner.

Figure~\ref{EF_Diagram} shows the formation of \actd\ via this three-body excited formation (3-BEF) mechanism as an example. The original reaction network included the direct association of $\ce{CH3}$ and HCO, mediated by radical diffusion, as the main formation process for surface \actd. The chance to form the \actd\ in that purely diffusive model is small, because it would require immobile heavy radicals to meet at low temperature. The new 3-body process described above, as well as the photodissociation-induced and Eley-Rideal processes, would allow this reaction to occur non-diffusively. 
However, the excited production of CH$_3$ could also allow reaction with abundant surface CO. In the first step, an H atom meets and then reacts with a $\ce{CH2}$ radical that is adjacent to a CO molecule. This reaction is exothermic by 4.80 eV ($\sim$55,700~K), sufficient to overcome the barrier to the CH$_3$ + CO reaction (nominally 2,870~K, see below). Once this follow-up reaction has occurred, the product CH$_3$CO, which is a precursor to acetaldehyde, can easily be converted into a stable species via hydrogenation by another H atom. The entire process is described as follows:
\begin{eqnarray}
\rm H~+~CH_{2} & \rightarrow & \rm CH_{3}^{*}~  \nonumber \\
\rm CH_{3}^{*}~+~CO & \xrightarrow{\text{E$_{\mathrm{A}}$}} & \rm CH_{3}CO~ \\
\rm H~+~CH_{3}CO &\rightarrow & \rm CH_{3}CHO~   \nonumber
\end{eqnarray}

\noindent where an asterisk indicates an excited species. Similar reactions for the production of $\ce{CH3OCH3}$ and $\ce{CH3OCHO}$ through the 3-BEF process are as follows: 
\begin{eqnarray}
\rm H~+~CH_{2} & \rightarrow & \rm CH_{3}^{*} \nonumber \\
\rm CH_{3}^{*}~+~H_{2}CO & \xrightarrow{\text{E$_{\mathrm{A}}$}} & \rm CH_{3}OCH_{2} \\
\rm H~+~CH_{3}OCH_{2} &\rightarrow & \rm CH_{3}OCH_{3} \nonumber 
\end{eqnarray}

\begin{eqnarray}
\rm H~+~H_{2}CO &\rightarrow & \rm CH_{3}O^{*} \nonumber \\
\rm CH_{3}O^{*}~+~CO & \xrightarrow{\text{E$_{\mathrm{A}}$}} & \rm CH_{3}OCO \\
\rm H~+~CH_{3}OCO &\rightarrow & \rm CH_{3}OCHO \nonumber
\end{eqnarray}

The 3-BEF process technically concerns only the first two reactions out of the three, in each case; the final hydrogen-addition step most typically occurs through the usual Langmuir-Hinshelwood mechanism that is already included in the model, although non-diffusive mechanisms may also act to add the final H atom.

Due to the more complicated requirement to consider the energy of formation in each case, the three new 3-BEF processes shown above were individually coded into the model, rather than constructing a generic mechanism. For this reason, the 3-BEF mechanism is included only in the first round of three-body processes. The production rate of the standard diffusive process for the initiating reaction in each case is responsible for the entire value of $R_{\mathrm{app}}(i)$, and only one term is required in Eq. (6). Crucially, the reaction efficiency for the second reaction in the process (i.e. the reaction whose rate is actually being calculated with the 3-BEF method) is initially set to unity, to signify that the activation energy barrier is immediately overcome.

Unfortunately, the activation energies of the above reactions between the radicals and CO or H$_2$CO are not well constrained. The chemical network of \citet{garrod13} included the CH$_3$ + CO and CH$_3$O + CO reactions, adopting a generic activation energy barrier of 2,500~K, based on the approximate value for the equivalent reactions of atomic H with CO and H$_2$CO. A reaction between CH$_3$ and H$_2$CO was also present in that network, with products CH$_4$ and HCO, and $E_{A}$=4440~K; this reaction is retained here in addition to the new pathway.

For the present network, we calculate an approximate activation energy of 2,870~K for the CH$_3$ + CO reaction using the Evans-Polanyi (E-P) relation \citep[e.g.][]{deanandbozzelli00}; this would be well below the energy produced by the initiating reaction ($\sim$55,700~K). Due to the lack of comparable reactions for a similar E-P estimate for the activation energy of the CH$_3$ + H$_2$CO reaction, the same value of 2,870~K might be assumed, placing it also comfortably less than the energy produced by the H + CH$_2$ $\rightarrow$ CH$_3$ reaction. Few determinations exist for the activation energy of the CH$_3$O + CO reaction, although an experiment places it at 3,967~K \citep[][for temperatures 300-2500~K]{huynh08}. This also is less than the energy produced by the initiating reaction, H + H$_2$CO $\rightarrow$ CH$_3$O ($\sim$10,200~K). In any case, the activation energies involved in each of the three reactions mentioned here are sufficiently large that they should be of no importance without the inclusion of the 3-BEF mechanism to provide the energy required, while the 3-BEF mechanism itself is assumed to go at maximum efficiency. However, the latter assumption may not necessarily be accurate, depending on the form of the energy released by the reaction, and whether there is any substantial loss prior to reaction actually occurring (see \S \ref{3bef-best}).

\begin{deluxetable}{llllll}
\tablewidth{0pt}
\tabletypesize{\footnotesize}
\tablecolumns{6}
\tablecaption{Non-diffusive mechanisms included in model setups \label{modeltickbox}}
\scriptsize
\tablehead{
\colhead{Model} & \colhead{\rotatebox{90}{Eley-Rideal}} & \colhead{\rotatebox{90}{Photodissociation-Induced}} & \colhead{\rotatebox{90}{Three-body}} & \colhead{\rotatebox{90}{Three-body Excited Formation}} & \colhead{\rotatebox{90}{Adjusted 3-BEF efficiency for MF}}
}
\startdata
Control & & & & & \\
E-R & $\checkmark$ & & & & \\ 
PD-Induced & & $\checkmark$ & & & \\ 
3-B & & & $\checkmark$ & & \\
3-B+3-BEF & & & $\checkmark$ & $\checkmark$ & \\ 
3-BEF Best & & & $\checkmark$ & $\checkmark$ & $\checkmark$ \\
\enddata
\end{deluxetable}

\subsection{Physical conditions}

{\em MAGICKAL} is a single-point model, but a spatially-dependent picture of the chemistry of \object{L1544} can be achieved by running a set of models with different physical conditions at specific positions within the prestellar core. Recently, \citet{chacon-tanarro19} determined the parameterized density and temperature structure of \object{L1544} as follows, considering the optical properties of dust grains as a function of radius:
\begin{equation}
T(r) \, \mathrm{[K]} =12 - \frac{12 - 6.9}{1+\left(\frac{r}{28.07''}\right)^{1.7}} 
\end{equation}
\begin{equation}
n_{\mathrm{H}_2}(r) \, \mathrm{[cm}^{-3}\mathrm{]} =\frac{1.6\times10^6}{1+\left( \frac{r}{17.3''}\right)^{2.6}} 
\end{equation}
where $r$ is measured in arcseconds. Based on this density structure, we determine 15 densities at which the chemical models are to be run, ranging logarithmically from the minimum of $n_\textrm{H}=4.4\times10^{4}~{\rm cm^{-3}}$ ($\sim$11,000 AU) to the maximum of $n_\textrm{H}=3.2\times10^{6}~{\rm cm^{-3}}$ (core center). An additional eight density points are then placed to achieve better spatial resolution toward the core center (where the density profile is relatively flat). The appropriate temperature for each point is then chosen from the profile, based on density/radius.

In order to take account of the gradual collapse of the gas into this final density profile, the density used for each chemical model in the set is independently evolved using a simple modified free-fall collapse treatment. (The radial position of each model point is thus not explicitly considered during this evolution). Each point begins with a gas density of $n_\textrm{H,0}=3\times10^{3}~{\rm cm^{-3}}$, with an initial $\ce{H}/\ce{H2}$ ratio of $5\times10^{-4}$. The density evolution stops once each model reaches its specified final density, resulting in a marginally different evolutionary time for each density point. The collapse treatment is based on that used by \citet{rawlings92}. Magnetic fields can play an important role in the equilibrium of dense cores, significantly slowing down the collapse process. Estimating an accurate collapse timescale is challenging, although the ratio of the ambipolar diffusion time~($\tau_\textrm{ap}$) and the freefall time~($\tau_\textrm{ff}$) is typically assumed to be $\tau_\textrm{ap}/\tau_\textrm{ff}\thicksim10$~\citep[see, e.g.,][]{hennebelle19}. Thus, the magnetic retardation factor for collapse, B, is adopted here to control the collapse timescale. This parameter takes a value between 0 (static) and 1 (free-fall) and is technically density-dependent. In our model, this value is set for simplicity to 0.3 for all density points, which results in a collapse timescale approximately 3 times longer than the free-fall timescale. A time of a little over $3\times10^{6}$ year is therefore required to reach the final density at each point, although much of this time is spent under relatively low-density conditions as the collapse gradually ramps up.

The density evolution for each model is accompanied by increasing visual extinction, which evolves according to the expression $A_\textrm{V}=A_\textrm{V,0}(n_{H}/n_{H,0})^{2/3}$~\citep{garrodandpauly11}. The initial extinction values are set such that the values at the end of the chemical model runs correspond to the linear integration of the density profile, converted to visual extinction using the relationship $N_{H}=1.6 \times 10^{21}A_\textrm{V}$. An additional background visual extinction of 2 is added for all positions and times, under the assumption that \object{L1544} is embedded in a molecular cloud \citep[e.g.][]{vasyunin17}. In contrast to density and visual extinction, the temperature is held steady throughout the chemical evolution for each density point, with the same value adopted for both the gas and the dust. Temperatures range from approximately 8 to 14~K depending on radius, which is consistent with the observational features \citep{crapsi07}.

\section{Results} 

\begin{figure*}
\gridline{\fig{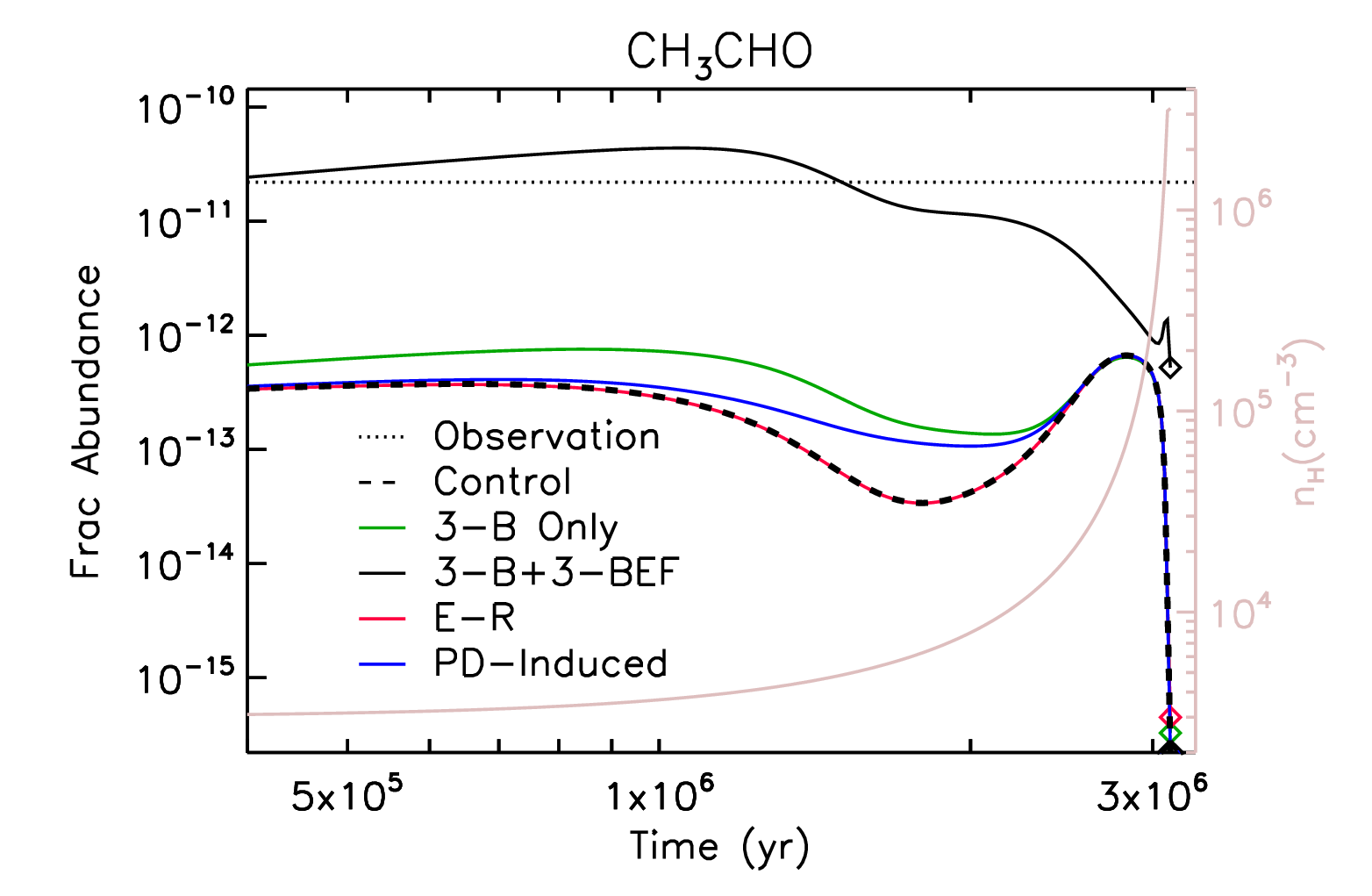}{0.5\textwidth}{}
             \fig{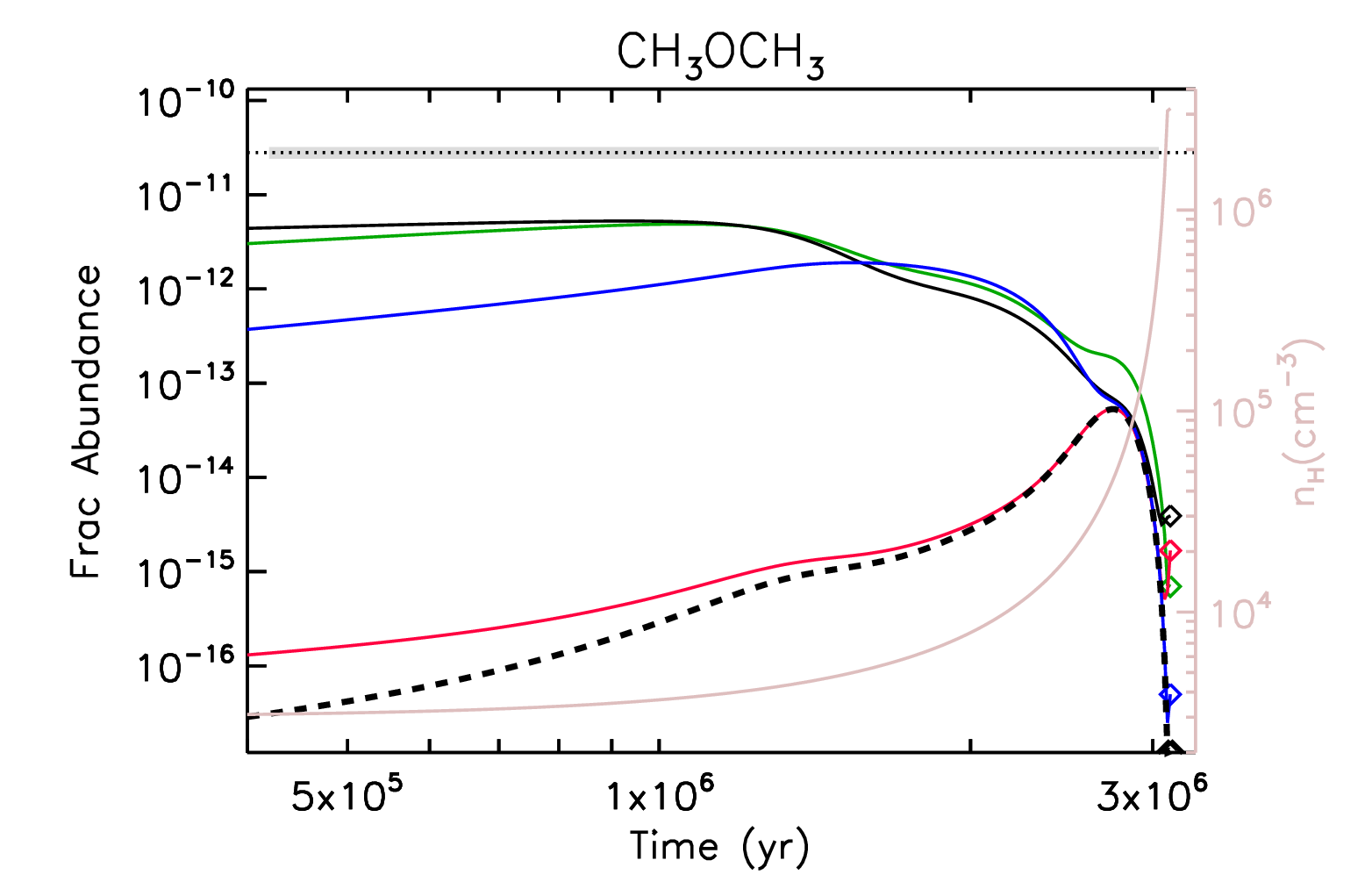}{0.5\textwidth}{}}
\gridline{\fig{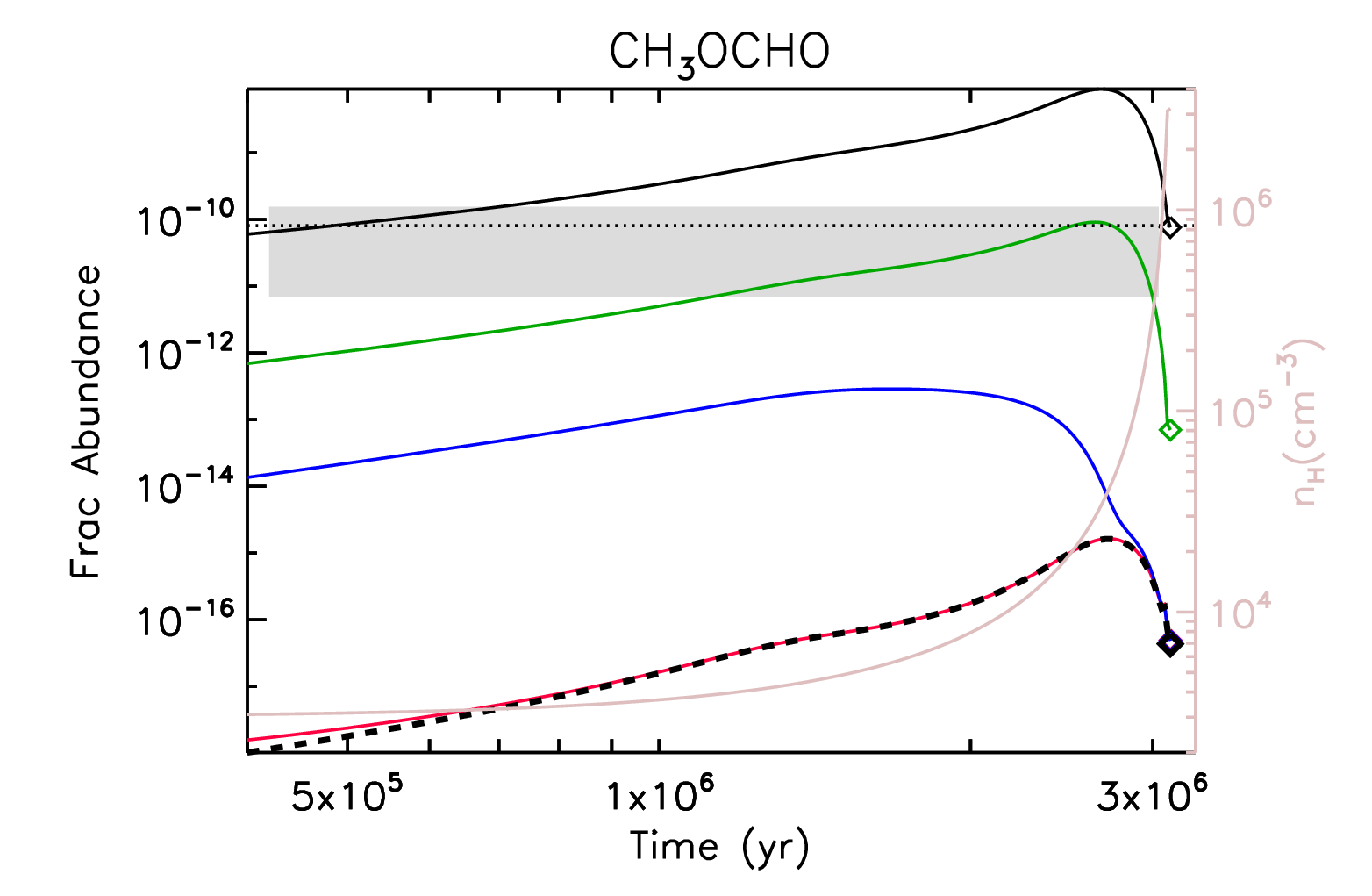}{0.5\textwidth}{}}
\caption{
Time evolution of the gas-phase abundances of the three O-bearing COMs, at the core center, for models using each of the new mechanisms. The abundance from the control model is denoted as a black dashed line. Diamonds indicate the abundances at the end of each model run. The black dot-dash lines and the gray shaded regions indicate the observational abundances and their uncertainties, obtained from \citet{jimenez-serra16} for \object{L1544}. The gas density is indicated by the right-hand vertical axis and the similarly colored line.
\label{timeVSgas-abundance}}
\end{figure*}

\begin{figure*}
\gridline{\fig{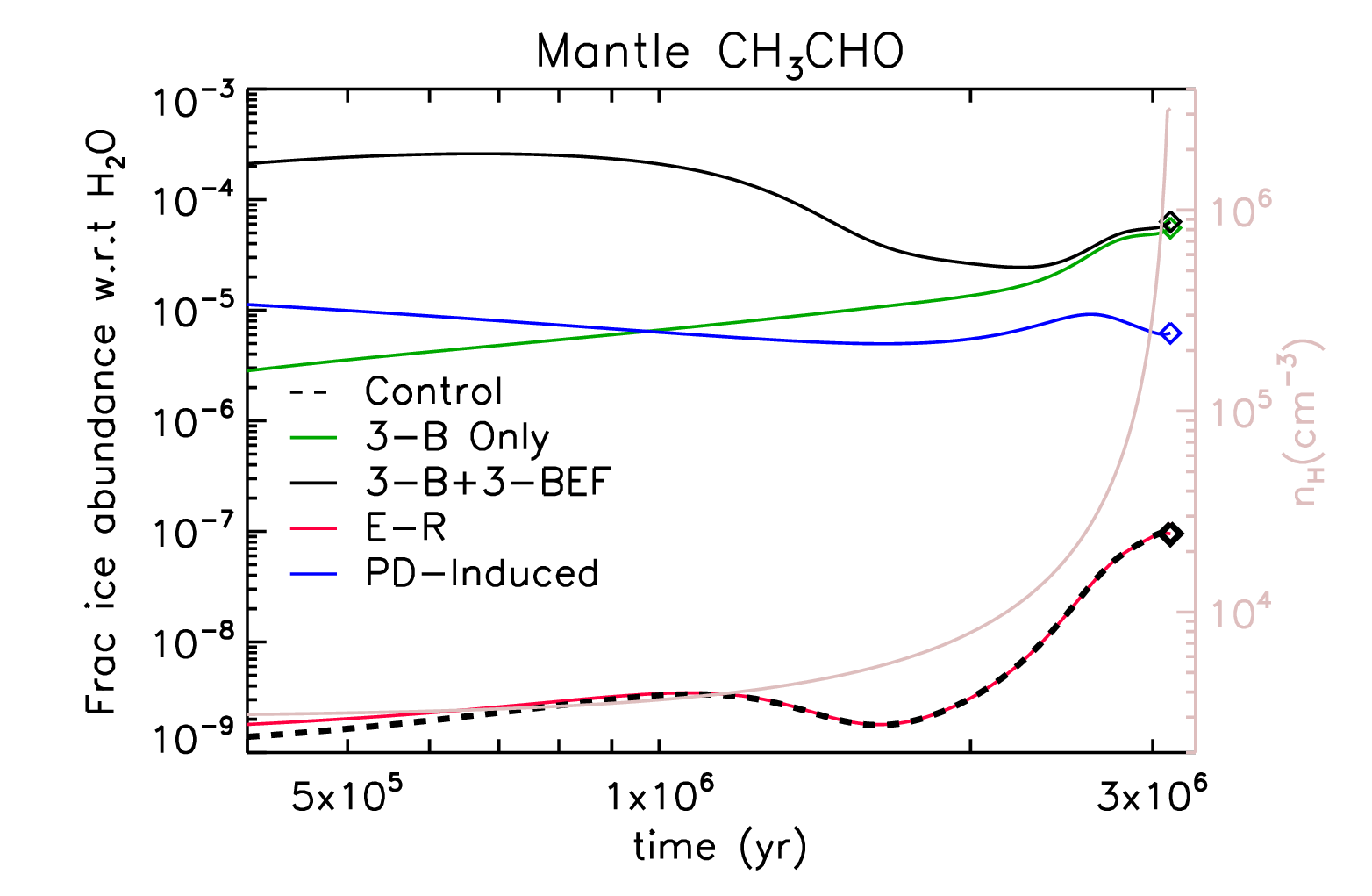}{0.5\textwidth}{}
             \fig{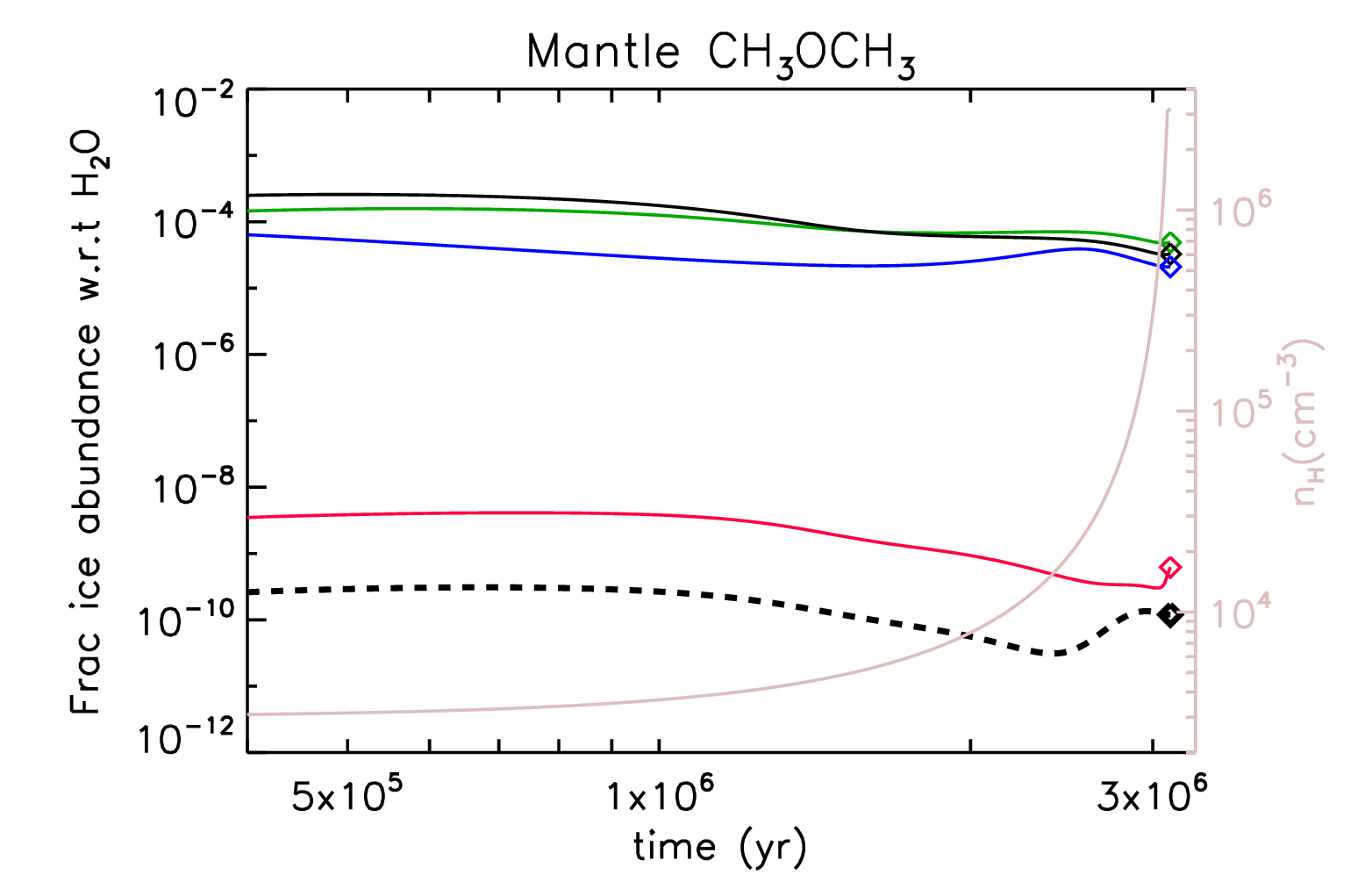}{0.5\textwidth}{}}
\gridline{\fig{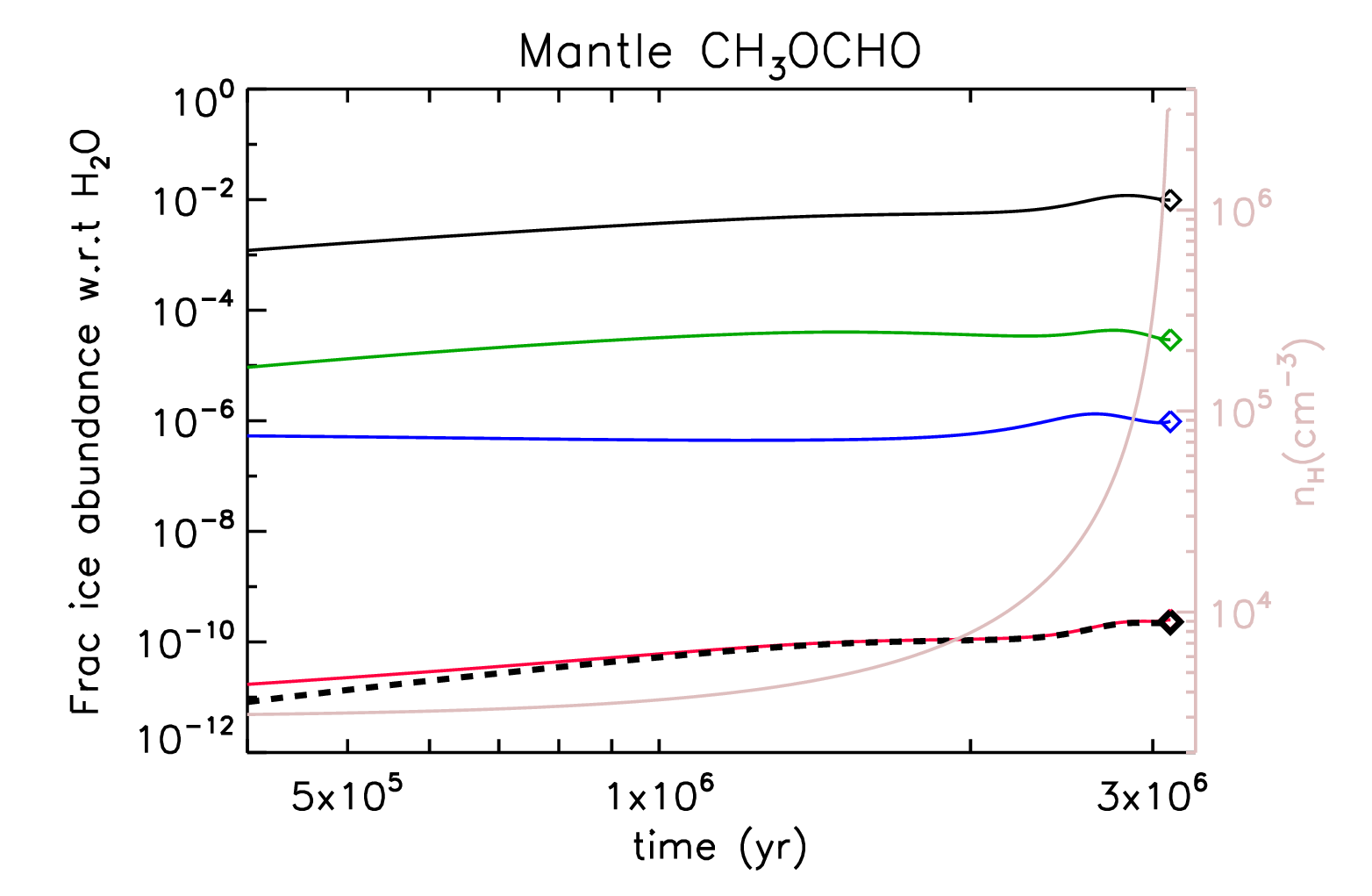}{0.5\textwidth}{}}
\caption{
Time evolution of the abundances in the dust-grain ice mantles of the three O-bearing COMs at the core center, for models using each of the new mechanisms. The abundance from the control model is denoted as a black dashed line. Diamonds indicate the abundances at the end of each model run.
The gas density is included for reference, indicated by the right-hand vertical axis and the similarly colored line.
\label{timeVSmantle-abundance}}
\end{figure*}

The time-evolution of the fractional abundances at the core-center position is presented in Fig.~\ref{timeVSgas-abundance} (gas phase) and Fig.~\ref{timeVSmantle-abundance} (solid phase), for each of the main chemical model setups. In the control model, no new mechanisms are added. In each of the other model setups, a single new mechanism is added to the control model setup, except for model 3-B+3-BEF, in which it is assumed that the 3-BEF mechanism could not occur without the 3-B mechanism also being active.

As seen in Fig.~\ref{timeVSgas-abundance}, every new mechanism introduced here, excluding E-R, significantly increases the abundances of $\ce{CH3OCH3}$ and $\ce{CH3OCHO}$ in the gas phase during core evolution, while $\ce{CH3CHO}$ is only substantially increased via 3-BEF. However, it should be noted that the increased fractional abundances rapidly drop as density increases toward the end-time of all the models, mostly converging to the control-model values. This indicates that the new mechanisms may hardly affect the gas-phase COMs at the core center, but may be more effective at more distant radii (i.e. lower density regions); this would nevertheless result in higher abundances toward the core-center position when averaged over the line of sight to include lower-density gas.

The presence of the COMs in the gas phase following their formation on grain surfaces is the result of chemical desorption. All surface reactions that form a single product have a small possibility of returning that product to the gas phase. The upper limit on the ejection probability per reaction is 1\%.

Similar to the gas phase, every mechanism excluding E-R significantly increases the solid-phase populations of the COMs (Fig.~\ref{timeVSmantle-abundance}). Note that the solid-phase population of $\ce{CH3CHO}$, whose gas-phase abundance is only strongly increased by the 3-BEF mechanism, increases even in the 3-B (only) and PD-Induced models. The 3-B and 3-B+3-BEF models converge to essentially the same value at the end-time. Dimethyl ether in the mantle is produced in similar quantities by each of the three effective mechanisms, while 3-BEF and then 3-B are more important than PD-Induced formation in the case of methyl formate. The E-R mechanism produces only marginal increases in mantle abundances of acetaldehyde and methyl formate. The increase in dimethyl ether production caused by E-R is around an order of magnitude throughout most of the evolution, although this is dwarfed by the effects of the other mechanisms.

\begin{figure*}
\gridline{\fig{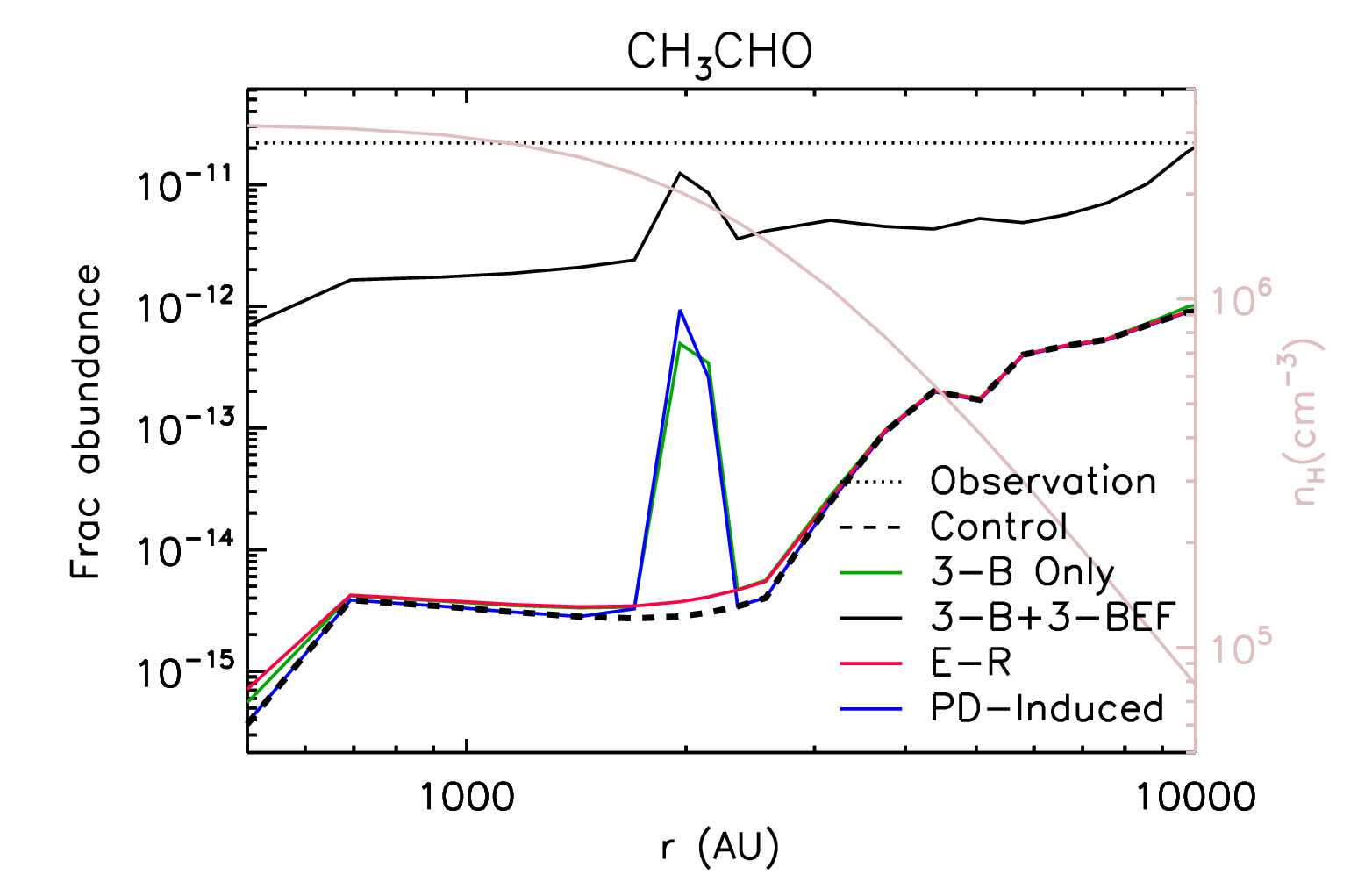}{0.5\textwidth}{}
             \fig{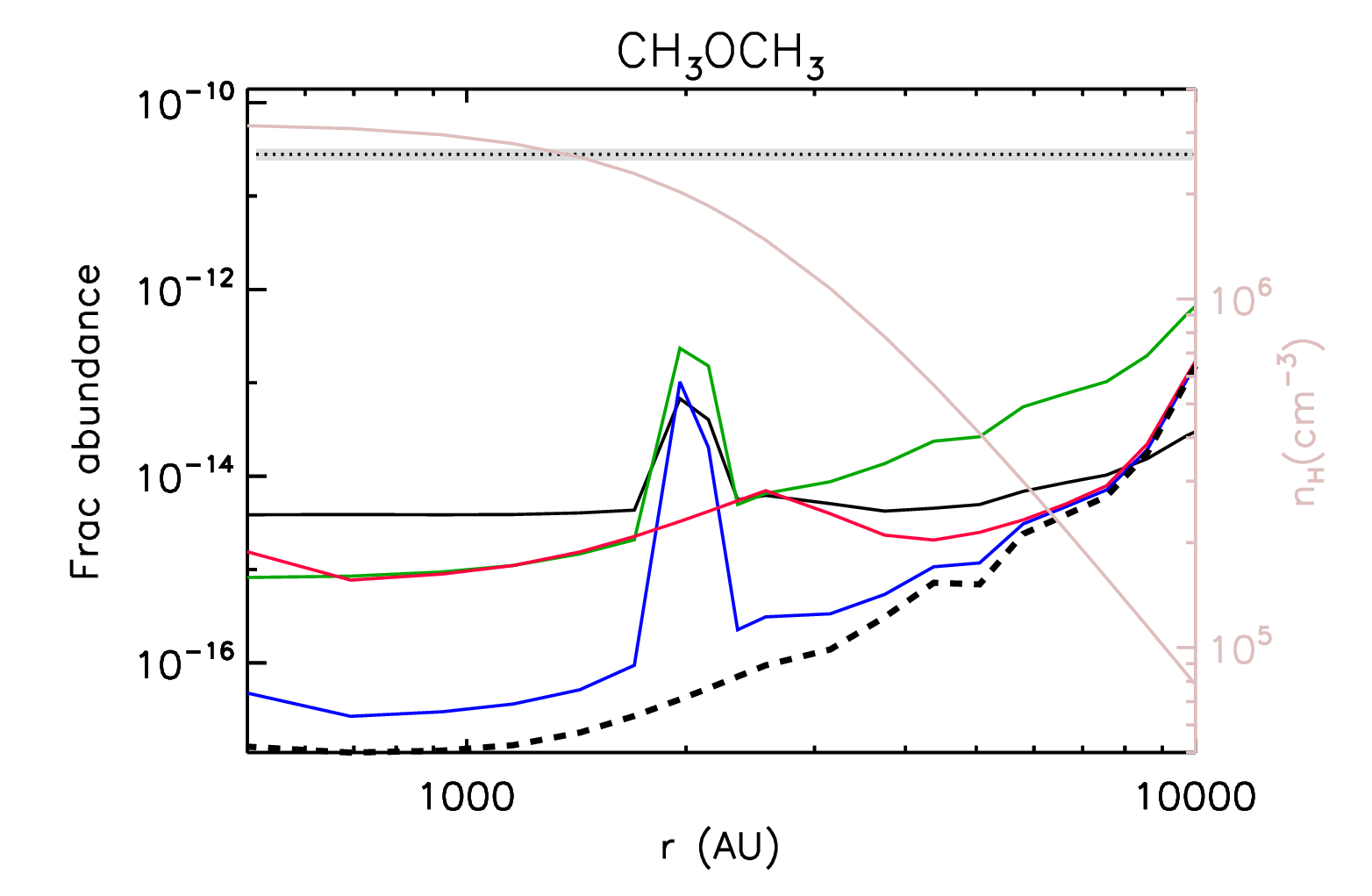}{0.5\textwidth}{}}
\gridline{\fig{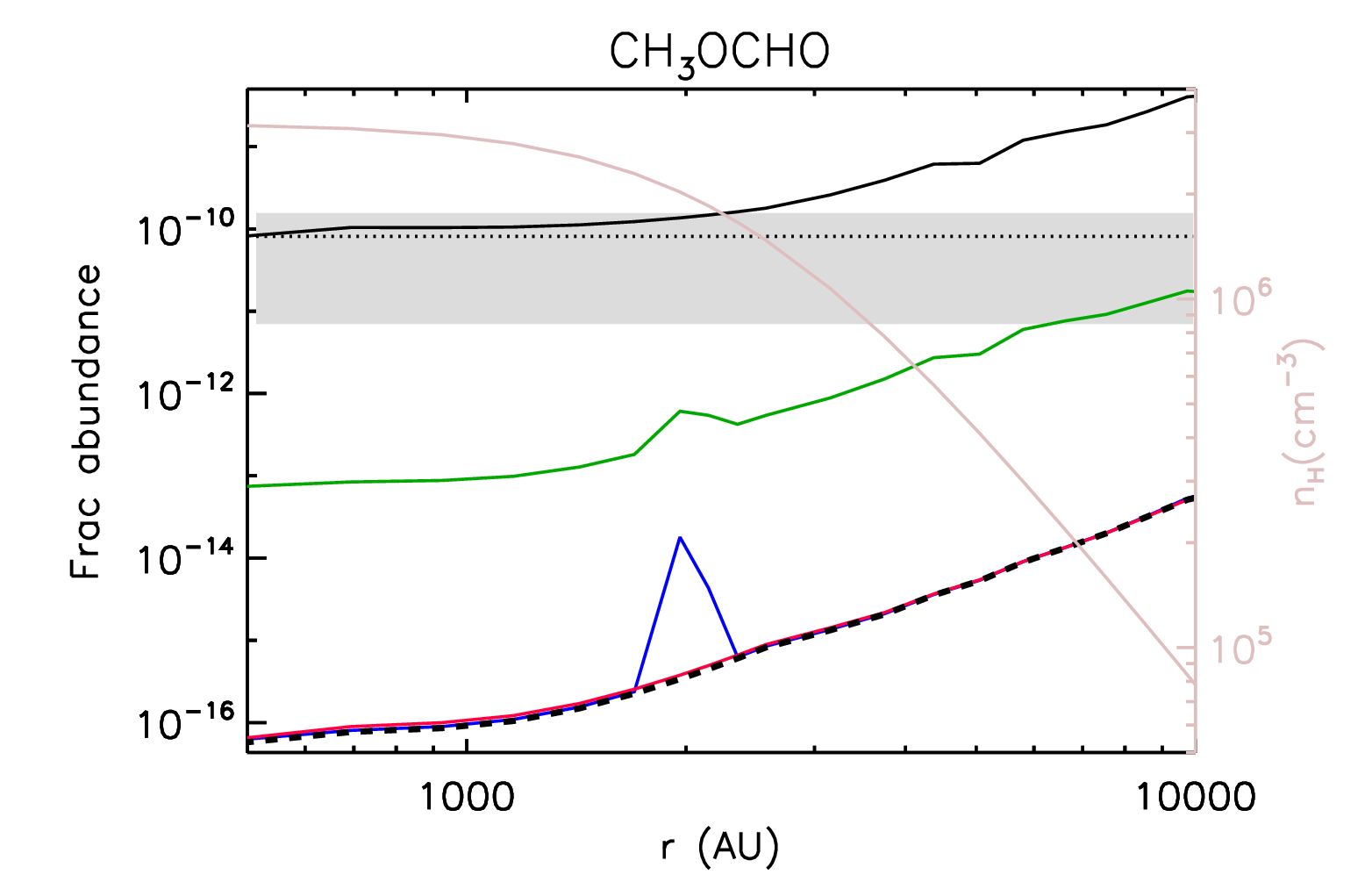}{0.5\textwidth}{}}
\caption{
Radial distribution of the gas-phase abundances of the three O-bearing COMs depending on the mechanisms. 
The black dotted line and grey shaded region indicates the abundance from the observation and error, respectively. The observational data and its error is referred from the \citet{jimenez-serra16}. The observational error of the $\ce{CH3CHO}$ is not provided in the reference. The abundance from the control model is denoted as a black dashed line and a black dotted line, respectively. While the result from the 3-B combined with the 3-BEF is denoted as a black solid line, a green line represent the 3-B only model. The red and blue solid line indicates the abundances from the E-R model and PD-Induced model, respectively.
\label{radiVSgas-abundance}}
\end{figure*}

\begin{figure*}
\gridline{\fig{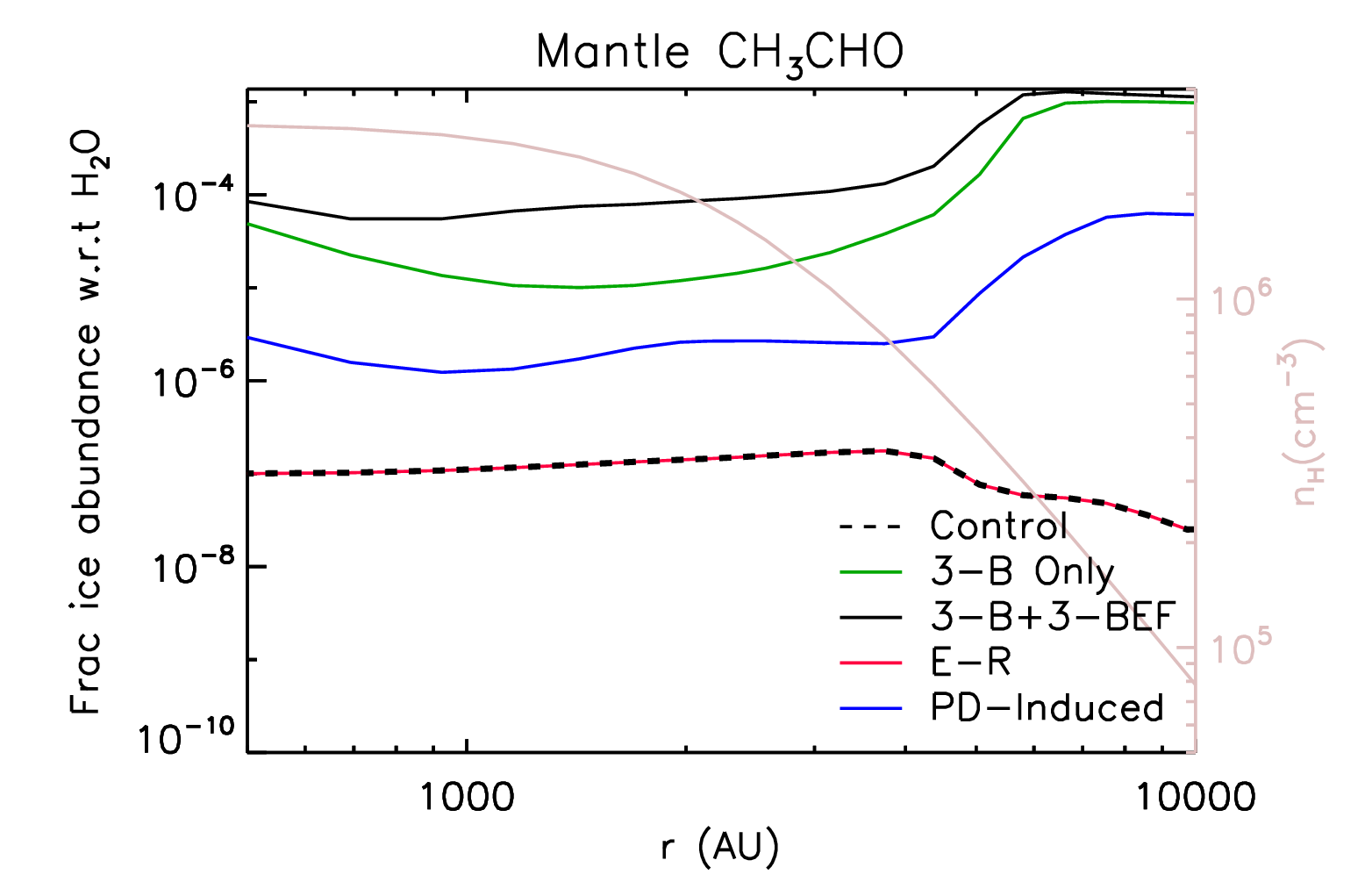}{0.5\textwidth}{}
          \fig{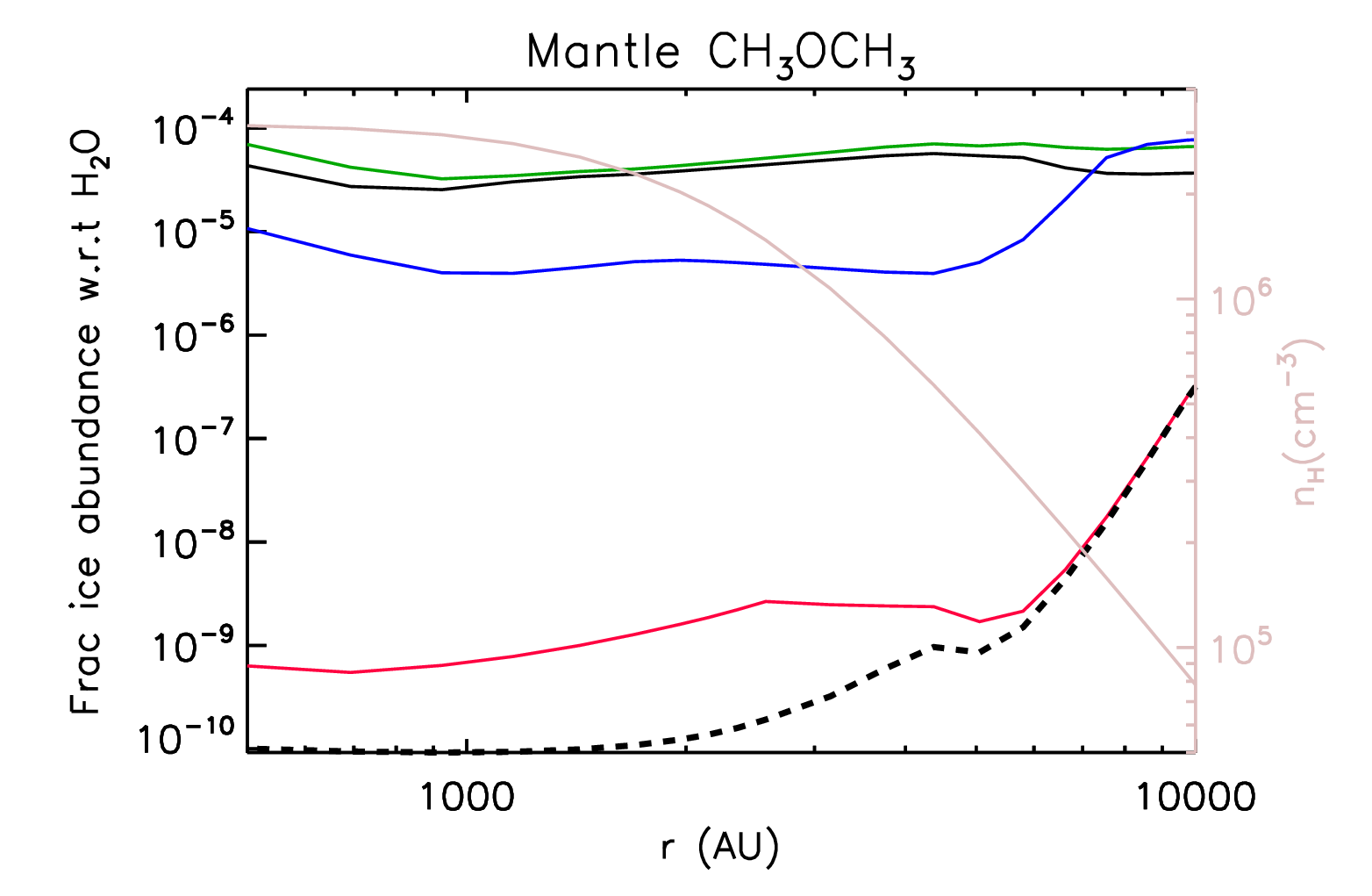}{0.5\textwidth}{}}
\gridline{\fig{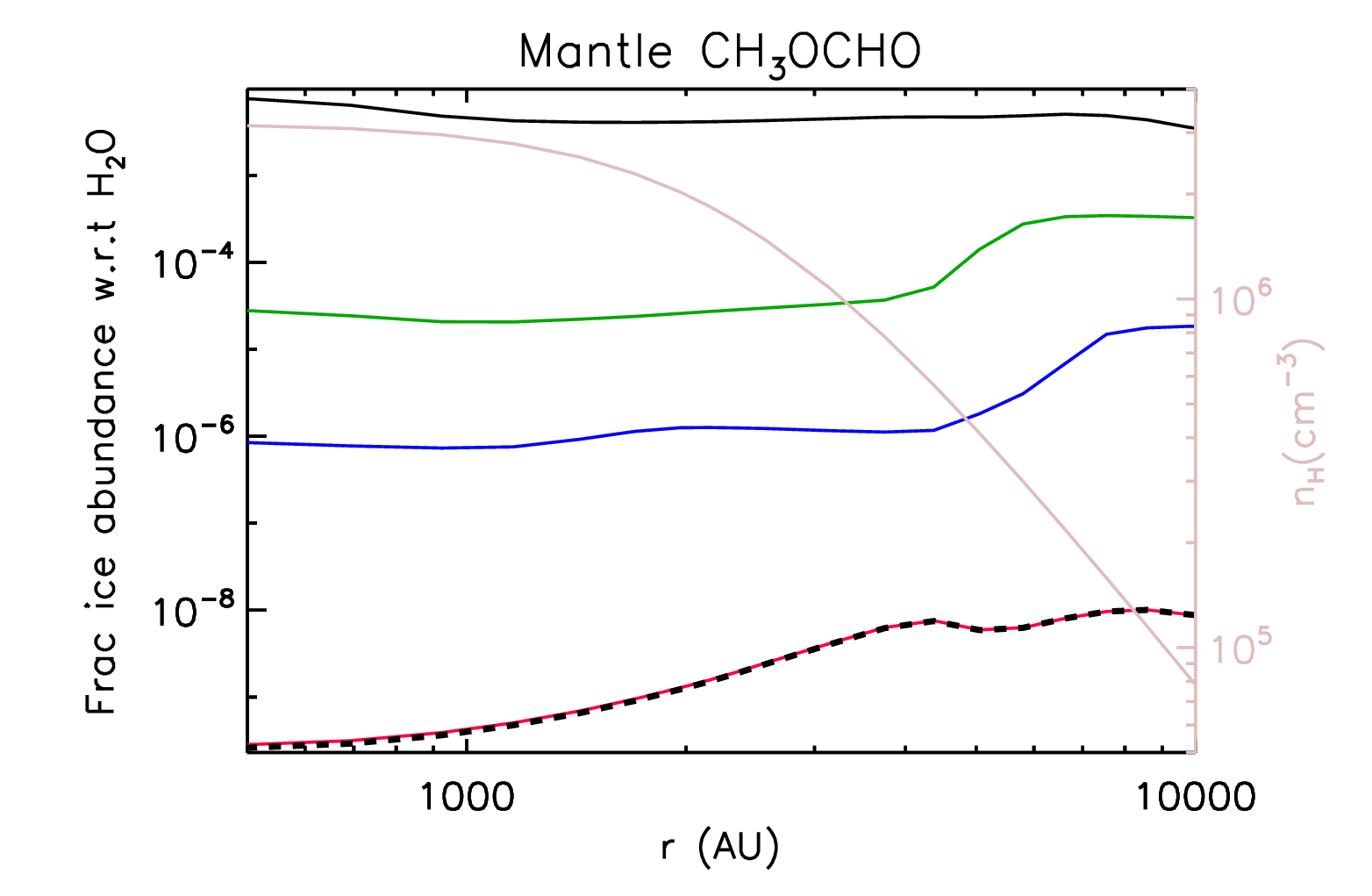}{0.5\textwidth}{}}
\caption{
Radial distribution of the solid-phase abundances of the three O-bearing COMs depending on the mechanisms. 
The black dotted line and grey shaded region indicates the abundance from the observation and error, respectively. The observational data and its error is referred from the \citet{jimenez-serra16}. The observational error of the $\ce{CH3CHO}$ is not provided in the reference. The abundance from the control model is denoted as a black dashed line and a black dotted line, respectively. While the result from the 3-B combined with the 3-BEF is denoted as a black solid line, a green line represent the 3-B only model. The red and blue solid line indicates the abundances from the E-R model and PD-Induced model, respectively.
\label{radiVSsolid-abundance}}
\end{figure*}

Fig.~\ref{radiVSgas-abundance} shows the radial distribution of gas-phase COM abundances using the full radius--density--temperature profile model results; abundances shown correspond to the end-time abundances, at which the final density profile is achieved. The observational values that are also indicated in the figure are for the core-center position; however, those observations correspond to a beam of radius $\thicksim$1900 AU, and would also sample a range of physical conditions along the line of sight -- some caution should therefore be taken in directly comparing them with the local fractional abundance values.

It may be seen that the general trend, even for the control model, is for COM abundances to increase toward greater radii.
The 3-B+3-BEF model produces maximum molecular abundances for acetaldehyde and methyl formate similar to the observational values. For the latter molecule, the modeled fractional abundance exceeds the observational values at radii greater than around 2500 AU, although the absolute gas density begins to fall off at these positions, so that they should contribute less to the total column density of the molecule. At the largest radii modeled, the 3-B (only) model produces methyl formate sufficient to match the observed abundance (although, again, perhaps with little contribution to total column density). Acetaldehyde also reaches its peak abundance at large radii, although it reaches a similar abundance at smaller radii.

For each of the new models, the local fractional abundance of $\ce{CH3OCH3}$ is greatest at positions away from the core center, but a significant increase in abundance is achieved at almost all positions for every model, versus the control. However, the maximum value achieved (for the 3-B model) is still at least two orders of magnitude lower than the observations, both in the inner regions and at the outer edge. Curiously, for dimethyl ether, the most effective model is the 3-B (only) model, whereas the 3-B+3-BEF model is the most productive for the other two COMs. 

In each of the 3-B, 3-B+3-BEF and PD-Induced models, acetaldehyde and dimethyl ether abundances show a prominent peak feature at around 2000 AU. This feature is also present for methyl formate in the PD-Induced and 3-B (only) models. Observations of \object{L1544} by \citet{jimenez-serra16} show higher fractional abundances of $\ce{CH3CHO}$ and $\ce{CH3OCH3}$ toward an off-center position at $r\simeq $4000 AU, versus those at the core-center, with $\ce{CH3OCHO}$ arguably showing similar behavior. Our 3-B, 3-B+3-BEF and PD-Induced models all show this general behavior (methyl formate in the 3-B+3-BEF model notwithstanding), albeit at a somewhat different radius from the observations. The origin of this peak and its similarity to the observations are discussed in more detail in \S~\ref{COMdistribution_3b-best}.

Figure~\ref{radiVSsolid-abundance} shows the radial distribution of the ice-mantle abundances at the end-time of each model, plotted as a function of the water abundance in the ice at each position. 
In contrast with the gas-phase, all the new mechanisms but E-R significantly increase the solid-phase abundances of COMs at all radii. This is partly because the ice mantle preserves the earlier surface layers during the evolution of the prestellar core, when significant enhancement of the COM abundances in the gas-phase is found (see Fig.~\ref{timeVSgas-abundance}), which is itself caused by increased efficiency in the production of COMs on the grain surfaces. However, the PD-Induced model permits substantial ongoing processing of mantle material itself.

While COM production is not especially important in the control or E-R models, the others attain substantial COM abundances in the ices, comparable with gas-phase values observed in {\em hot} molecular cores. The maximum abundance achieved by methyl formate in the 3-B+3-BEF model is close to 1\% of water abundance at the core center, i.e around $10^{-6}$ with respect to total hydrogen. This value may thus be {\em too} high to agree with observations of hot cores/corinos, if the abundances achieved in the prestellar stage should be preserved intact to the later, warmer stages of evolution.

It is also noteworthy that the COM fraction in the ices is in general somewhat greater at larger radii, although the dimethyl ether abundance is fairly stable through the core in the 3-B and 3-B+3-BEF models, and methyl formate is also stable throughout in the 3-B+3-BEF model.

\begin{figure*}
\gridline{\fig{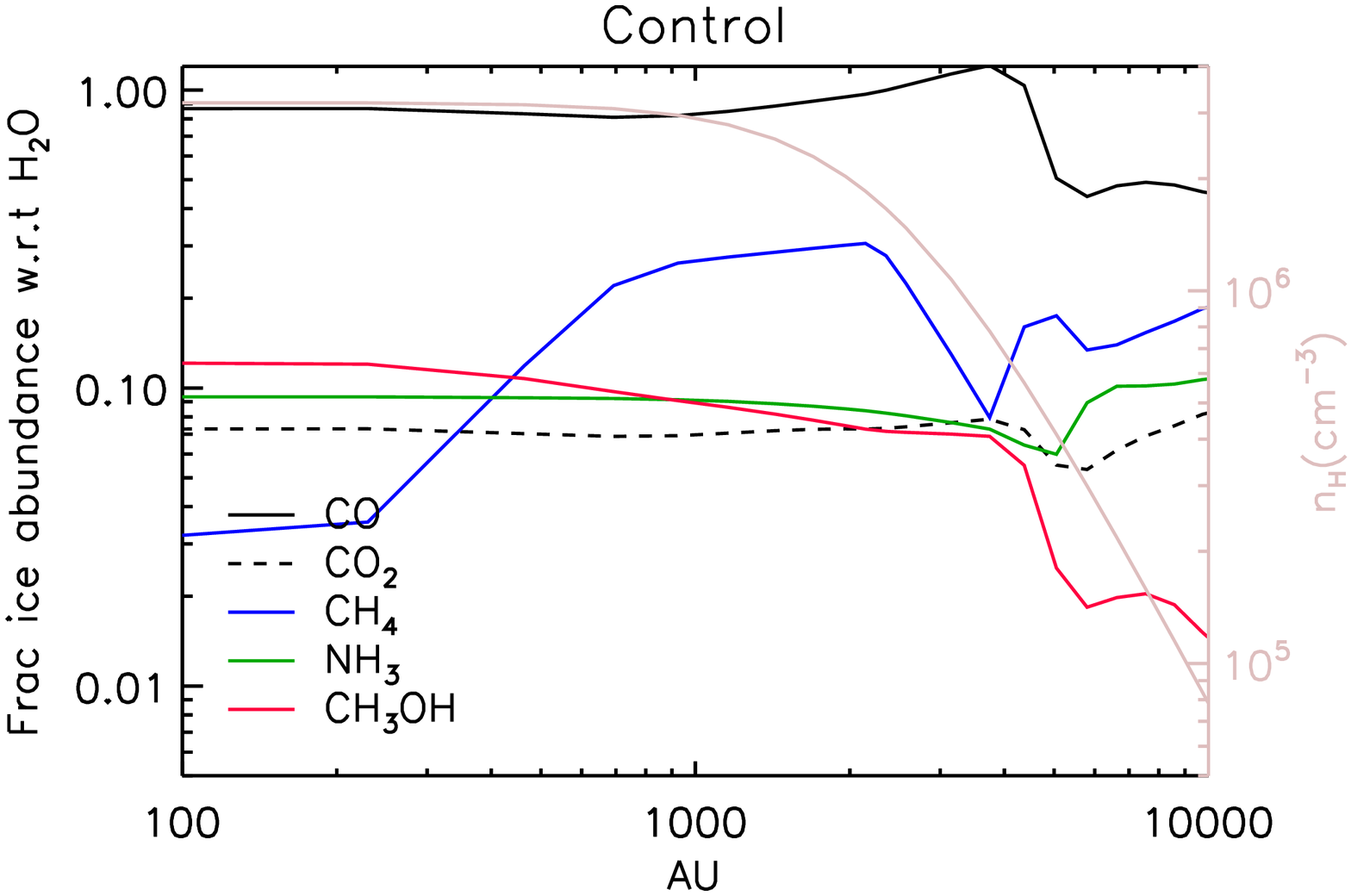}{0.5\textwidth}{}
          \fig{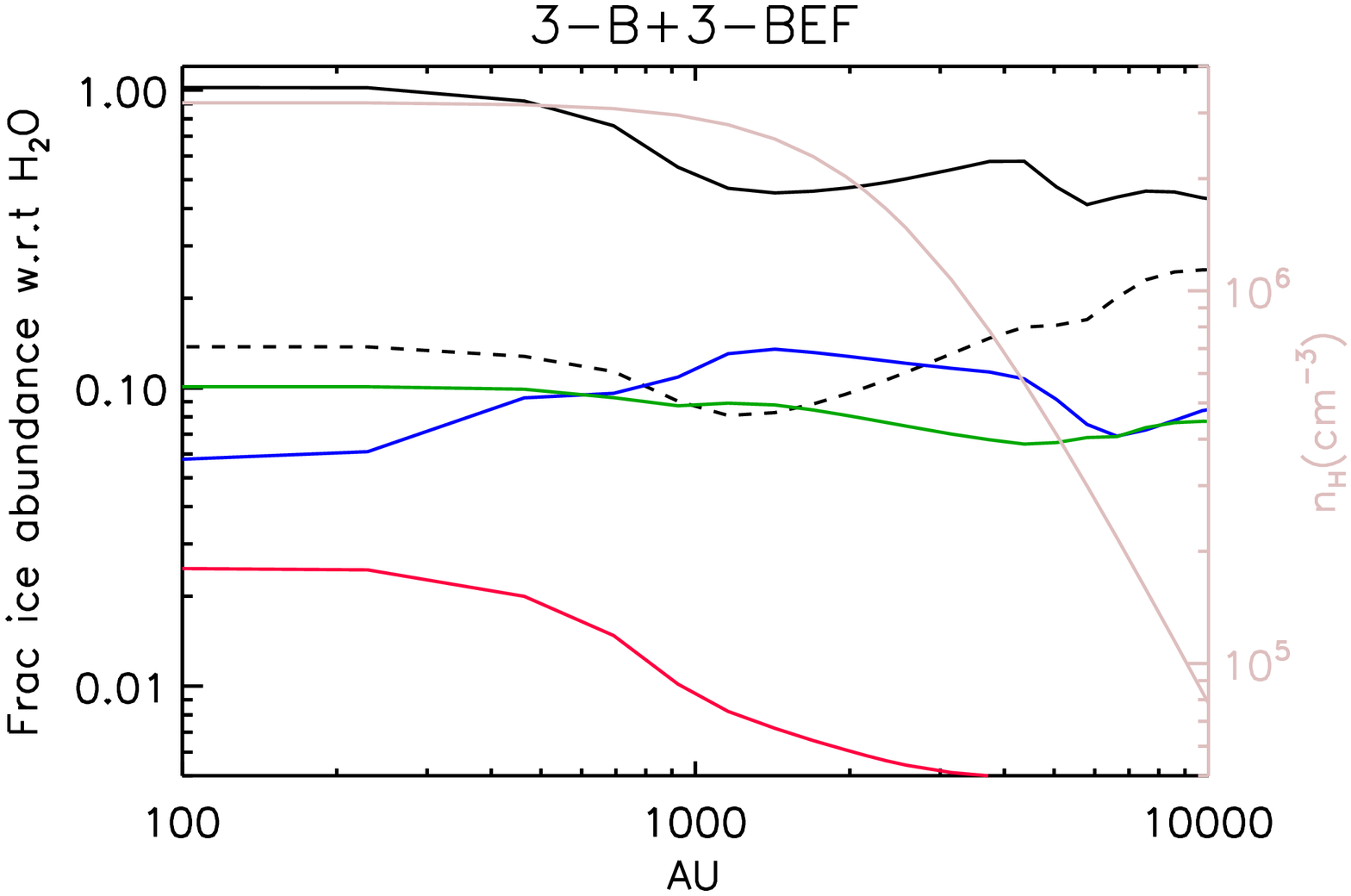}{0.5\textwidth}{}}
\caption{Radial distribution of the ice abundances of the main ice constituents. The abundance of CO and $\ce{CO2}$ is denoted as a black solid and a black dashed line, respectively. A blue, green, and red solid line represents the abundances of $\ce{CH4}$, $\ce{NH3}$, and $\ce{CH3OH}$, respectively.
\label{main_ice_rVSXi}}
\end{figure*}

Figure~\ref{main_ice_rVSXi} shows the final radial distribution of the main ice constituents as a fraction of the local water ice abundance, for the control model and for the 3-B+3-BEF model. The absolute abundance of profile of water ice is fairly constant across the profiles. The latter is taken as representative of all the new models, due to their similarity, except for the E-R model, which is rather similar to the control. Based on the absolute abundance profiles, column densities for each ice species are derived by integrating along the line of sight (without beam convolution); the resulting abundances with respect to $\ce{H2O}$ ice column density are summarized in Table~\ref{main_ice_colden}. Comparable observational ice abundances are also shown, taken from \citet{boogert15}, who provided median values with respect to $\ce{H2O}$, along with the full range of the observed abundances (from subscript to superscript value). Both of our model setups produce a centrally-peaked distribution of $\ce{CH3OH}$ ice, while CO ice is approximately as abundant as $\ce{H2O}$ ice, especially toward inner radii where the most extreme depletion occurs. %The abundance of $\ce{CO2}$ ice in the control model is somewhat lower than the median value, but is still within the range of observations for the 3-B+3-BEF model. 
With the new mechanisms, a more gently-sloped distribution appears, and a better match with observational abundances of CO and $\ce{CO2}$ is achieved. The other ice components in the 3-B+3-BEF model are within the observational range as well.

\begin{deluxetable}{cccccc}
\tablewidth{0pt}
\tabletypesize{\footnotesize}
\tablecolumns{6}
\tablecaption{Abundances relative to the $\ce{H2O}$ ice column density. \label{main_ice_colden}}
\scriptsize
\tablehead{
\colhead{Model} & \colhead{$\ce{CO}$} & \colhead{$\ce{CO2}$} & \colhead{$\ce{CH4}$} & \colhead{$\ce{NH3}$} & \colhead{$\ce{CH3OH}$}}
\startdata
Observation \tablenotemark{a} &  $0.21^{0.85}_{(<0.03)}$ & $0.28^{0.50}_{0.12}$ & $0.05^{0.11}_{0.01}$ & $0.06^{0.10}_{0.03}$ & $0.06^{0.25}_{(<0.01)}$\\
Control     &  0.83 & 0.07  &  0.19 & 0.09 & 0.07 \\
%PD-Induced  &  0.52 & 0.16  & 0.09  & 0.08 & 0.05 \\
%3-B  &  0.49 & 0.11  & 0.13  & 0.08 & 0.04 \\
3-B+3-BEF & 0.57 & 0.12 & 0.11 & 0.08 & 0.01 \\
\enddata
\tablenotetext{a}{\citet{boogert15}; values correspond to low-mass YSOs.}
\end{deluxetable}

\subsection{Column density analysis}

Observational abundances may not accurately represent the true local abundances within a source. This is because the observational intensities are not only averaged over the line of sight, but are also affected by the excitation characteristics of each observed species, and by the response of the telescope beam. Considering this, it is indispensable to perform spectral simulations for better comparison with the observations. The spectral model used here simulates molecular lines (of COMs) that are expected to be observable, and uses chemical abundances shown in Fig.~\ref{radiVSgas-abundance} as the underlying distribution. The 1-D chemical/physical model is treated as spherically symmetric, so that molecular emission can be simulated along lines of sight passing through the core at various offsets (including directly on-source), assuming local thermodynamic equilibrium. Each line of sight passes through a range of gas densities, temperatures and chemical abundances. The resulting 2-D simulated intensity maps for each frequency channel are then convolved with a Gaussian telescope beam of appropriate size, dependent on frequency and the telescope in question. \cite[For a more detailed description of the spectral model, see][]{garrod13}. The FWHM of the molecular lines is assumed here to be 1~km/s, with a spectral resolution of 250 kHz, although the simulations are quite insensitive to the precise choice of parameters.

The integrated intensities of the ensemble of molecular lines is used in a rotational diagram analysis \citep{goldsmithandlanger99} to obtain column densities ($N_\textrm{tot}$) and rotational temperatures ($T_\textrm{rot}$) for each molecule. These quantities can then be compared directly with those obtained from observations. Beam sizes were assumed to be $\thicksim$28$''$-31$''$ between 79-87~GHz and $\thicksim$24$''$-26$''$ between 94-103~GHz, based on the size of the observing beam of the IRAM 30~m telescope. The distance to the model prestellar core is assumed to be 140pc~\citep{elias78}.

\begin{deluxetable}{ccccc}
\tablewidth{0pt}
\tabletypesize{\footnotesize}
\tablecolumns{5}
\tablecaption{List of targeted transitions with their spectroscopic properties\label{COMlines}}
\scriptsize
\tablehead{
\colhead{Molecule} & \colhead{Transition} & \colhead{Frequency (GHz)} & \colhead{$E_\textrm{up}$ (K)} & \colhead{$A_\textrm{ij}$ (s$^-1$)}
}
\startdata
A-$\ce{CH3CHO}$ & 2$_{1,2}$-1$_{0,1}$ & 84.21976 & 4.96 & $2.4 \times 10^{-6}$ \\
E-$\ce{CH3CHO}$ & 5$_{0,5}$-4$_{0,4}$ & 95.94744 & 13.9 & $3.0 \times 10^{-5}$ \\
A-$\ce{CH3CHO}$ & 5$_{0,5}$-4$_{0,4}$ & 95.96346 & 13.8 & $3.0 \times 10^{-5}$ \\
E-$\ce{CH3CHO}$ & 5$_{1,4}$-4$_{1,3}$ & 98.86331 & 16.6 & $3.0 \times 10^{-5}$ \\ 
$^{*}$A-$\ce{CH3CHO}$ & 5$_{1,4}$-4$_{1,3}$ & 98.90094 & 16.5 & $3.0 \times 10^{-5}$ \\
\hline
AA-$\ce{CH3OCH3}$ & 3$_{2,1}$-3$_{1,2}$ & 84.63680 & 11.0 & $4.4 \times 10^{-6}$ \\
AA-$\ce{CH3OCH3}$ & 2$_{2,0}$-2$_{1,1}$ & 86.22872 & 8.36 & $3.5 \times 10^{-6}$ \\
AA-$\ce{CH3OCH3}$ & 2$_{2,1}$-2$_{1,2}$ & 89.70281 & 8.36 & $3.8 \times 10^{-6}$ \\
AA-$\ce{CH3OCH3}$ & 6$_{0,6}$-5$_{1,5}$ & 90.93754 & 19.0 & $5.7 \times 10^{-6}$ \\ 
AA-$\ce{CH3OCH3}$ & 3$_{2,2}$-3$_{1,3}$ & 91.47931 & 11.1 & $4.9 \times 10^{-6}$ \\
AA-$\ce{CH3OCH3}$ & 4$_{2,3}$-4$_{1,4}$ & 93.85964 & 14.7 & $5.6 \times 10^{-6}$ \\
AA-$\ce{CH3OCH3}$ & 5$_{2,4}$-5$_{1,5}$ & 96.85246 & 19.3 & $6.2 \times 10^{-6}$ \\
$^{*}$AA-$\ce{CH3OCH3}$ & 4$_{1,4}$-3$_{0,3}$ & 99.32600 & 10.2 & $8.8 \times 10^{-6}$ \\
\hline
A-$\ce{CH3OCHO}$ & 7$_{2,6}$-6$_{2,5}$ & 84.45475 & 19.0 & $8.0 \times 10^{-6}$ \\
A-$\ce{CH3OCHO}$ & 7$_{3,4}$-6$_{3,3}$ & 87.16129 & 22.6 & $7.8 \times 10^{-6}$ \\
A-$\ce{CH3OCHO}$ & 8$_{1,8}$-7$_{1,7}$ & 89.31664 & 20.1 & $1.0 \times 10^{-5}$ \\
A-$\ce{CH3OCHO}$ & 7$_{2,5}$-6$_{2,4}$ & 90.15647 & 19.7 & $9.8 \times 10^{-6}$ \\ 
A-$\ce{CH3OCHO}$ & 9$_{1,9}$-8$_{1,8}$ & 100.0805 & 24.9 & $1.5 \times 10^{-5}$ \\
A-$\ce{CH3OCHO}$ & 8$_{1,7}$-7$_{1,6}$ & 100.4907 & 22.8 & $1.5 \times 10^{-5}$ \\
A-$\ce{CH3OCHO}$ & 9$_{0,9}$-8$_{0,8}$ & 100.6834 & 24.9 & $1.5 \times 10^{-5}$ \\
$^{*}$A-$\ce{CH3OCHO}$ & 8$_{2,6}$-7$_{2,5}$ & 103.4787 & 24.6 & $1.5 \times 10^{-5}$ \\
\enddata
\tablecomments{Acetaldehyde line data from JPL catalogue based on the data set of \citet{bauder76}. Dimethyl ether line data from JPL catalogue based on the data set of \citet{lovas79, neustock90}. Methyl formate line data from JPL catalogue based on the data set of \citet{ilyushin09, plummer84}. The representative molecular transition used for the normalized convolved intensity analysis in \S~\ref{methanol_distribution} is denoted with $^{*}$.}
\end{deluxetable}

The radiative transfer and rotational diagram analysis is performed toward the on-source position, and toward two offset positions: (i) the peak of the COM abundances (2000 AU), and (ii) the low-density outer-shell (9000 AU). By considering these three positions, we may compare the modeled COM peaks with the observational ones, and determine the dependence of the chemical reactions on the local physical conditions in the prestellar core.

One strategy to apply this radiative transfer and rotational diagram technique would be to simulate precisely the same molecular lines used in individual observational datasets for \object{L1544}. However, since the present aim is to determine a well-defined column density (and rotational temperature) based on the models, with which observed column densities may be directly compared, we instead choose a selection of lines that may plausibly be (or indeed have been) detected toward cold sources, and which include a range of upper energy levels. Emission lines of \actd\ and \mf\ recently detected toward the cold dark cloud B5 \citep{taquet17} are chosen for this analysis. While \citet{taquet17} detected a relatively large number of molecular lines for \actd\ and \mf, only four transitions of \dme\ with a limited range of $E_\textrm{up}$ (10--11 K) were detected by those authors. Our adoption of {\em only} those lines could therefore cause substantial uncertainty in the determination of $N_\textrm{tot}$(\dme ). For this reason, we choose eight bright (i.e. high A$_{ij}$) AA-transitions of \dme\, with $E_\textrm{up}$ ranging from 8--19~K, using the {\em Splatalogue} web tool \footnote{\url{http://www.cv.nrao.edu/php/splat}}. The spectroscopic data originate from the JPL line list \footnote{\url{https://spec.jpl.nasa.gov}} \citep{bauder76, lovas79, neustock90, ilyushin09, plummer84}; the COM transitions considered in this analysis are listed in Table~\ref{COMlines}.

\begin{deluxetable}{cccc}
\tablewidth{0pt}
\tabletypesize{\footnotesize}
\tablecolumns{4}
\tablecaption{The molecular column densities of the COMs at the core center \label{columndensities_on}}
\scriptsize
\tablehead{
\colhead{Model} & \colhead{$\ce{CH3CHO}$} & \colhead{$\ce{CH3OCH3}$} & \colhead{$\ce{CH3OCHO}$} \\
\colhead{} & \colhead{$(\textrm{cm}^-2)$} & \colhead{$(\textrm{cm}^-2)$} & \colhead{$(\textrm{cm}^-2)$}}
\startdata
Observation \tablenotemark{a} & 1.2$\times10^{12}$ & 1.5$\times10^{12}$ (4.0$\times10^{11}$) & 4.4$\times10^{12}$(8.0$\times10^{12}$)\\
Control &   2.6$\times10^{10}$ (4.3$\times10^{7}$) & 1.2$\times10^{10}$ (1.9$\times10^{7}$)  &1.1$\times10^{9}$ (9.9$\times10^{6}$) \\
3-B   & 3.8$\times10^{10}$ (1.1$\times10^{8}$)  & 2.6$\times10^{10}$ (1.8$\times10^{8}$)  & 4.7$\times10^{11}$ (4.4$\times10^{9}$)\\
3-B+3-BEF  & 1.2$\times10^{12}$ (4.1$\times10^{9}$) & 2.6$\times10^{9}$ (7.0$\times10^{7}$) & 1.1$\times10^{14}$ (1.3$\times10^{12}$) \\
E-R  & 2.7$\times10^{10}$ (4.5$\times10^{7}$) & 1.3$\times10^{10}$ (1.1$\times10^{7}$)  & 1.1$\times10^{9}$ (9.8$\times10^{6}$)\\ 
PD-Induced  & 4.0$\times10^{10}$ (1.2$\times10^{8}$) &1.4$\times10^{10}$ (6.0$\times10^{7}$)  &1.3$\times10^{9}$ (9.6$\times10^{6}$)  \\
\enddata
\tablenotetext{a}{\citet{jimenez-serra16}}
\tablecomments{Values in parentheses indicate observational or rotational diagram line-fitting (model) errors.}
\end{deluxetable}

\begin{deluxetable}{cccc}
\tablewidth{0pt}
\tabletypesize{\footnotesize}
\tablecolumns{4}
\tablecaption{The molecular column densities of the COMs toward the off-center COM peak \label{columndensities_off_2000}}
\tablehead{
\colhead{Model} & \colhead{$\ce{CH3CHO}$} & \colhead{$\ce{CH3OCH3}$} & \colhead{$\ce{CH3OCHO}$} \\
\colhead{} & \colhead{$(\textrm{cm}^-2)$} & \colhead{$(\textrm{cm}^-2)$} & \colhead{$(\textrm{cm}^-2)$}}
\startdata
Observation \tablenotemark{a} &  3.2$\times10^{12}$ & 7.7$\times10^{11}$ (3.2$\times10^{11}$) & 2.3$\times10^{12}$ (2.8$\times10^{12}$) \\
Control &  3.3$\times10^{10}$ (3.7$\times10^{7}$) &  2.8$\times10^{10}$ (3.6$\times10^{7}$) & 1.4$\times10^{9}$ (6.9$\times10^{6}$)\\
3-B   &  3.6$\times10^{10}$ (4.2$\times10^{7}$) & 4.2$\times10^{10}$ (5.7$\times10^{7}$)  & 5.3$\times10^{11}$ (2.1$\times10^{9}$)\\
3-B+3-BEF  & 1.1$\times10^{12}$ (4.6$\times10^{9}$) & 1.8$\times10^{9}$ (1.2$\times10^{7}$) &1.3$\times10^{14}$(1.3$\times10^{12}$) \\
E-R  & 3.3$\times10^{10}$ (3.9$\times10^{7}$) &  2.9$\times10^{10}$ (2.9$\times10^{7}$) & 1.4$\times10^{9}$ (6.8$\times10^{6}$)\\ 
PD-Induced  & 3.3$\times10^{10}$ (3.7$\times10^{7}$) & 2.8$\times10^{10}$ (3.6$\times10^{7}$) & 1.5$\times10^{9}$ (7.3$\times10^{6}$) \\
\enddata
\tablenotetext{a}{\citet{jimenez-serra16}}
\tablecomments{The off-position for the observations is 4000 AU, corresponding to the observational methanol peak of \object{L1544} \citep{jimenez-serra16}. In the models, the fractional abundance peak occurs around 2000 AU. Values in parentheses indicate observational or rotational diagram line-fitting (model) errors.}
\end{deluxetable}

\begin{deluxetable}{cccc}
\tablewidth{0pt}
\tabletypesize{\footnotesize}
\tablecolumns{4}
\tablecaption{The molecular column densities of the COMs toward the outer-shell (radius 9000 AU) \label{columndensities_off_9000}}
\tablehead{
\colhead{Model} & \colhead{$\ce{CH3CHO}$} & \colhead{$\ce{CH3OCH3}$} & \colhead{$\ce{CH3OCHO}$}\\
\colhead{} & \colhead{$(\textrm{cm}^-2)$} & \colhead{$(\textrm{cm}^-2)$} & \colhead{$(\textrm{cm}^-2)$}}
\startdata
Control &  2.5$\times10^{10}$ (3.8$\times10^{6}$) &  3.9$\times10^{10}$ (8.8$\times10^{7}$) & 1.3$\times10^{9}$ (1.4$\times10^{7}$) \\
3-B   & 3.0$\times10^{10}$ (3.1$\times10^{6}$)  &  5.7$\times10^{10}$ (4.4$\times10^{7}$) & 3.7$\times10^{11}$ (3.5$\times10^{9}$)\\
3-B+3-BEF  & 7.5$\times10^{11}$ (2.0$\times10^{8}$) & 1.5$\times10^{9}$ (2.2$\times10^{5}$) & 9.5$\times10^{13}$ (6.0$\times10^{11}$)\\
E-R  & 2.6$\times10^{10}$ (3.8$\times10^{6}$) & 4.0$\times10^{10}$ (8.7$\times10^{7}$)  & 1.3$\times10^{9}$ (1.4$\times10^{7}$)\\ 
PD-Induced  & 2.5$\times10^{10}$ (3.7$\times10^{6}$) & 3.9$\times10^{10}$ (8.8$\times10^{7}$) & 1.5$\times10^{9}$ (1.5$\times10^{7}$) \\
\enddata
\tablecomments{Values in parentheses indicate rotational diagram line-fitting errors.}
\end{deluxetable}

\begin{figure*}
\gridline{\fig{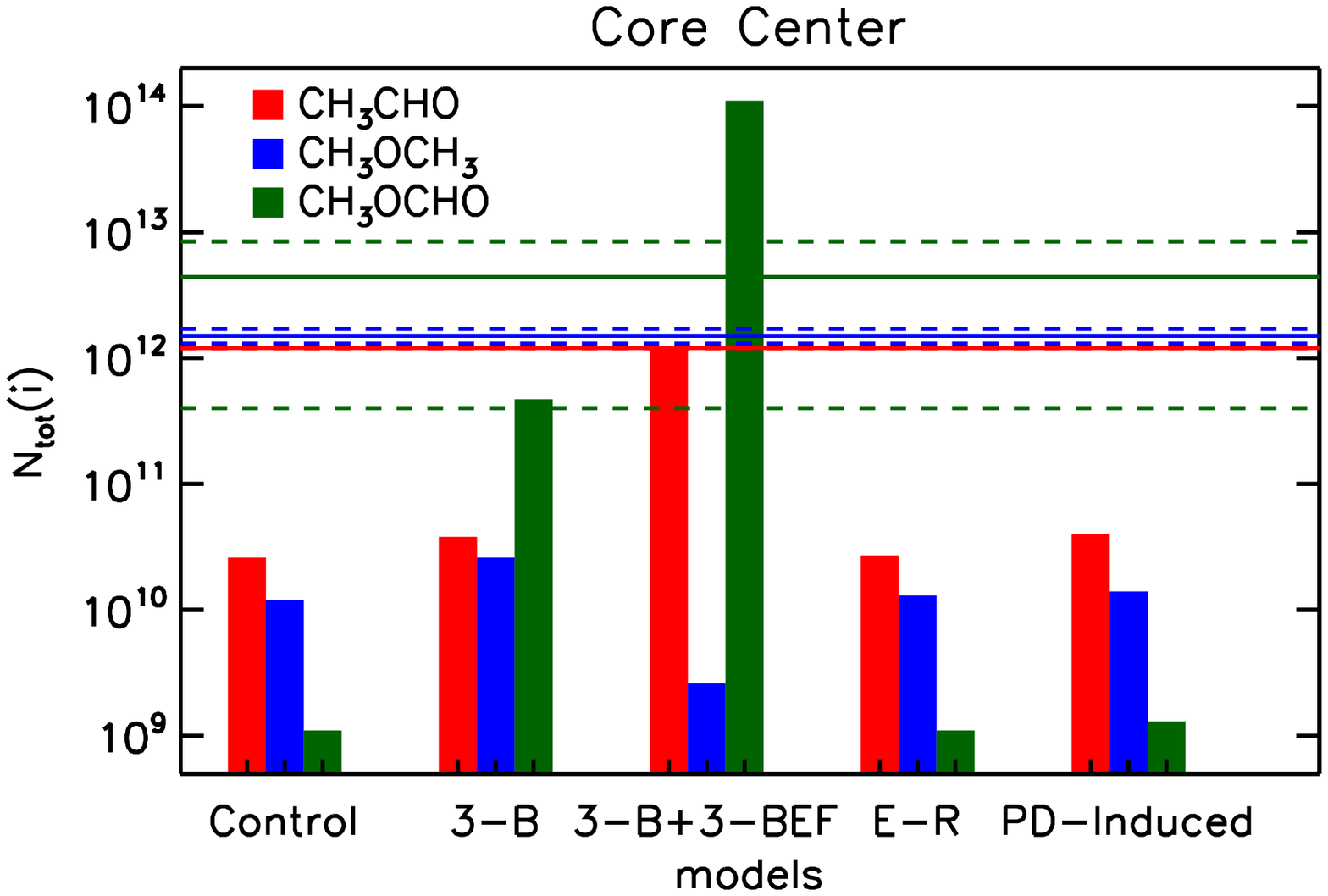}{0.5\textwidth}{}}
\gridline{\fig{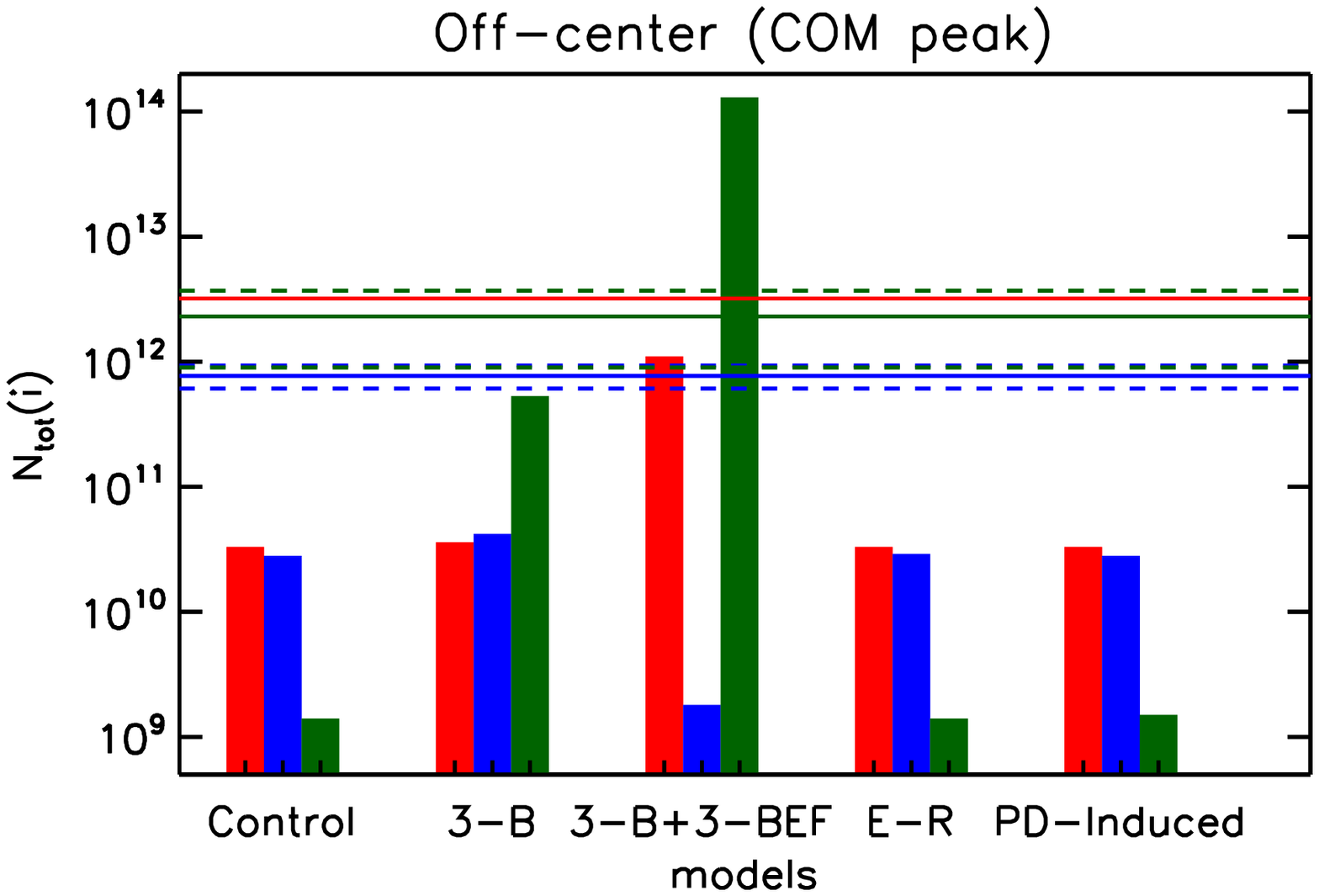}{0.5\textwidth}{}
          \fig{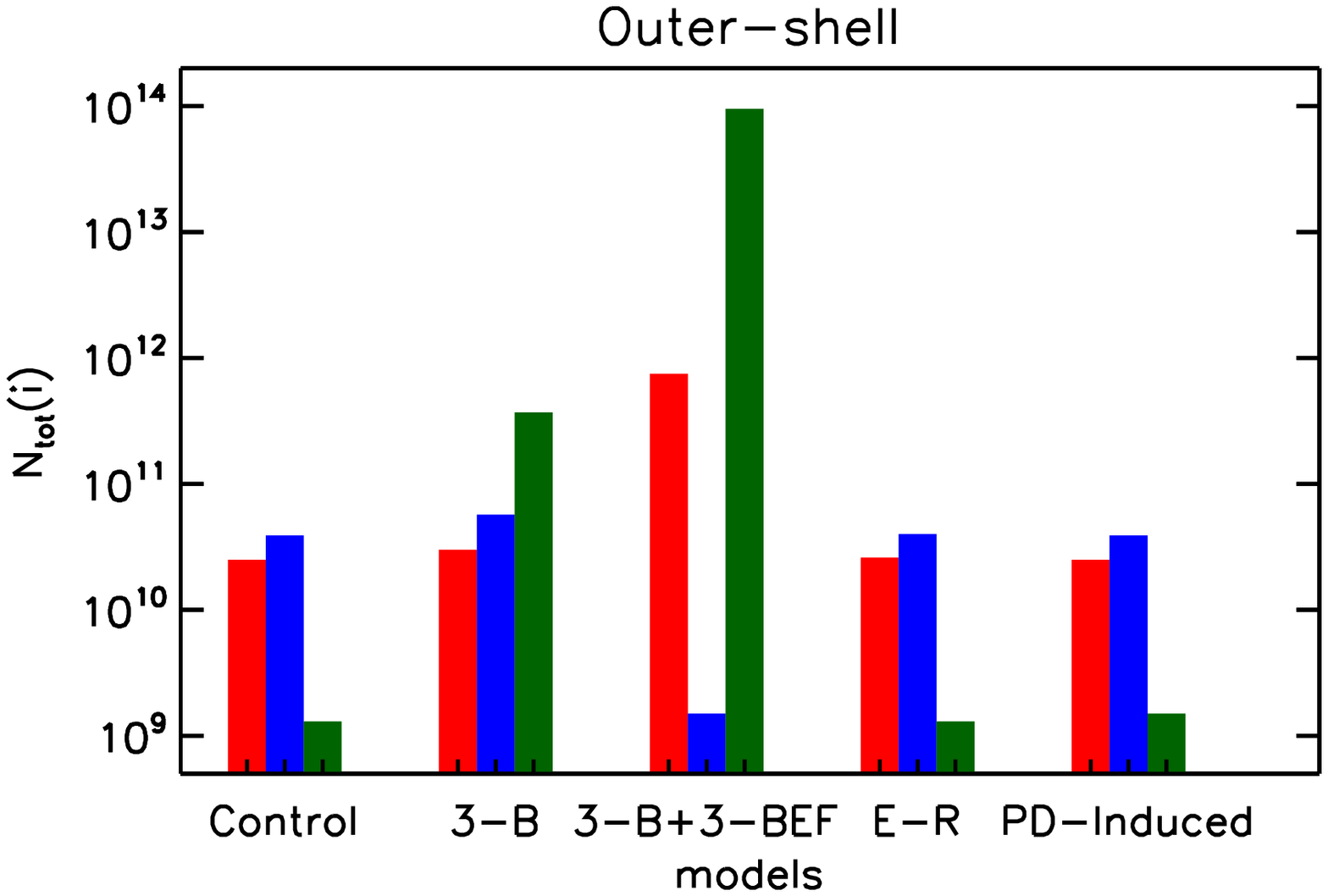}{0.5\textwidth}{}}
\caption{Column densities calculated from each model. Red, blue, and green histogram represents the molecular column densities of AA, DME, and MF, respectively. The errors of the modeled column densities are not presented here as those values are relatively small ($< 3\%$) compared to the bar size. The observed value and its error bound are presented together with solid and dashed horizontal lines respectively. For the COM peak position (lower-left panel), the the observations was performed towards 4000 AU, corresponding to the observational methanol peak of \object{L1544} \citep{jimenez-serra16}, while the off-position for the model is 2000 AU, where fractional abundance peak occurs.
\label{colden_histogram}}
\end{figure*}

\subsection{Column densities of O-bearing COMs toward the core center} \label{sec-col-den}

Tables~\ref{columndensities_on} -- \ref{columndensities_off_9000} (core center, 2000 AU, 9000 AU) compare the molecular column densities obtained from the RD analysis of different chemical models with observational values from the literature; observational errors and rotational diagram line-fitting error estimates are given in parentheses. Figure~\ref{colden_histogram} shows the molecular column densities for each model at three different positions as histograms. The observed value (a solid horizontal line) and its error bounds (dashed horizontal lines) are presented together for comparison. While every chemical model introduced here significantly underproduces \dme , both the 3-B and 3-B+3-BEF models result in meaningful differences from the control model for the other two COMs at the core center (Table~\ref{columndensities_on}). One thing to note is that the COMs are more actively formed via 3-BEF than solely by the 3-B mechanism. For example, while 3-B+3-BEF significantly increases the column density of $\ce{CH3CHO}$ as well as $\ce{CH3OCHO}$, 3-B substantially increases $\ce{CH3OCHO}$ only. Furthermore, even though both the 3-B and 3-B+3-BEF mechanisms enhance the $\ce{CH3OCHO}$ population significantly, the increment is much higher in the 3-B+3-BEF model (3-B+3-BEF even substantially overproduces $\ce{CH3OCHO}$ -- see also \S~\ref{3bef-best}).

In either the 3-B or 3-BEF models, the key quantities through which the production rates of COMs (on the grains or in the gas phase) may be understood are the surface abundance of reactants, and the production (i.e. {\em appearance}) rates of their reaction partners. The latter quantity is an explicit component of the new expressions for non-diffusive processes, whereas in the regular L-H formulation it does not appear. The higher formation rates in the 3-BEF model can be explained by the fact that this mechanism involves the addition of radicals to stable compounds (which are thus more abundant on the grain surface) in the second step of the consecutive reaction chain (Eqs. 18-20), while the 3-B process involves the addition of sparse radicals (Eqs. 13, 15, and 17).

\begin{figure*}
\centering
\includegraphics[width=0.75\textwidth]{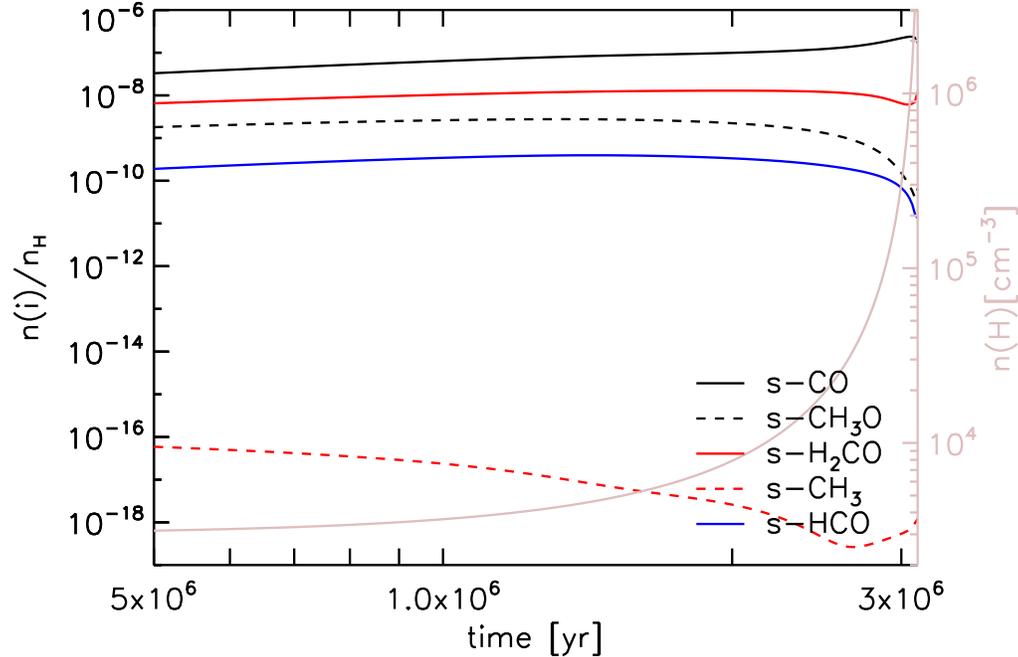}
\caption{Time evolution of the fractional ice composition of the reactants on the grain-surface related to the 3-B+3-BEF mechanisms forming \dme (red lines) and \mf (black lines). Note that the abundances shown refer specifically to species on the outer grain/ice surface and not within the ice mantles. A fractional abundance of $\sim$ $1.3 \times 10^{-12}$ corresponds to one particle per grain.}
\label{EFspecies}
\end{figure*}

The greater rate of formation of $\ce{CH3OCHO}$ over that of $\ce{CH3CHO}$ in either the 3-B or 3-BEF model can also be understood in the same context. In the 3-B model, the reactants $\ce{CH3}$ and $\ce{CH3O}$ are technically competing with each other to form either $\ce{CH3CHO}$ or $\ce{CH3OCHO}$ by reacting with HCO radicals on the grain surface. As seen in Fig.~\ref{EFspecies}, the fractional grain-surface abundance of $\ce{CH3O}$ (shown for the core-center position) is much higher than that of $\ce{CH3}$. The production rate of \ce{CH3O} is also much greater than that of \ce{CH3}, which is partly why its surface abundance is higher. Similarly, in the 3-BEF model, $\ce{CH3}^{*}$ and $\ce{CH3O}^{*}$ are competing with each other to react with CO on the grain surface to form either $\ce{CH3CHO}$ or $\ce{CH3OCHO}$; CO is abundant, and the appearance rates of $\ce{CH3}^{*}$ and $\ce{CH3O}^{*}$ directly determine the formation rates of $\ce{CH3CHO}$  and $\ce{CH3OCHO}$.

Note that only a fraction of newly-formed methyl radicals can take part in the formation of the COMs through the 3-BEF mechanism, because only the excited methyl radicals formed via hydrogenation of $\ce{CH2}$ have this mechanism available; abstraction of H from $\ce{CH4}$ by other H atoms is slightly endothermic, so it should not produce $\ce{CH3}^{*}$.

Thus, the radical $\ce{CH3}$ acts as a bottleneck to the formation of the COMs in the non-diffusive models. Also, the gradual depletion of C and related hydrocarbons from the gas phase, while CO remains abundant, means that production of \ce{CH3} cannot keep up with the production of CO-related radicals. The reaction of \ce{CH3} with H or H$_2$ from the gas phase to re-form \ce{CH4} also keeps the average grain-surface \ce{CH3} abundance low. The production of HCO and \ce{CH3O} radicals continues to be effective as methanol builds up; while the net direction of the CO chemistry is to convert it gradually to methanol, there are backward reactions at every step, including H-abstraction from \ce{CH3OH}, that allow the intermediate radicals to maintain some level of surface coverage and a sustained rate of production/appearance.

Given that the formation of $\ce{CH3OCH3}$ is related to $\ce{CH3}$ in both the 3-B or 3-BEF models, the lower column density of $\ce{CH3OCH3}$ in those models can be explained. 
CO and $\ce{H2CO}$ are competing to form either $\ce{CH3CHO}$ or $\ce{CH3OCH3}$ by reacting with the excited methyl radicals on the grain surface, but CO is much more abundant than $\ce{H2CO}$ (Fig.~\ref{EFspecies}). The small amount of excited methyl radicals on the surface are thus preferentially consumed to form $\ce{CH3CHO}$.

The E-R mechanism does not make a substantial difference to the gas-phase abundances versus the control~(see Fig.~\ref{radiVSgas-abundance} and Table~\ref{columndensities_on}). This is because the E-R process requires high surface coverage of the reactive species on the grains to be effective. This result is not exactly consistent with the results from \citet{ruaud15}. They find that the combination of E-R and their complex-induced reaction mechanisms is efficient enough to reproduce the observed COM abundances at temperatures as low as 10 K. Beyond the uncertainty in the level of contribution of either mechanism, the different model parameters of both studies should be noted: \citet{ruaud15} mainly focus on the accretion of carbon atoms and assume a much higher binding energy (3600--8400~K) than ours (800~K). This may cause higher concentration of reactive species on the grain surface, allowing the E-R process to be efficient.   

The PD-induced reaction process is ineffective in increasing the population of COMs in the gas phase at the core center. However, the PD-induced model significantly increases (more than 2 orders of magnitude) the amount of COMs in the ice mantles throughout the core's evolution (see figure~\ref{radiVSsolid-abundance}). 
Other studies suggest indeed that the bulk ice is where the majority of physico-chemical changes caused by radiation chemistry are likely to occur \citep{johnson90, spinksandwoods90, shingledecker17}.
The enhanced population of the COMs in the ice mantle does not actively affect the population in the gas phase, because the COM products are preserved in the mantle rather than diffusing to the grain-surface, which is directly coupled to the gas phase. Even though this process does not make a prominent difference in the gas-phase abundance for the prestellar core, it would significantly affect the chemistry during the warm-up period of a protostellar core in which accumulated mantle manterial is ejected from the grains.

\subsection{Optimization of the 3-BEF model} \label{3bef-best}

As discussed in \S \ref{sec-col-den}, the formation of methyl formate (\mf) through the 3-BEF mechanism is so efficient 
that \mf\ is significantly overproduced, while this is not the case for acetaldehyde (\actd). The 3-BEF mechanism as described in \S \ref{subsec:3-bef} is assumed to proceed with 100~\% efficiency; however, the appropriate value in individual cases could be lower if the exothermic energy available from the initiating reaction ($E_{\mathrm{reac}}$) is similar in magnitude to the activation energy barrier ($E_{\mathrm{A}}$) of the subsequent reaction. Assuming the energy is initially released into the vibrational modes of the excited species, it may not be available in the required mode for reaction with an adjacent species to occur before that energy is lost to the surface, or indeed that the excited species diffuses away entirely from its reaction partner. If the excited product has $s$ internal vibrational modes, the 3-BEF process would be expected to have substantially sub-optimal efficiency in the case where $E_{\mathrm{A}} > E_{\mathrm{reac}} / s$, while it would not occur at all in the case where $E_{\mathrm{A}} > E_{\mathrm{reac}}$. The former condition would appear to hold for the reactions shown in Eqs. (20), in which methyl formate is produced; here, $s$(CH$_3$O)=9, $E_{\mathrm{reac}}\simeq10,200$~K, and $E_{\mathrm{A}}$=3,967~K for the CH$_3$O + CO $\rightarrow$ CH$_3$OCO reaction \citep{huynh08}.

Rice-Ramsperger-Kassel (RRK) theory may be introduced to obtain a statistical estimate of the efficiency. Using the same formulation that is employed to determine the probability of chemical desorption in the model \citep{garrod07}, the probability of a successful 3-BEF process would be:
\begin{equation}
P=\left[1-\frac{E_\textrm{A}}{E_\textrm{reac}}\right]^{s-1}
\end{equation}
where $s$ now includes an additional vibrational mode representing the reaction coordinate (i.e. $s$(CH$_3$O)=10). For the reactions forming \mf, the values provided above give a probability of 1.2\%, while for the reactions producing \actd\ and \dme\ the probability would be 73\%. This shows that the \textit{P}(\mf ) of 1 originally introduced in our 3-B+3-BEF model was too high, explaining the overproduction of that species.

For the present models, in which only three 3-BEF processes are explicitly considered, we empirically test a selection of efficiencies for the reaction to form \mf\, ranging incrementally from 100\% to 0.1\% in factors of 10; the other two 3-BEF reactions are assumed to operate at maximum efficiency. It is found that a probability of 0.1\% best reproduces the molecular column densities from the observations.

The empirically-determined optimal efficiency is clearly lower than the simple RRK treatment above would suggest. However, the latter does not include competition between reaction and diffusion of the excited species, which could account for at least a factor of a few, representing several diffusion directions. Likewise, additional translational degrees of freedom of the excited species could be considered in Eq. (23), rather than just one reaction coordinate. We note also that these modifications would reduce the efficiency of the other two 3-BEF reactions considered here, perhaps bringing them closer to around 10\%. The molecular dynamics study by \citet{fredon17} of reaction-induced non-thermal diffusion would indeed suggest that translational motion would be a necessary factor to consider in a detailed treatment of the 3-BEF process (although it should be noted that those authors assumed all of the energy to be immediately released into translational modes, rather than distributed also into internal vibration and/or rotation.

\begin{figure*}
\gridline{\fig{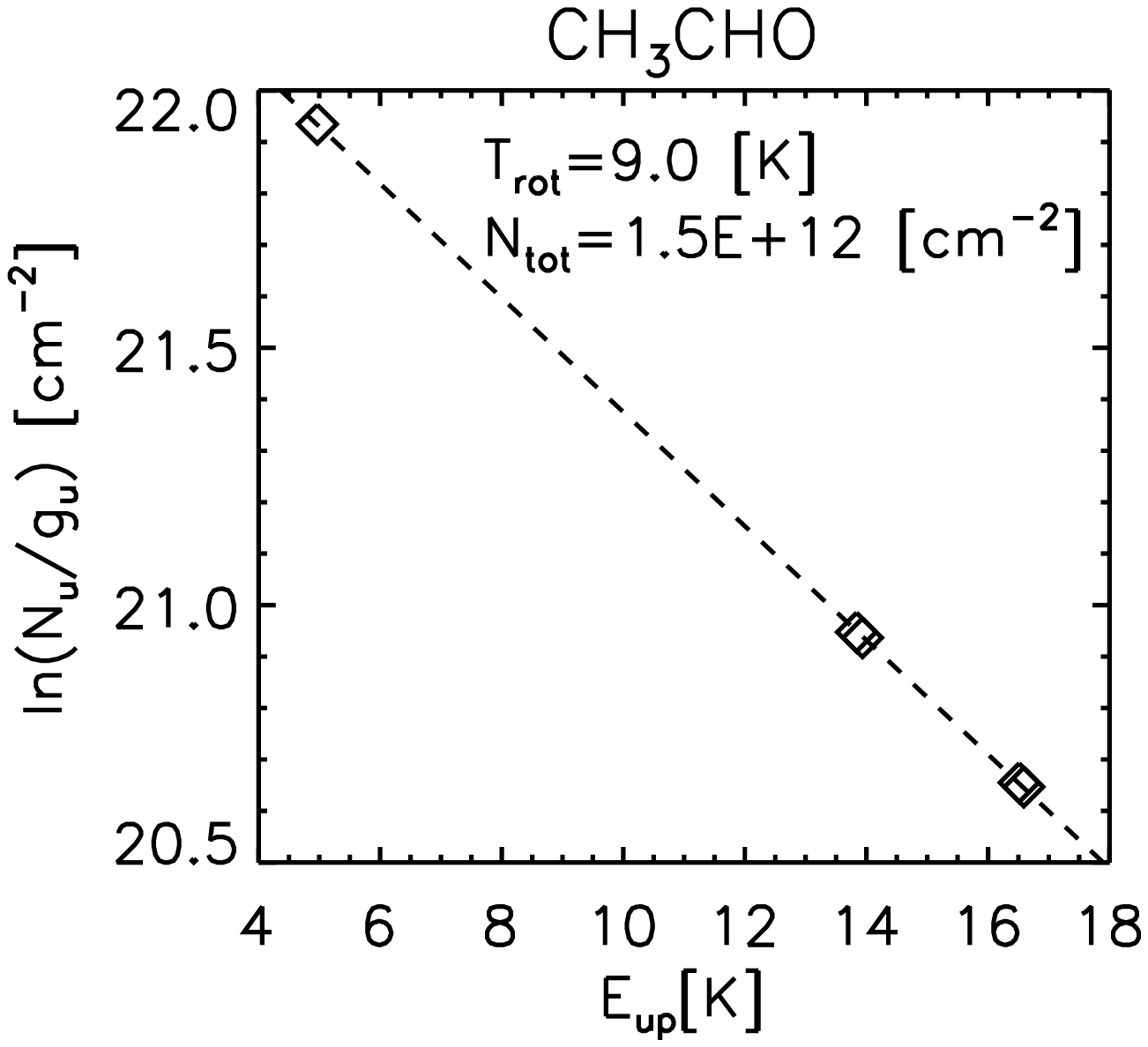}{0.3\textwidth}{}
          \fig{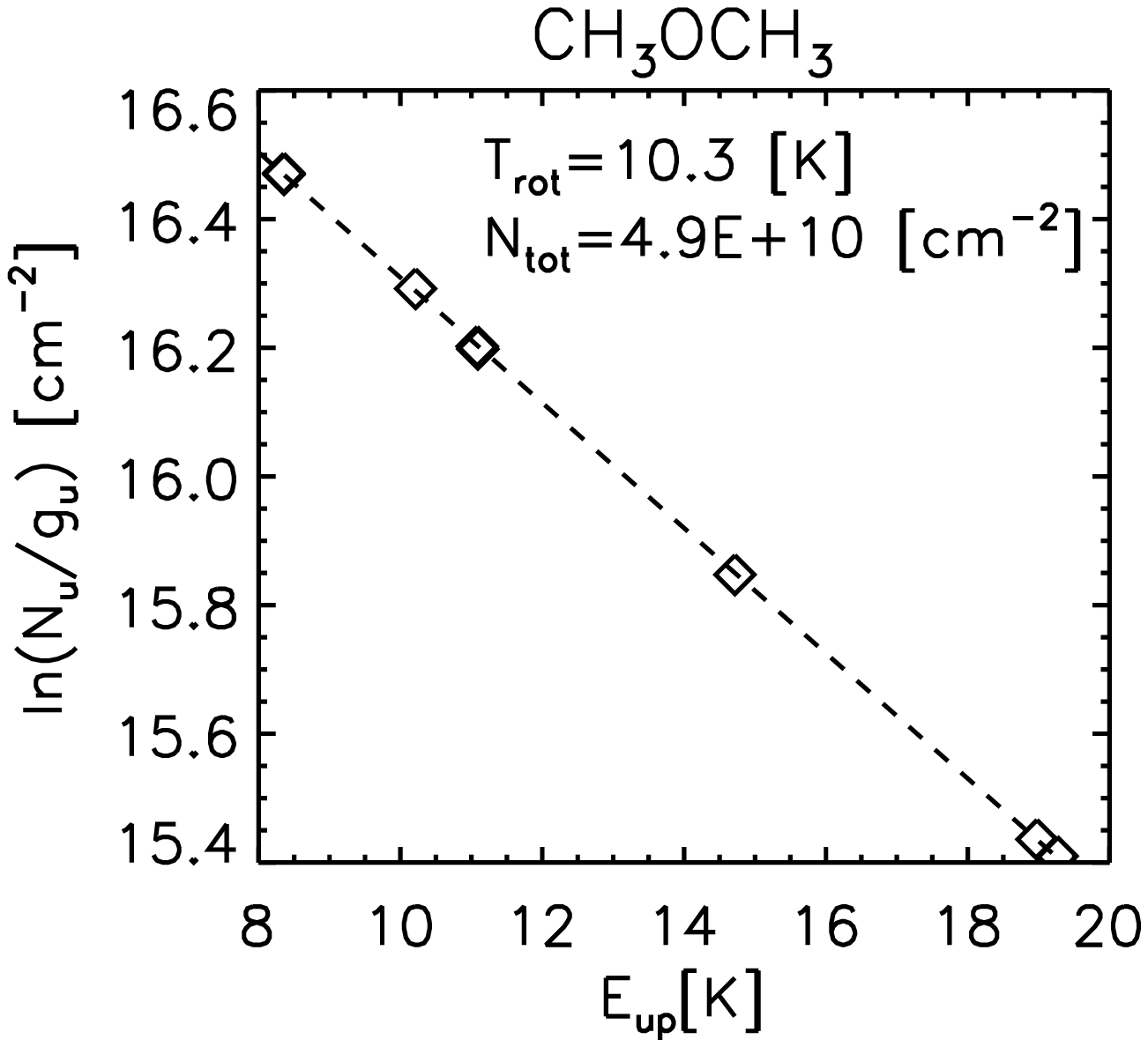}{0.3\textwidth}{}
          \fig{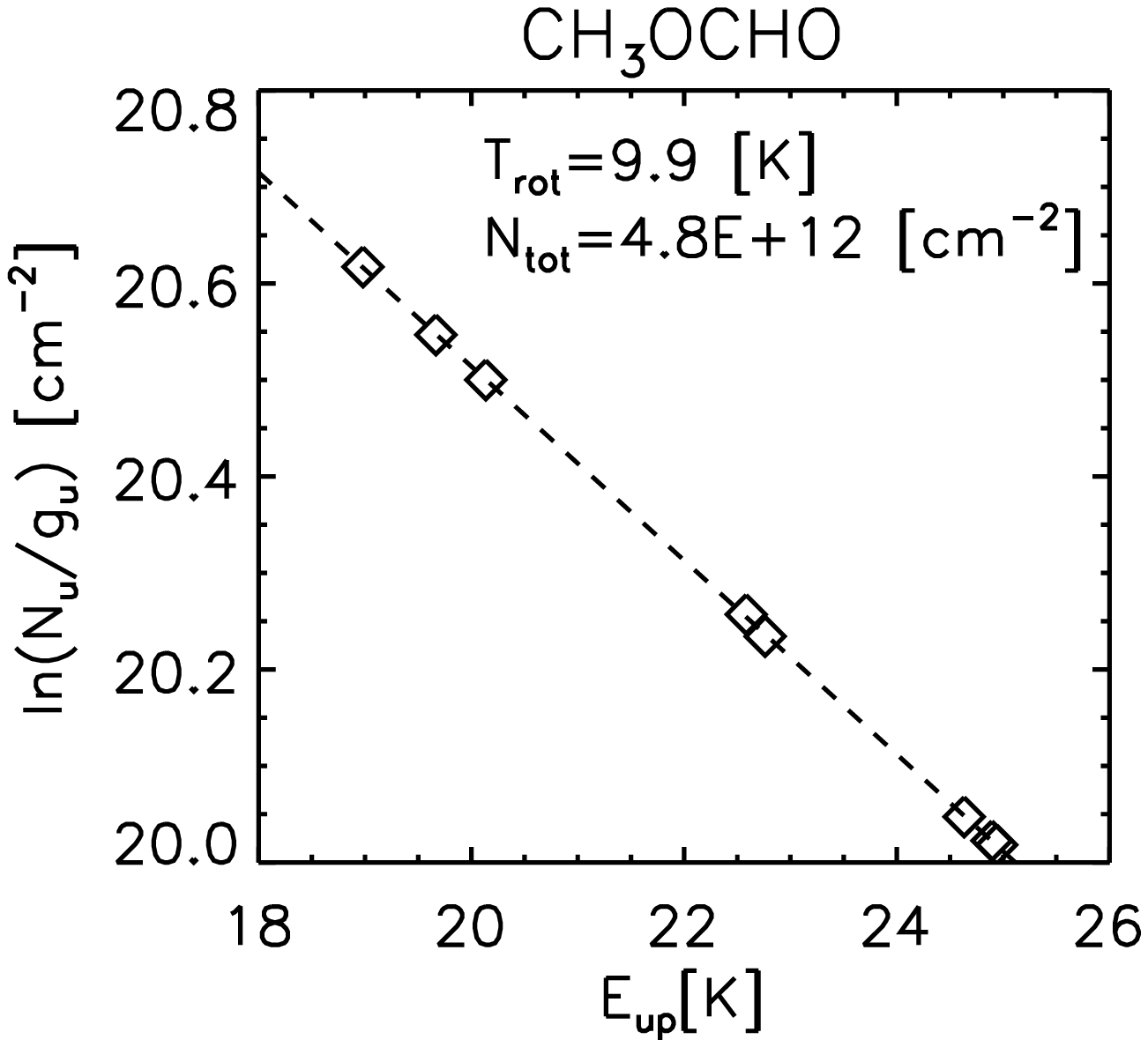}{0.3\textwidth}{}}
\caption{
Rotation diagrams for the three O-bearing COMs at the core center. The black dashed lines represent the fit.
\label{RD_COMs_on}
}
\end{figure*}

Figure \ref{RD_COMs_on} shows rotational diagrams obtained from LTE radiative transfer calculations based on the 3-BEF Best model molecular profiles, with the beam directed toward the core center. Table~\ref{comparisoncolumndensities} compares the molecular column densities towards the core center from this model with the observational literature values. The errors (in parentheses) for modeled column densities are derived from the standard deviation of linear regression fitting in rotation diagrams. The three-body mechanisms introduced here are efficient enough to reproduce the amount of \mf\ and \actd\ in the prestellar core when an appropriate efficiency for the 3-BEF mechanism is adopted.

\begin{deluxetable}{cccc}
\tablewidth{0pt}
\tabletypesize{\footnotesize}
\tablecolumns{4}
\tablecaption{Column densities of COMs in the 3-BEF Best model\label{comparisoncolumndensities}}
\scriptsize
\tablehead{
\colhead{Species} & \colhead{$\ce{CH3CHO}$} & \colhead{$\ce{CH3OCH3}$} & \colhead{$\ce{CH3OCHO}$} \\
\colhead{} & \colhead{$(\textrm{cm}^-2)$} & \colhead{$(\textrm{cm}^-2)$} & \colhead{$(\textrm{cm}^-2)$}}
\startdata
Observation (Core Center) \tablenotemark{a} 	& 1.2 $\times 10^{12} $ 				& 1.5 $\times 10^{12} (2.0 \times 10^{11}) $	& 4.4 $\times 10^{12} (4.0 \times 10^{12})$ \\
Core Center 							& 1.7 $\times 10^{12} (5.7 \times 10^{9})$ & 3.4 $\times 10^{10} (3.3 \times 10^{8})$ 		& 4.0 $\times 10^{12} (2.7 \times 10^{10})$ \\
2000 AU 								& 1.4 $\times 10^{12} (5.4 \times 10^{9})$ & 4.8 $\times 10^{10} (1.1 \times 10^{8})$ 		& 4.8 $\times 10^{12} (1.1 \times 10^{10})$ \\
9000 AU 								& 8.5 $\times 10^{11} (3.3 \times 10^{8})$ & 5.7 $\times 10^{10} (3.0 \times 10^{7})$ 		& 4.0 $\times 10^{12} (3.8 \times 10^{10})$ \\
\enddata
\tablecomments{Values in parentheses indicate observational or rotational diagram line-fitting (model) errors.}
\tablenotetext{a}{\citet{jimenez-serra16}}
\end{deluxetable}

\begin{figure*}
\gridline{\fig{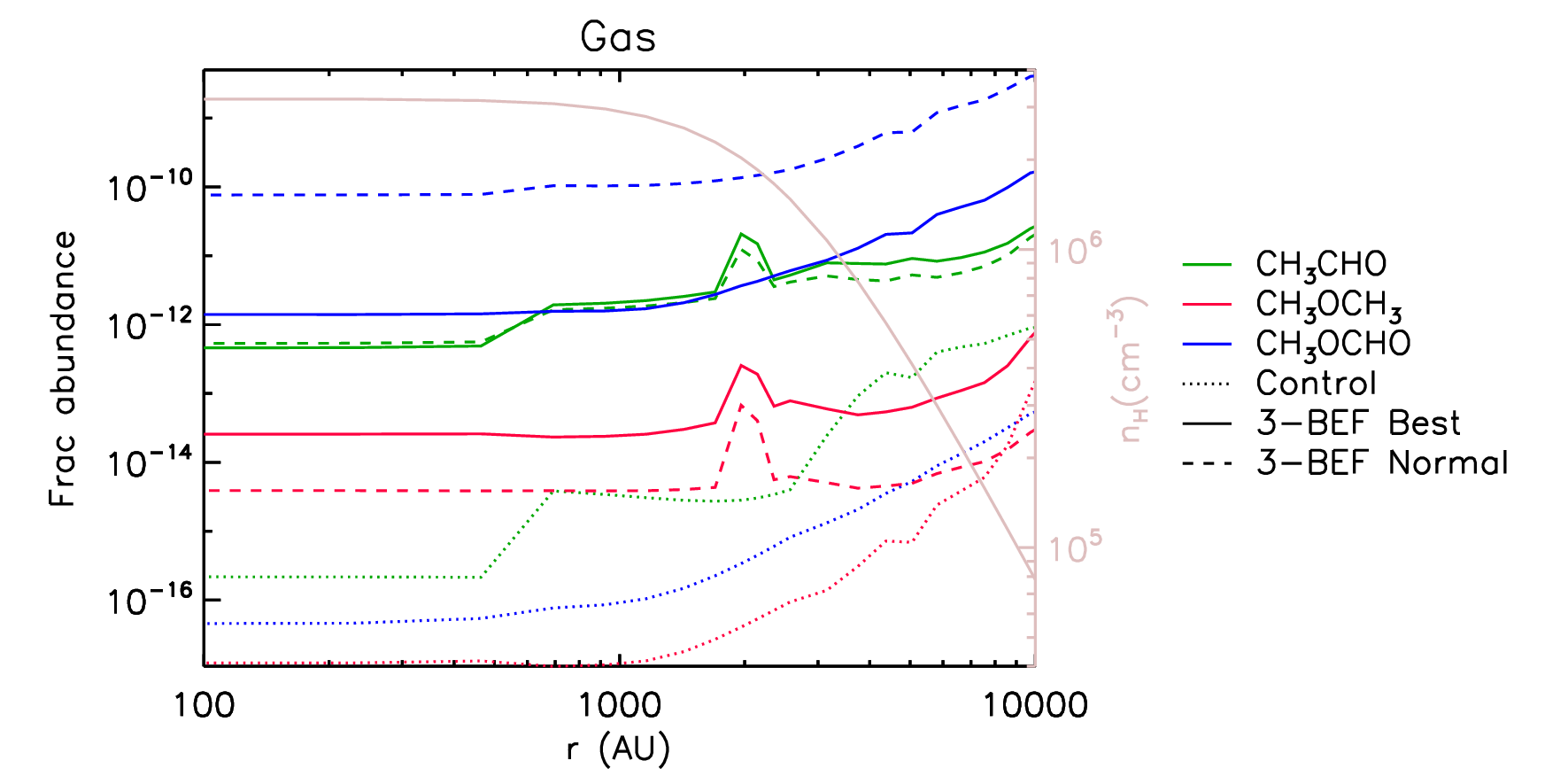}{0.75\textwidth}{}}
\gridline{\fig{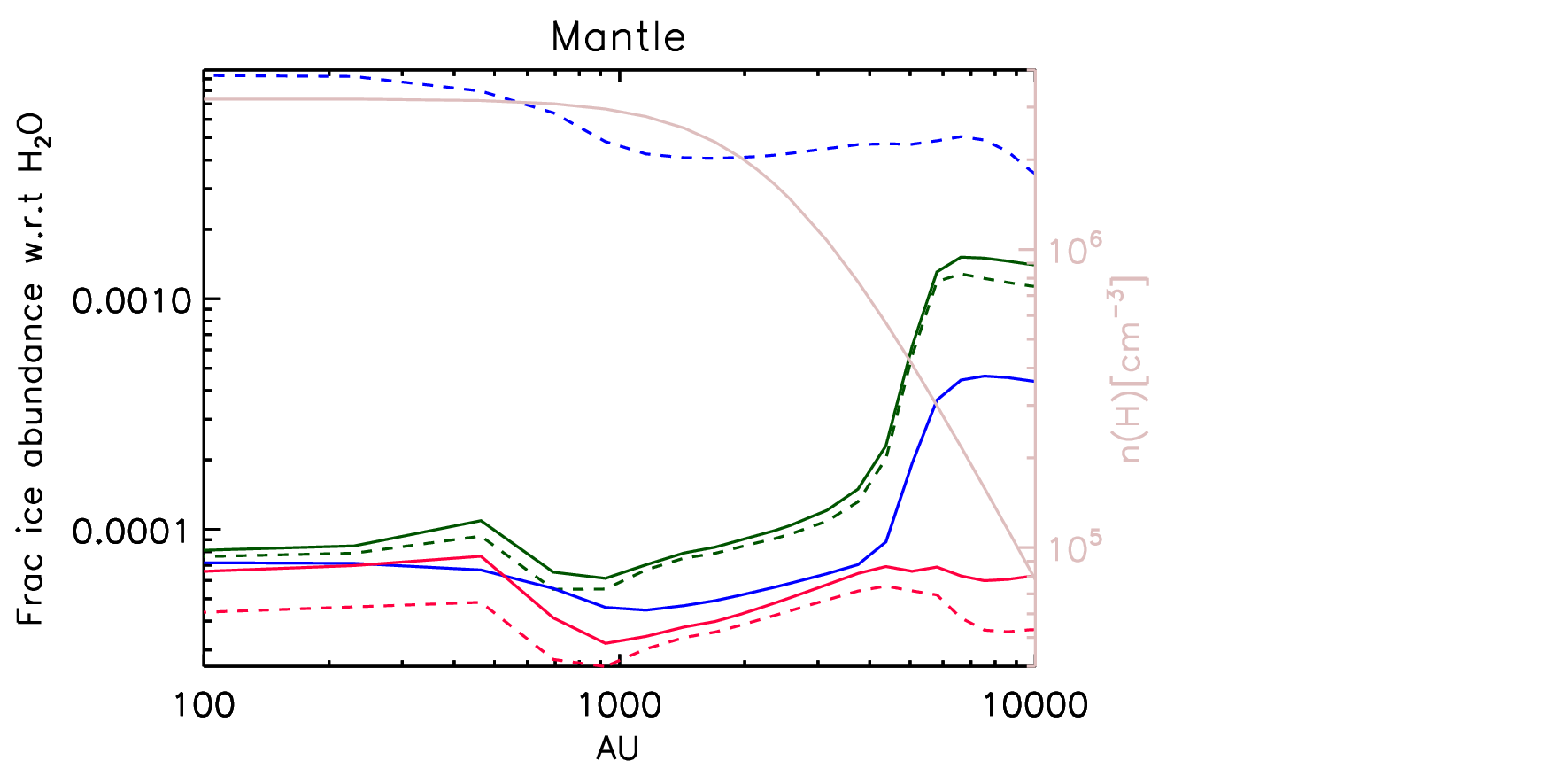}{0.75\textwidth}{}}
\caption{Comparison of chemical distribution of the 3-BEF Best (solid lines) in the gas with those of the normal 3-B+3-BEF (dashed lines). Control model results are also shown (dotted lines).
\label{EffComp}}
\end{figure*}

Figure \ref{EffComp} compares the chemical distribution of the 3-BEF Best results~(solid lines) with those of the control (dotted lines) and the normal 3-BEF (dashed lines) models. While the amount of \mf\ is significantly reduced in the 3-BEF Best model compared to the normal 3-BEF in both gas- and solid-phase, the population of \dme\ increases (by roughly an order of magnitude in the gas); the weakening of the 3-BEF mechanism for methyl formate production leaves more of the CH$_3$O radical available to participate in other reactions, including the regular 3-B mechanism (CH$_3$ + CH$_3$O) that produces dimethyl ether. A commensurate increase is seen in the column density values.

The adjustment to the efficiency of MF production through the 3-BEF process also reduces the solid-phase abundance of that molecule with respect to water back to more plausible values that are in line with the maximum typical values observed in hotter sources (i.e. around 10$^{-8}$ with respect to H$_2$). The fraction is higher beyond around 5,000 AU, but the total ice abundance at these positions would also be somewhat lower.

\subsection{$\ce{CO}$ hydrogenation and $\ce{CH3OH}$ abundances} \label{methanol_distribution}

The $\ce{CH3OH}$ map of \citet{bizzocchi14} shows a highly asymmetric non-uniform ring surrounding the dust peak of \object{L1544}. This morphology is consistent with central depletion and preferential release of methanol in the region where CO starts to freeze out significantly. \citet{jimenez-serra16} shows that COMs are actively formed and already present in this methanol peak.

\begin{figure*}
\gridline{\fig{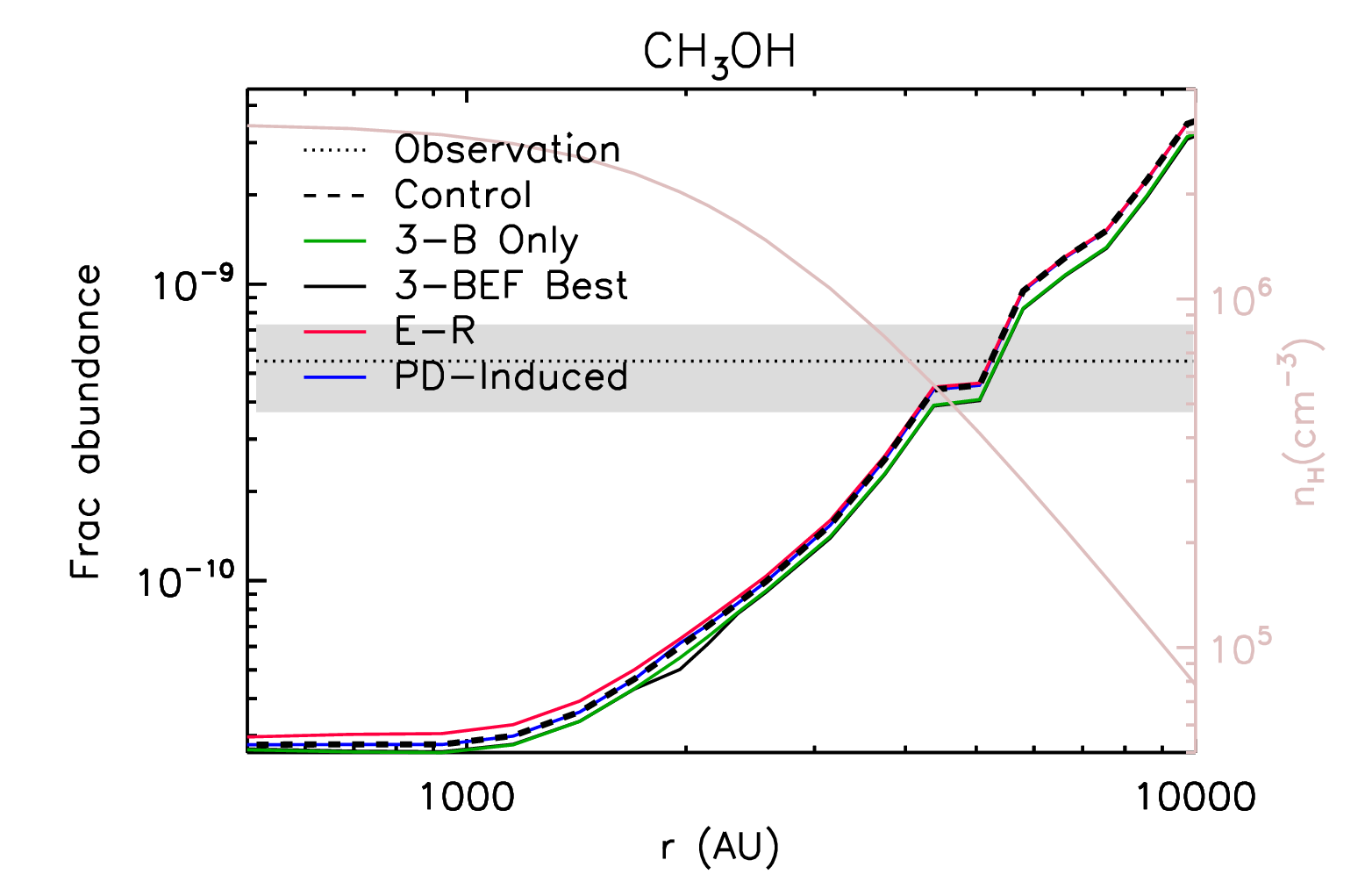}{0.55\textwidth}{(a)}}
\gridline{\fig{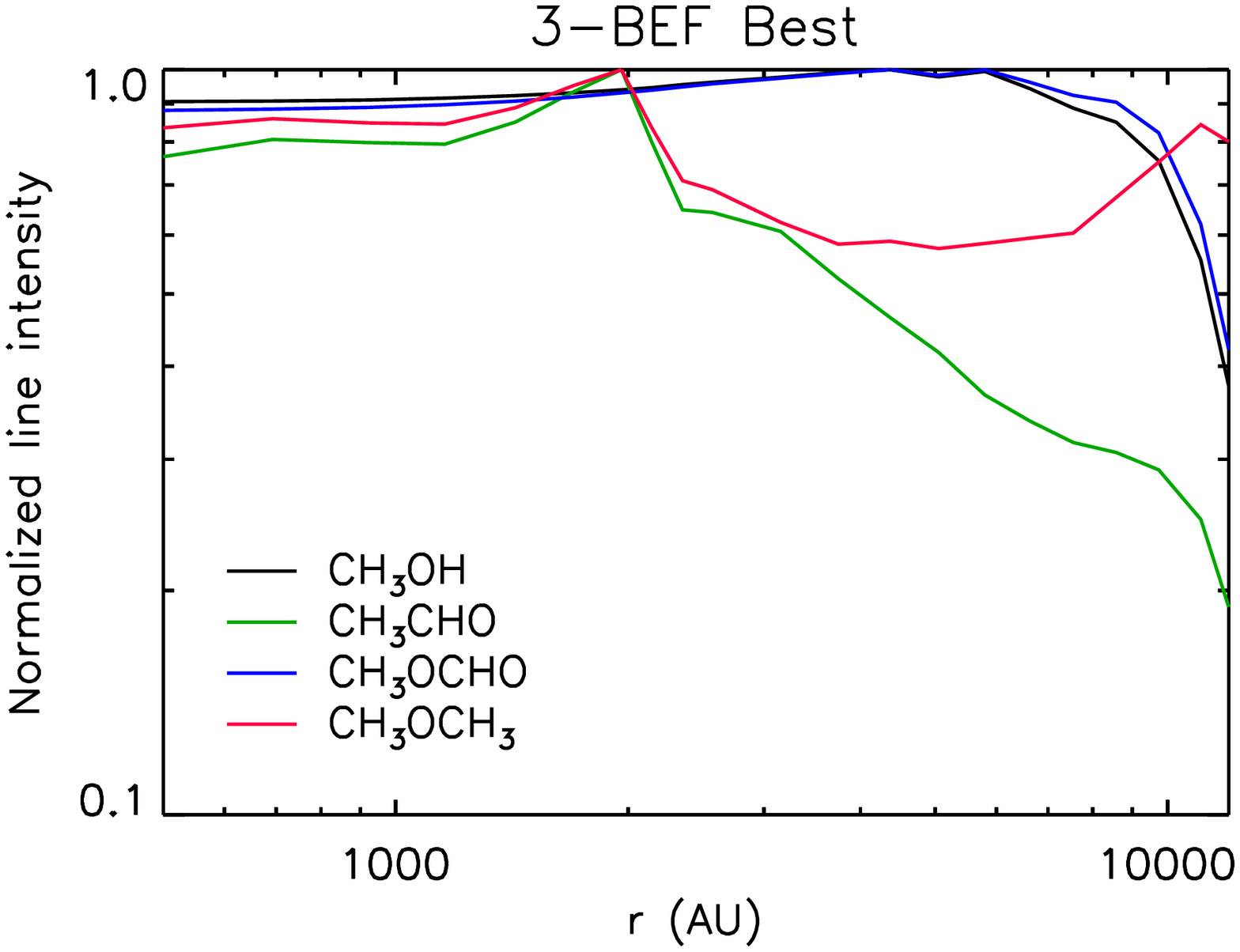}{0.55\textwidth}{(b)}}
\caption{
(a) Radial distribution of fractional abundance of $\ce{CH3OH}$ for each model. The abundance value from the observations toward the core center is denoted a black dotted line. Gas density as a function of radius is also indicated. (b) The normalized, convolved intensity of a representative emission line for methanol and for each of the three COMs of interest, shown as a function of the offset of the beam from the on-source position.
\label{CH3OH}}
\end{figure*}

The upper panel of Fig.~\ref{CH3OH} shows the radial distribution of $\ce{CH3OH}$ fractional abundance for each of the chemical models; abundances are very similar for all models at all positions. Methanol in the gas is mainly formed as the result of the hydrogenation of grain-surface CO all the way to CH$_3$OH, followed by chemical desorption. The radial distribution of the {\em fractional abundance} of gas-phase methanol has its peak well beyond where the observations would suggest. However, it should be noted that the gas density in these more distant regions drops off significantly, according to the physical profile. The location of the peak in {\em absolute} abundance would provide a better comparison directly with observations, although the best method is to consider the column density structure of methanol explicitly.

\begin{figure*}
\centering
\includegraphics[width=0.5\textwidth]{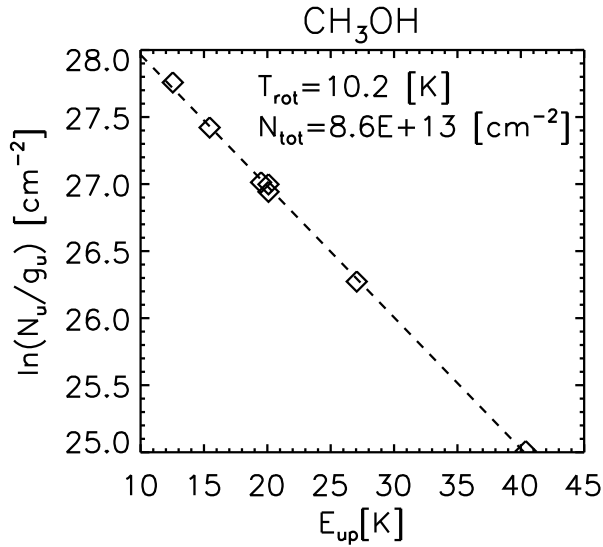}
\caption{The rotation diagram for $\ce{CH3OH}$ . The black dashed lines represent the fit.
\label{RD_CH3OH}}
\end{figure*}

The lower panel of Fig.~\ref{CH3OH} shows the normalized convolved intensity of a representative emission line of methanol, as a function of the beam offset from the center, using the radiative transfer model already described with the 3-BEF Best model data. Since the lines are optically thin and are well represented by an LTE treatment (see Fig.~\ref{RD_CH3OH} and Table \ref{CH3OHlines}), the line intensity profile scales well with the column density profile along each line of sight. The modeled methanol emission shows a peak near to 4000 AU as reported in the observations, even though this feature is not so obvious, as the slope is quite gentle. The same treatment is shown for the other three COMs of interest. Methyl formate shows a fairly similar distribution of emission to that of methanol, while the other two COMs show peaks at 2000~AU as seen in the fractional abundances. The representative molecular transition used for this analysis is denoted with an asterisk in Tables~\ref{COMlines} and \ref{CH3OHlines}.

\begin{deluxetable}{ccccc}
\tablewidth{0pt}
\tabletypesize{\footnotesize}
\tablecolumns{5}
\tablecaption{List of methanol transitions with their spectroscopic properties  \label{CH3OHlines}}
\scriptsize
\tablehead{
\colhead{Molecule} & \colhead{Transition} & \colhead{Frequency (GHz)} & \colhead{$E_\textrm{up}$ (K)} & \colhead{$A_\textrm{ij}$ (s$^-1$)}
}
\startdata
E-$\ce{CH3OH}$ & 5$_{-1}$-4$_{0}$ & 84.52117 & 40.4 & $2.0 \times 10^{-6}$ \\
E-$\ce{CH3OH}$ & 2$_{-1}$-1$_{-1}$ & 96.73936 & 12.5 & $2.6 \times 10^{-6}$ \\
E-$\ce{CH3OH}$ & 2$_{0}$-1$_{0}$ & 96.74455 & 20.1 & $3.4 \times 10^{-6}$ \\
E-$\ce{CH3OH}$ & 3$_{0}$-2$_{0}$ & 145.09375 & 27.1 & $1.2 \times 10^{-5}$ \\ 
E-$\ce{CH3OH}$ & 3$_{-1}$-2$_{-1}$ & 145.09744 & 19.5 & $1.1 \times 10^{-5}$ \\
E-$\ce{CH3OH}$ & 1$_{0}$-1$_{-1}$ & 157.27083 & 15.4 & $2.2 \times 10^{-5}$ \\ 
$^{*}$E-$\ce{CH3OH}$ & 2$_{0}$-2$_{-1}$ & 157.27602 & 20.1 & $2.2 \times 10^{-5}$ \\
\enddata
\tablecomments{Methanol line data from JPL catalogue based on the data set of \citet{xu08}. The representative molecular transition used for the normalized convolved intensity analysis in \S~\ref{methanol_distribution} is denoted with $^{*}$.}
\end{deluxetable}

The full RD analysis is performed for methanol as for the other COMs. The seven E-transition lines of $\ce{CH3OH}$ that were detected by \citet{taquet17} are chosen for this (Table~\ref{CH3OHlines}). A single fit to all lines provides a column density 8.6$\times 10^{13}~\textrm{cm}^{-2}$ at the core center. This value is roughly consistent with the observation  \citep[2.6$\times 10^{13} ~\textrm{cm}^{-2}$, ][]{bizzocchi14}. The precise value, as with those of the other COMs, will be dependent on the fidelity of the chemical desorption treatment used here.

\section{discussion}

Of the several new non-diffusive processes tested here, the Eley-Rideal mechanism appears to have the least effect, due largely to the low surface coverage of reactive species. Those reactive species that might benefit from the spontaneous arrival of a reaction partner from the gas phase always maintain low fractional surface coverage due to their reactivity with highly diffusive surface species e.g.~atomic H. Species that do build up a large surface coverage, like CO, tend to have large barriers to reaction, so that incoming species are more likely to diffuse away than to react spontaneously. The importance of the E-R process to typical surface reactions is unlikely to be substantial under any physical conditions as long as atomic H remains mobile.

Photodissociation-induced reactions, in which the PD process acts spontaneously to bring a reactive radical into contact with some other species, has no significant influence on the gas-phase abundances of complex organics, but has a strong effect on the COM content of the ice mantles. The basic three-body process provides substantial improvement in the gas-phase abundances of COMs, notably methyl formate and dimethyl ether, by allowing the products of diffusive reactions (in some fraction of cases) to find a reaction partner themselves without requiring further diffusion. However, the excited formation mechanism, which allows the reaction of excited, newly-formed radicals with stable species (in spite of activation energy barriers) has the strongest effect, and is again most important for methyl formate and acetaldehyde. An adjustment to the efficiency of these processes, based on the available energy from the initiating reaction, appears to provide the best match with observational column densities of those molecules. 

It is important that the process that seems to reproduce most effectively the gas-phase abundances of the COMs (3-BEF) is one that occurs on the grain/ice surface itself, rather than deep within the mantle, allowing chemical desorption to return some fraction of the product to the gas phase.

The details of the various mechanisms and their implications are discussed in more detail below.

\subsection{H-abstraction/recombination as an amplifier of chemical desorption} \label{H-abstraction}

The models show substantial success in reproducing observed gas-phase column densities, through molecular production mechanisms operating on the surfaces of the icy dust grains. Consideration should therefore be given to the efficiency of the desorption mechanism that releases surface molecules into the gas phase. Although photo-desorption is included in all of the models presented here (with the explicit assumption of fragmentation of methanol as the result of this process), the most important ejection mechanism for grain-surface COMs is chemical desorption. In these models, this occurs with a {\em maximum} efficiency per reaction of 1\%; this efficiency is further lowered according to the RRK-based treatment described by Garrod et al. (2007).

Thus the formation of, for example, acetaldehyde, through Eqs. 18, culminating in the addition of an H-atom to the CH$_3$CO radical, may sometimes produce gas-phase CH$_3$CHO. However, the immediate desorption following its formation is not the only factor in ejecting those molecules into the gas. The chemical desorption effect is considerably amplified by the {\em abstraction} of H atoms from existing surface COMs, followed rapidly by recombination of the resulting radical with another H atom, inducing the ejection into the gas of some fraction of the product molecules. In the case of methanol, for instance, once it is formed on the grain surface through the repetitive addition of H to CO, the abstraction of H from CH$_3$OH by another H-atom allows it to be transformed back to its precursor ($\ce{CH3O/CH2OH}$), providing additional chances for chemical desorption -- indeed, this process of addition and abstraction was suggested by \citet{minissale16a} as a mechanism by which the depletion of CO from the gas phase could be slowed and its grain-surface conversion to methanol delayed. Similar H-abstraction/addition processes are present for each of the larger COMs of interest in our models.

To understand how significantly this process takes part in the overall chemical desorption scheme, four additional test models were run for conditions appropriate to the core center, turning off the H-abstraction reaction for each molecule (the three larger COMs plus methanol). The local fractional abundances of COMs from each test model are compared with the control in Table \ref{h-abstraction_core_center}. When the H-abstraction reaction of a specific COM is turned off, the gas-phase abundance of that molecule decreases by $\thicksim$1 order of magnitude. Furthermore, when H abstraction from methanol is switched off, it reduces the fractional abundance of other COMs such as $\ce{CH3OCHO}$ and $\ce{CH3OCH3}$, whose surface production is closely related to the $\ce{CH3O}$ radical. The abstraction of H from methanol by other H atoms in fact strongly favors the production of the CH$_2$OH radical; the network employed here uses surface reaction rates for these processes calculated by F. Goumans and S. Andersson \citep[see][]{garrod13} based on harmonic quantum transition state theory. However, as per the network of Garrod (2013), the recombination of CH$_2$OH with H is assumed to produce either methanol or H$_2$CO+H$_2$ with a branching ratio of 1:1. The production of formaldehyde in this way can then lead to reaction with H atoms again; this forward process strongly favors production of the CH$_3$O radical, thus influencing the production of DME and MF.

\begin{deluxetable}{cccccc}
\tablewidth{0pt}
\tabletypesize{\footnotesize}
\tablecolumns{6}
\tablecaption{Local fractional abundances of COMs at the core center when the H-abstraction reaction listed is {\em switched off} in the 3-BEF Best. \label{h-abstraction_core_center}}
\tablehead{
\colhead{} & \colhead{All on} & \colhead{$\ce{CH3OH}+\ce{H}$} & \colhead{$\ce{CH3CHO}+\ce{H}$} & \colhead{$\ce{CH3OCHO}+\ce{H}$} & \colhead{$\ce{CH3OCH3}+\ce{H}$}
}
\startdata
$\ce{CH3OH}$ & $2.9\times10^{-11}$  &  $5.4\times10^{-12}$  & $2.9\times10^{-11}$  & $2.8\times10^{-11} $ & $2.9\times10^{-11}$   \\
$\ce{CH3CHO}$ & $4.5\times10^{-13}$  &  $4.5\times10^{-13}$  & $3.8\times10^{-14} $ & $4.5\times10^{-13}$  & $4.5\times10^{-13}$   \\
$\ce{CH3OCHO}$ & $1.4\times10^{-12}$  & $5.5\times10^{-13}$  & $1.4\times10^{-12}$  & $7.2\times10^{-14}$  & $1.4\times10^{-12}$   \\
$\ce{CH3OCH3}$ & $2.6\times10^{-14}$  &$ 8.4\times10^{-15}$  & $2.6\times10^{-14}$  & $2.6\times10^{-14}$  & $1.4\times10^{-15}$   \\ 
\enddata
\end{deluxetable}

\subsection{COM distribution and COM peaks} \label{COMdistribution_3b-best}

As seen in Fig.~\ref{EffComp} for the 3-BEF Best model, the COMs in the gas phase have their lowest fractional abundances at the core center, gradually increasing toward the outer shell of the prestellar core. This general feature is observed regardless of model type (Fig.~\ref{radiVSgas-abundance}). Interestingly, a local fractional abundance peak for COMs is found at around 2000~AU, especially for \actd\ and \dme. This result suggests at least qualitative agreement with the observational result of \citet{jimenez-serra16}; those authors performed deep observations of the COMs toward the low-density outer shell (4000 AU) as well as the core center of \object{L1544}. While they observed higher abundances for all three COMs at the outer position, the level of enhancement for \mf\ was ambiguous, due to its large observational error.

The behavior seen in the models indicates that there are {\em two} possible peak features (or two plausible causes for observed peaks) that could become apparent in column densities or line intensities (e.g. lower panel of Fig.~\ref{CH3OH}) as opposed to fractional abundances. The first of these relates simply to the increased fractional abundances of COMs at large radii, combined with the drop-off in overall gas density at the greatest extents, producing a peak in the {\em absolute} molecular abundances that manifests in the resulting column density or line intensity profiles. This behavior is especially apparent for methyl formate (which does not show the bump-like feature at around 2000 AU). This peak seems to be in reasonably good agreement with the observational peak position; Fig.~\ref{CH3OH} indicates peak line intensities around 4000--6000 AU. A major cause of the lack of COMs in the gas phase at small radii (in terms of fractional abundance, e.g. Fig.~\ref{EffComp}) is that most of the gas-phase material at those locations has already accreted onto the grains and become locked into the ice mantles by the end-time of the models; little CO exists in the gas phase (on the order 10$^{-7}$ with respect to total H), thus grain-surface chemistry involving CO-related products is limited. At the greatest radii, freeze-out is incomplete and CO chemistry is still active with the accretion of new CO. Somewhat greater gas-phase abundances of atomic H at large radii, caused by the density slope, also encourage H-abstraction from COMs on the surfaces, followed by recombination and chemical desorption. 

The local peak at 2000~AU in the fractional abundances of acetaldehyde and dimethyl ether occurs particularly in models that use the 3-B and 3-BEF processes, and manifests also in the resultant column density profiles of those molecules (lower panel of Fig.~\ref{CH3OH}). The gas density at the 2000~AU position is at least three times higher than at the outer-peak region (4000--6000~AU), so the inner peak tends to dominate over the outer in its contribution to column densities, for the models/molecules in which that inner peak occurs.

What is the origin of the inner peak at 2000~AU? It is related to the freeze-out of gas-phase material through the core. It  traces a position at which the net rate of accretion of gas-phase material onto the grains is close to zero, caused by the high degree of depletion that has already occurred for most major gas-phase species other than hydrogen. For example, the gas-phase abundance of CO reaches a local minimum at this position. At radii internal and external to the 2000~AU peak region, there is slightly more gas-phase material remaining to be accreted onto dust grains at the end of the model runs, thus the ice mantles still continue (slowly) to grow. This local peak in freeze-out is due to the combined density and temperature profiles used in the models. The adsorption rates of neutrals scale with gas density, which is greatest at the core center, but they also scale with the square root of the gas temperature, which is greater at larger radii. The 2000~AU position is the point where the two profiles combine to give the largest total adsorption rate. The position of the maximum freeze-out position is thus strongly dependent on the density and temperature profiles. Furthermore, given a slightly longer model run time, the freeze-out peak would likely widen, as other positions reached a state of near-zero net accretion onto grains.

The stronger production of COMs (acetaldehyde and dimether ether) around this 2000~AU position peak is a consequence of the changing freeze-out conditions described above. Once the net rate of freeze-out reaches zero, it indeed undergoes a reversal in which there is a small, net rate of {\em loss} of material from the grains. This loss is caused by the desorption of molecular hydrogen from the grain surface, which is slowly replenished by the gradual outward diffusion of H$_2$ molecules embedded deep in the ice mantles. This H$_2$-loss process occurs throughout all the model runs, but is of little importance until the adsorption of non-volatile species diminishes, when gas-phase species become depleted. Once this net loss of material from the grains starts to occur, some molecules embedded in the upper layer of the ice mantles are ``uncovered'', becoming available for surface chemical processing. Most importantly, this includes CH$_4$, from which an H-atom may be chemically abstracted through several mechanisms, increasing both the production rate of CH$_3$ and its surface abundance. This drives up the three-body production of acetaldehyde and dimethyl ether (Eqs. 12--15), which are chemically desorbed into the gas phase -- either directly, or as the result of H-abstraction and recombination on the surface.

The behavior of the inner peak in certain COMs should therefore be treated with a degree of skepticism. Not only does its position depend on the interplay of the observationally-determined physical profiles, but its strength must be time-dependent. Furthermore, the ability of the chemical model to treat accurately the return (``uncovering'') of mantle material to the ice surface is limited by the use of only a single mantle phase, rather than the consideration of distinct layers within the ice \citep[cf.][]{taquet14}. If most of the methane residing in the mantles is present mainly in the deepest layers, the inner-peak effect described above would be overestimated here. It is also the case that, even with this mechanism in play, the gas-phase abundance of dimethyl ether is insufficient to reproduce observed column densities in \object{L1544} (although see \S~4.6). If such a mechanism is active, considering the uncertainty in its precise position (based on models), it may not be easily distinguished from the outer peak at 4000+~AU.

The peak in methanol column density occurs at the outer peak position, caused again by a peak in {\em absolute} abundance of that molecule. It is noteworthy that in the present models, the local fractional abundance of methanol does not need to exceed a value of a few 10$^{-9}$ to be able to reproduce the observed column density. Again, the strength of the methanol peak will be dependent on the efficiency of chemical desorption for that molecule, which is not well constrained through purely experimental means.

\subsection{The effect of diffusion barriers} \label{bindingenergy}

\begin{figure*}
\centering
\includegraphics[width=0.75\textwidth]{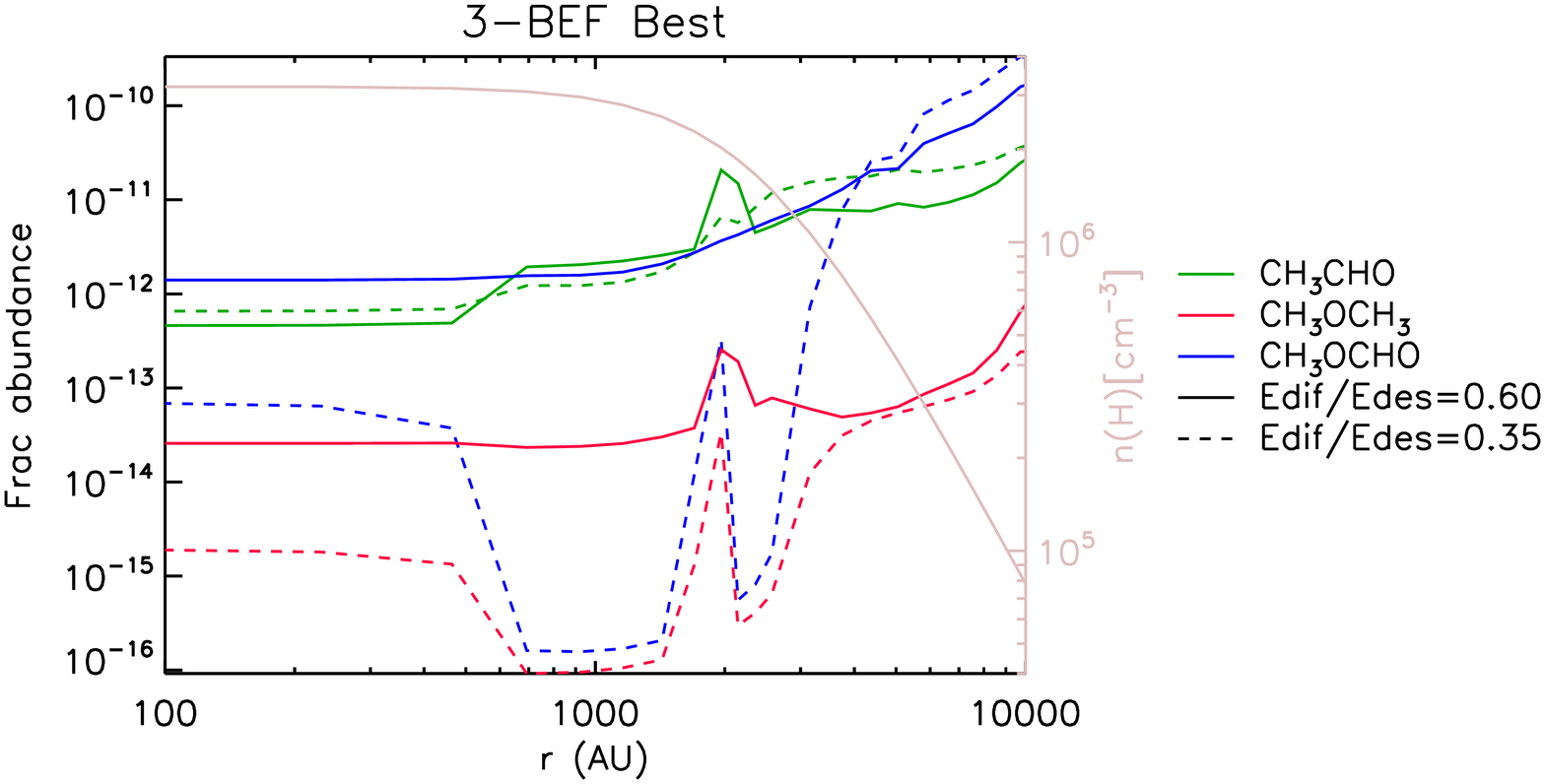}
\caption{Comparison of chemical distribution of the COMs in the 3-BEF Best models with different values of $E_{\mathrm{dif}}$:$E_{\mathrm{des}}$ for atoms.
\label{COMEdifComp}}
\end{figure*}

\begin{figure*}
\centering
\includegraphics[width=0.75\textwidth]{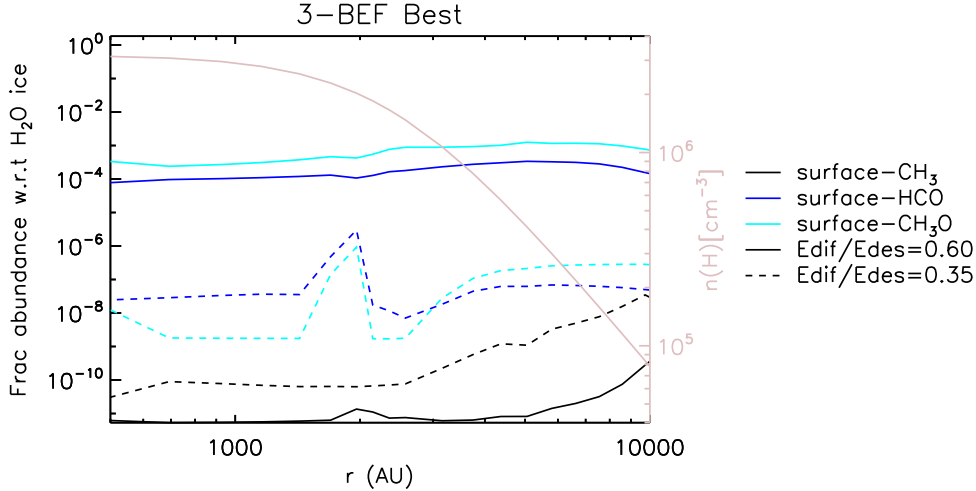}
\caption{Comparison of chemical distribution of the reactants in the 3-BEF Best models with different values of $E_{\mathrm{dif}}$:$E_{\mathrm{des}}$ for atoms.
\label{ReactantsEdifComp}}
\end{figure*}

In many astrochemical models including {\em MAGICKAL}, chemistry on the grains is governed by the diffusion of surface species via thermal hopping (any non-diffusive processes notwithstanding). The energy required for a particular species to hop from one surface binding site to another is given by the diffusion barrier $E_{\mathrm{dif}}$; this value is parameterized in the chemical model as some fraction of the desorption energy, i.e. $E_{\mathrm{dif}}$/$E_{\mathrm{des}}$. Even though this is a key parameter to describe the mobility of species on grain surfaces, the exact value has not historically been well constrained, broadly ranging from 0.3 to 0.8. In the present models, this parameter was set to $E_{\mathrm{dif}}$/$E_{\mathrm{des}}$=0.6 for all atomic species, which leans toward a high value based on recent experimental estimates by \citet{minissale16b}, who suggested 0.55 for atoms. In our past models (e.g. Garrod 2013), atoms and molecules were assigned the same fractional barrier value of 0.35, based on the optimum value for CO. All {\em molecular} species in the present models retain the 0.35 value.

At the very low surface temperatures that are found in prestellar cores, the diffusion of atoms in particular is of great importance. For this reason, test models were also run using the previous fractional diffusion barrier of 0.35 for atoms.
Figure~\ref{COMEdifComp} shows a comparison of COM abundances for the two cases, shown for the end-time of the 3-BEF Best model run using the \object{L1544} physical profiles as before. Using the higher $E_{\mathrm{dif}}$/$E_{\mathrm{des}}$ ratio, the COMs typically show much higher abundances at positions near the core center. This result is somewhat contradictory to the expectations of \citet{vasyunin17}, who suggest that the $E_{\mathrm{dif}}$/$E_{\mathrm{des}}$ ratio would not play a crucial role in cold environments, as diffusion of H and $\ce{H2}$ via tunneling is dominant. In our model,while tunneling through chemical barriers is included, surface diffusion via tunneling is not, as the barriers are assumed to be too broad for tunneling to be effective. In this case, the higher diffusion barrier for the atomic species means that the time taken for H atoms to reach and react with surface radicals is increased. This consequently raises the lifetimes of those radicals on the surface (see Fig.~\ref{ReactantsEdifComp}), which in turn renders the non-thermal mechanisms explored here more effective, increasing the production of COMs.

It should be noted that the higher $E_{\mathrm{dif}}$/$E_{\mathrm{des}}$ ratio does not always result in a larger amount of COMs (or radicals) on the grain surface. For example, the discrepancy in the COM abundances between the two models decreases at large radii, and methyl formate and acetaldehyde here are even a little more abundant in the case where atomic diffusion barriers are lower, due to slightly more effective H-abstraction from methane to produce CH$_3$.

The variation of the $E_{\mathrm{dif}}$/$E_{\mathrm{des}}$ ratios thus has an important effect on the chemical model results; the higher value for atomic species, and for H in particular, reproduces COM abundances more effectively, through the increase in radical lifetimes. \cite{sen17} calculated the distribution of binding energies and diffusion barriers for H on an amorphous water surface, suggesting representative values for each; although their H binding energy (661~K) is higher than the value used in our models (450~K), their diffusion barrier (243~K) is close to the value we use here (270~K) for the $E_{\mathrm{dif}}$/$E_{\mathrm{des}}$=0.6 models. We note also that \cite{sen17}, based on their calculations of diffusion rates using quantum transition-state theory, aver that tunneling (as opposed to the thermal mechanism) is likely of limited importance under most temperature conditions in dark clouds; our use of purely thermal diffusion rates in {\em MAGICKAL} is thus broadly consistent with that work.

In other models, in which non-diffusive chemical processes were {\em not} included, the variation of the H diffusion barrier might have less of an effect, as most of the active chemistry in that case would involve {\em only} atomic H. The lifetime of the radicals would therefore be of less relevance, since H would still be the dominant reaction partner. In the present models, the mobility of atomic hydrogen is a major determinant of the effectiveness of non-thermal processes in producing complex species.

\subsection{Gas-phase processes}

Due perhaps to the generally low abundances of DME that our chemical models provide, they do not appear to reproduce the correlation between \mf\ and \dme\ sometimes observed in various evolutionary stages of star-forming regions~\citep{jorgensen11, brouillet13, jaber14}. As a means by which such a relationship might arise, \citet{brouillet13} suggested protonated methanol $\ce{CH3OH2+}$ in the gas-phase as the common precursor to form \mf\ and \dme\ via reactions with $\ce{HCOOH}$ and $\ce{CH3OH}$. As a test, the proposed reactions were incorporated into our chemical network; however, they were too slow to be effective in producing \mf\ and \dme\ in our model, due to the low abundance of protonated methanol in the gas.

Recently, potentially influential gas-phase reactions were proposed by \citet{shannon13,shannon14}, who found that reactions of either OH or C($^{3}\textrm{P}$) with methanol are efficient at low temperatures, due to quantum tunneling:
\begin{eqnarray}
\ce{OH}~+~\ce{CH3OH}~&\rightarrow &~\ce{CH3O}~+~\ce{H2O} \nonumber \\
\ce{C} (^{3}\textrm{P})~+~\ce{CH3OH}~&\rightarrow &~\ce{CH3}~+~\ce{HCO}\nonumber
\end{eqnarray}
The gas-phase methanol reactions could act not only as an efficient loss process for gas-phase methanol, but also produce more radicals that would be available as reactants to form other COMs when they accrete onto grain surfaces, or directly in the gas phase if such processes are efficient. \citet{vasyuninandherbst13} suggested a gas-phase radiative association reaction between the radicals that are produced by the above mechanisms, to form DME:
\begin{eqnarray}
\ce{CH3O}~+~\ce{CH3} &\rightarrow& \ce{CH3OCH3} + h\nu \nonumber
\end{eqnarray}
To understand how significantly these reactions would affect the overall formation of COMs in our chemical model, we ran a test model that included all three (with rate coefficients on the order of $10^{-10}$ cm$^{-3}$). However, gas-phase methanol was still predominantly destroyed by ion-molecule reactions at the core center. The contribution of the above neutral-neutral reactions to the loss of gas-phase methanol was minor ($\thicksim 2\%$), hardly changing the abundances of methanol and the three COMs, while the radiative association reaction also showed minimal influence.

\citet{balucani15} proposed a gas-phase mechanism that would form methyl formate from dimethyl ether through the radical $\ce{CH3OCH2}$. The dimethyl ether itself would form through the efficient radiative association of the radicals CH$_3$ and CH$_3$O :
\begin{gather}
\ce{CH3OCH3}~+~\ce{(F,Cl)} \rightarrow \ce{CH3OCH2}+~\ce{(HF,HCl)} \nonumber \\
\ce{CH3OCH2}~+~\ce{O} \rightarrow \ce{HCOOCH3}~+~\ce{H} \nonumber
\end{gather}
Although our network does not include fluorine, the incorporation of the other reactions into our model did not make a meaningful difference to the results, because they involve a one-way process where \dme\ is converted into \mf. In our chemical model, neither the radiative association of the radicals CH$_3$ and CH$_3$O nor any other processes were efficient enough to form abundant \dme . As such, several key reactions concerning gas-phase chemistry of COMs do not affect our chemical model significantly. 

Thus, at least under the conditions tested in our physical model, we found no efficient gas-phase mechanisms that could produce either DME or MF.

\subsection{Other surface processes}

The 3-body excited-formation mechanism included here is especially efficient for the initiating reaction H + CH$_2$ $\rightarrow$ CH$_3$, which is highly exothermic, but which also results in a small product, CH$_3$, that has only a limited number of vibrational modes in which the resulting energy may be stored. The models suggest that when this is coupled with highly abundant CO on the grain surface, the subsequent reaction between the two proceeds at a sufficient pace to produce enough CH$_3$CO (and thence CH$_3$CHO) to be able to explain the gas-phase abundance of the latter molecule (given an adequate desorption mechanism). The production of CH$_3$O via the hydrogenation of formaldehyde is also exothermic, but not sufficiently so to allow the subsequent reaction with abundant CO to proceed at high efficiency. Nevertheless, this low-efficiency mechanism is capable of producing enough CH$_3$OCO (and thence CH$_3$OCHO) to account for the presence of methyl formate in the gas phase.

As noted in Section 3.3, a more detailed treatment of the 3-BEF mechanism should include not only the energy partition between bonds, but also translational degrees of freedom of the excited species. This would impact the RRK calculation, but could also provide an alternative outcome to the process. The RRK treatment as formulated in Section 3.3 assumes that the efficiency of the process is determined solely by the competition between energy going into the ``reaction mode'' and energy being lost to the surface. However, if diffusion spontaneously occurred, moving the two reactants apart, then the process would automatically end (unless another reactant were present in this new site), regardless of the energy status of the excited species. We would expect this effect to reduce efficiency by a factor of say 4 (on the basis of there being four available diffusion directions), even for the otherwise efficient 3-BEF mechanism that produces CH$_3$CO/CH$_3$CHO. The production of CH$_3$CHO may therefore be somewhat less efficient than the simple 100\% approximation used in the treatment presented here.

It is also of interest to consider specifically the possible effects of reaction-induced diffusion, such as that studied by \citet{fredon17} for stable molecules including methane. If reactive species like CH$_3$ were able to undergo some non-thermal diffusion as the result of excitation caused by their formation, they could react with other radicals that they could not otherwise reach under low-temperature (i.e. non-diffusive) conditions. In fact, as we allude in Section 2.4, the standard (non-excited) 3-B mechanism that we already implement in the models will automatically include such processes to a first approximation. The treatment that we construct for 3-B processes does not explicitly require the reactants to be immediately contiguous, but rather to become so immediately following the initiating reaction. If one were to consider a newly-formed radical species, $A$, taking some finite and approximately straight-line trajectory across an ice surface, the probability of it encountering some reaction partner, $B$, along its path would still be given, to first order, by $N(B)/N_S$, as already included in Eq. (6). The simple 3-B mechanism is therefore broad enough to cover this specific case also.

While the 3-BEF mechanism for the production of the dimethyl ether precursor, CH$_3$OCH$_2$, should be highly efficient based on the statistical calculations in \S~\ref{3bef-best}, the lower abundance of H$_2$CO on the grain surfaces appears to be too low to allow this mechanism to account for gas-phase DME. It should also be noted that in this work it was assumed that DME is the only product of this reaction. It is possible, and perhaps favorable, for ethanol (C$_2$H$_5$OH) also ultimately to be formed, if the methyl radical attaches to the carbon atom in formaldehyde, producing a radical C$_2$H$_5$O. This would naturally limit the yield of DME through the suggested excited-formation mechanism.

Are there alternative surface processes that might produce sufficient dimethyl ether if the reactants could be brought together through some non-diffusive process? One possibility might be the reactions of the carbene CH$_2$ with methanol (CH$_3$OH). Methylene, CH$_2$, is a diradical in its ground (triplet) state. Reactions of triplet CH$_2$ with methanol could involve the abstraction of hydrogen from CH$_3$OH, followed by immediate radical-radical addition of the resultant CH$_3$ to the remaining CH$_3$O or CH$_2$OH. The review of \cite{tsang87} suggests gas-phase rate coefficients for the abstraction processes (without the subsequent recombination); the activation barrier for the CH$_3$O branch is marginally lower than that for CH$_2$OH, indicating that CH$_3$O (and thence DME) might be the preferred product. On a dust-grain/ice surface, the production of CH$_2$, either by H-addition to CH or by the barrier-mediated reaction of H$_2$ with atomic C, would likely be exothermic enough to allow the subsequent abstraction barriers to be overcome. However, abstraction might be fast in any case, even without (vibrationally) excited CH$_2$, due to hydrogen tunneling through the activation-energy barrier.

Another possibility is that the higher-energy {\em singlet} CH$_2$ could undergo a direct, barrierless insertion into the methanol molecule, producing either dimethyl ether or ethanol. \cite{bergantini18} investigated the action of singlet CH$_2$, produced through the irradiation of a mixed CH$_4$/CH$_3$OH ice, to produce DME and ethanol in this way; they found essentially equal production of the two branches. If, instead of the dissociation of methane, the hydrogenation of carbon on the grain surfaces were the means by which singlet CH$_2$ were produced, then this mechanism could occur effectively as a non-diffusive (i.e. three-body) process, although the short lifetime of the singlet methylene might make a diffusive meeting of the reactants unlikely. Although the dissociation of methane, as per those experiments, is an entirely plausible starting point for the production of COMs within ice mantles, it is an unlikely explanation for the {\em gas-phase} detection of COMs.

A further consideration, relating to the production of CH$_3$CHO via the 3-BEF mechanism, is the possible production of ketene, CH$_2$CO, through the reaction CH$_2$ + CO $\rightarrow$ CH$_2$CO. This process could also occur through the 3-BEF mechanism, following production of methylene through exothermic surface reactions. The more complex nature of the coding of the 3-BEF mechanism required us to include only the three 3-BEF mechanisms directly related to MF, MDE and AA in the present work, but the application of this mechanism to the full chemical network might impact ketene production. An immediate question would be whether the ketene production might also preclude the production of acetaldehyde, as the CH$_2$ used to produce ketene would otherwise be required to produce the CH$_3$ needed for AA production. Furthermore, one could argue that the production of CH$_3$ in the presence of CO, as needed for our 3-BEF route to AA, would first require contiguous CH$_2$ and CO, and that this CH$_2$ would also have to be formed in the presence of CO, making ketene the preferred product {\em instead} of AA. Such a view implicitly assumes that there is no reaction-induced diffusion occurring, when in fact, due to the large exothermicities of the reactions in question, it is highly likely that there is some form of diffusion following each reaction. As mentioned above, this diffusion does not make either the 3-B nor the 3-BEF treatments any less accurate, as we do not explicitly rule out such occurrences. Rather, it might be better to assume that, on a surface at least, such reaction-induced diffusion is the rule, rather than the exception, and thus that there is no expectation nor requirement that any newly-formed reaction product considered in the 3-B mechanisms necessarily meets it own reaction partner within its immediate surroundings. In that case, any conditionality in the production of one species from another, based on location, would be lost. On the topic of ketene in particular, it is also possible that it may be hydrogenated to acetaldehyde anyway; the reaction H + CH$_2$CO $\rightarrow$ CH$_3$CO is assumed in our network to have a barrier of 1320~K \citep{senosiain06}, which is lower than, for example, the typically-assumed barrier to hydrogenation of CO. In future work, we will apply the 3-BEF process to the entire surface network, allowing the relationship between acetaldehyde and ketene to be explored more deeply.

\cite{fedoseev15} conducted experiments investigating the production of COMs, specifically glycolaldehyde and ethylene glycol, through non-diffusive surface reactions of HCO radicals produced through H and CO co-deposition. Those experiments did not detect any methyl formate production, but follow-up work by \cite{chuang2016}, who co-deposited various combinations of CO, H$_2$CO, CH$_3$OH and H, demonstrated methyl formate production in the setups that involved direct deposition of formaldehyde (H$_2$CO). Thus, under the conditions of their experiments, the HCO and CH$_3$O radicals required for the radical-radical reactions to produce CH$_3$OCHO derived mainly or uniquely from H$_2$CO reactions with atomic H (either H addition or H-abstraction by H atoms). In the case of CO and H deposition alone, they suggested that the reaction between two HCO radicals dominates, producing glyoxal (HCOCHO) that can be further hydrogenated to glycolaldehyde (CH$_2$(OH)CHO) and ethylene glycol ($\ce{HOCH$_2$CH$_2$OH}$).

The reaction network we use here includes HCO--HCO reaction routes \citep{garrod08a}, with one branch producing glyoxal, and an equal branch producing CO and H$_2$CO through a barrierless H-abstraction process. Our network does not include the further hydrogenation of glyoxal, but the removal of HCO radicals should be well enough treated. 

In our astrochemical models (in which the overall system is much more complicated than the laboratory setups, with many more species and processes), grain-surface formaldehyde and methanol are produced through CO hydrogenation by H atoms; in the experiments, the CO + H system does not produce enough formaldehyde to allow substantial production of CH$_3$O and thence methyl formate. The outcomes of the models, which are run over astronomical timescales, should not therefore be expected to correspond directly with the experimental outcomes. However, the mechanism of non-diffusive radical chemistry that seems to produce methyl formate in the experiments (via HCO + CH$_3$O) is present in our models (the basic three-body mechanism).

The key comparison with the experiment concerns our implementation of the excited three-body formation mechanism for the reaction of CH$_3$O with CO (the other 3-BEF mechanisms tested in this work involve CH$_3$ and are therefore not tested in the laboratory experiments). The very low efficiency that we require (0.1\%) for immediate reaction is likely to be too small to have an important effect in a laboratory regime where the regular three-body process is presumably efficient (unlike in our prestellar core models). In this sense, it seems superficially consistent with the experiments, since there are no experimental setups in which methyl formate was not found in which our excited formation mechanism would predict it to be highly abundant. Indeed, our mechanism should only become important if other means of production (such as the regular three-body process) are already weak. Thus, it may be difficult to test the excited-formation mechanism for methyl formate through experimental means.

Another possibility exists for the production of all three COMs considered here: that is, that they are produced in the ice mantles, through UV processing (or some other means). The ice mantle material would then have to be removed by some violent process such as sputtering by cosmic rays. \cite[Such a process would also result in some degree of complex molecule production, e.g.][]{shingledecker18}. However, it is unclear whether such mechanisms would be capable of maintaining gas-phase abundances of COMs at the required levels.

A separate point of discussion concerns the experimental evidence surrounding the interaction of H atoms specifically with solid-phase acetaldehyde. \citet{bisschop07} studied the deposition of H onto a pre-deposited surface of pure CH$_3$CHO at temperatures ranging from 12.4 -- 19.3~K. They found reaction products C$_2$H$_5$OH, H$_2$CO, CH$_3$OH, and CH$_4$, which they posited to be formed either through repetitive hydrogenation (ethanol), or fragmentation into a stable molecule and a radical, which may be further hydrogenated to a stable species. In our model, it is assumed that H atoms interact with CH$_3$CHO by abstracting another hydrogen atom from the aldehyde end of the molecule. If the alternative mechanisms measured in the laboratory should compete strongly with this process, then the mechanism described in Section \ref{H-abstraction}, in which H-abstraction and re-hydrogenation work together to enhance reactive desorption, could become less effective, and the acetaldehyde produced on the surface could be converted to entirely different species. 

The \citet{bisschop07} data suggest production yields for ethanol of $\sim$20\%, with other products also on the order of 10\%. However, these yields are provided as a fraction of the acetaldehyde initially available in the surface layer of the ice; they do not indicate yields per hydrogen atom or per H--CH$_3$CHO interaction. Furthermore, the experiments would not appear to be sensitive to processes in which acetaldehyde were converted to CH$_3$CO, then re-hydrogenated to CH$_3$CHO. As a result, it is not possible to determine how strongly H-abstraction may dominate over hydrogenation or fragmentation, or vice versa. However, each of these processes would involve an activation energy barrier, and it is found that abstraction from aldehyde groups occurs more readily than H-addition. \citet{hippler02} calculated barriers to such processes, including the H + CH$_3$CHO $\rightarrow$ C$_2$H$_5$O addition reaction, finding an activation energy of 22.4 kJ mol$^{-1}$ (2690~K), versus the literature value for abstraction of 17.6 kJ mol$^{-1}$ \citep[2120~K][]{warnatz84}. Assuming the simple rectangular-barrier tunneling treatment used in our models, and assuming a 1~\AA\, barrier width, the abstraction process should go around 350 times faster than hydrogenation. The preferred gas-phase value in the more recent review by \citet{curran06} suggests an even higher barrier to hydrogenation of 26.8 kJ mol$^{-1}$ (3220~K), which would provide an abstraction/hydrogenation ratio closer to $10^5$. Fragmentation is more sparsely studied in the literature, but based on the Bisschop et al. study we presume those mechanisms to occur at similar rates to the hydrogenation mechanism. Since chemical desorption in our model is calculated to proceed in a little less than 1\% of cases, we would not expect our results for acetaldehyde to be strongly affected by the inclusion of alternative reaction branches, either on the grains or in the gas-phase.

\subsection{$\ce{O2}$ production} \label{o2incomet}

\begin{figure*}
\gridline{\fig{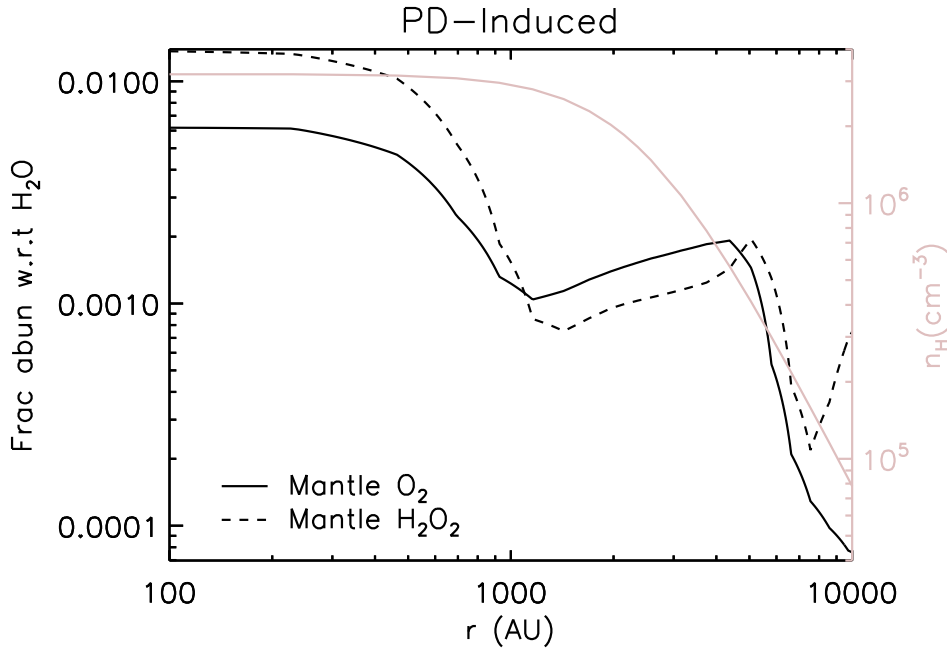}{0.75\textwidth}{}}
\caption{
Radial distribution of the solid-phase abundances of $\ce{O2}$~(solid lines) and $\ce{H2O2}$~(dashed lines) in the PD-Induced model. \label{MO2MH2O2}}
\end{figure*}

Aside from its effect on COM abudances in the ice mantles, the PD-induced reaction mechanism also produces a significant increase in $\ce{O2}$ ice abundance; this effect is noteworthy, as it may provide a clue to the origin of $\ce{O2}$ in comets. Gas-phase $\ce{O2}$ was recently observed toward comet 67P/C-G, as part of the Rosetta mission \citep{bieler15}. It was found that $\ce{O2}$ achieves a fractional abundance as high as $\thicksim$4\% with respect to water, indicating this compound as one of the most dominant species in cometary material. While the origin of molecular oxygen is still controversial because of its difficulty in observation, the strong correlation with $\ce{H2O}$ implies a connection to dust-grain ice chemistry rather than gas-phase chemistry in the coma. 

Many studies directly or indirectly suggest a primordial nature for $\ce{O2}$ in comets. For example, \citet{rubin15} confirmed the presence of $\ce{O2}$ in the Oort cloud comet 1P/Halley at a level similar to that seen in the Jupiter-family comet 67P/C-G. This suggests that $\ce{O2}$ may be common, regardless of dynamical history, indicating a primordial origin. \citet{mousis16} proposed that the radiolysis of water-containing interstellar ices in low-density environments such as molecular clouds could produce $\ce{O2}$ in high abundance. Meanwhile, \citet{taquet16} conducted a range of astrochemical models based on diffusive grain-surface chemistry, to investigate three possible origins for $\ce{O2}$ in comets: (i) dark cloud chemistry; (ii) formation in protostellar disks; and (iii) luminosity outbursts in disks. They concluded that dark clouds are the most plausible regime in which to form the $\ce{O2}$, through diffusive O-atom addition on grain/ice surfaces at temperatures around 20~K. However, as they noted, the temperature required in their models is rather higher than typical dark cloud values. \citet{garrod19} suggested that the upper layers of cold-storage comets could be processed to increase their O$_2$ content, as the result of photolysis and radiolysis by the galactic UV and cosmic ray fields. The chemical models presented by Garrod included the same PD-induced mechanism used in the present study.

Here, we propose non-diffusive, photodissociation-induced processing of interstellar ice mantles as the possible origin of the abundant $\ce{O2}$ in comets; this avoids the requirement for higher-temperature diffusive chemistry on dust-grain surfaces. 

Fig. \ref{MO2MH2O2} shows the radial distribution of the $\ce{O2}$ ice~(solid lines) in our model of \object{L1544} with the PD-induced process activated. The fractional abundance of $\ce{O2}$ ice in this model is as high as 0.6~$\%$ with respect to water ice toward the core center, which is 1-2 orders of magnitude higher than in the other models. The abundance of $\ce{O2}$ ice has been suggested to be low in prestellar core material, because $\ce{O2}$ is efficiently hydrogenated to form $\ce{H2O}$ and $\ce{H2O2}$ ices at low temperature~\citep{ioppolo08}. However, the PD-Induced process induces the association of O-atoms in the icy mantle (sourced from water molecules), resulting in the production of large amounts of $\ce{O2}$. The other models (without PD-induced reactions) accumulate the $\ce{O2}$ ice on the grain-surface over time rather than directly synthesizing it within the ice mantles. If interstellar $\ce{O2}$ ice formed via the PD-induced process were locked in until such ices were incorporated into comets, it could remain there to contribute to the $\ce{O2}$ population observed in 67P/C-G.

It should be noted that $\ce{H2O2}$ ice in the PD-induced model is more abundant than $\ce{O2}$ ice (dashed line; Fig. \ref{MO2MH2O2}), in contrast to observations. The $\ce{H2O2}/\ce{O2}$ ratio found in 67P/C-G by ROSINA was as low as $6.0\pm0.7\times10^{-3}$ \citep{bieler15}. The same discrepancy was found by \citet{garrod19}, and is related to the relative formation rates of other compounds in the ice as well as $\ce{O2}$. Photodissociation of water leads first to OH; two such radicals may recombine through the PD-induced processes to form $\ce{H2O2}$ in high abundance. The present chemical network may be missing some chemical reactions related to the destruction of $\ce{H2O2}$, or may overestimate the efficiency to form $\ce{H2O2}$ via the PD-Induced process. The efficiency of H-atom diffusion within the bulk ice to abstract another H atom from H$_2$O$_2$ may also be an important factor; Fig. \ref{MO2MH2O2} shows that the H$_2$O$_2$ abundance strongly dominates O$_2$ in the coldest (inner) regions of the core. The sum of the O$_2$ and H$_2$O$_2$ abundances collectively reach a value on the order of 1\% in total at the core center. Inclusion of O$_3$ and O$_2$H boosts this total further. The photodissociation that leads to O$_2$, H$_2$O$_2$ and other species' production from H$_2$O in these models is the result of the secondary, cosmic ray-induced UV field. Thus, the abundances of each would likely be enhanced by the adoption of a somewhat larger cosmic-ray ionization rate. A longer evolutionary timescale, or further processing of the ices during the later disk stage, could also lead to enhancement. Radiolysis, i.e. direct cosmic-ray impingement on the ice mantles could also act in concert with the photodissociation effect.

Further modeling to reproduce the amount of O$_2$ and \ce{H2O2} seen in 67P/C-G is outside the scope of this work, but may be a fruitful means to elucidate the origins of cometary \ce{O2}. The vast majority of this \ce{O2} could indeed be interstellar, produced by photolysis.

\section{Conclusions} 

Here we have introduced new rate formulations that allow astrochemical models to simulate a number of new, non-diffusive chemical mechanisms on interstellar dust-grain surfaces and within bulk ices. These formulations are fully compatible with existing model treatments for diffusive chemistry. Some of the non-diffusive mechanisms considered here, such as the Eley-Rideal process and three-body reactions, are automatically taken into account in microscopic Monte Carlo kinetics models of the same systems, but must be explicitly added in rate-based treatments such as the one used here. Others, such as the three-body excited-formation mechanism and photodissociation-induced reactions, are entirely new.

Crucially, it is shown that non-diffusive processes can affect the bulk-ice, the ice-surface, and -- indirectly -- the gas-phase composition, through a cyclic H-abstraction and addition process that amplifies the efficiency of chemical desorption. Eley-Rideal reaction processes appear not to have a strong effect in our implementation.

To place the new non-diffusive mechanisms into a context in which they could be directly tested, a physical model approximating the prestellar core \object{L1544} was adopted, with a focus on reproducing the observed gas-phase abundances of the complex organic molecules (COMs) acetaldehyde, dimethyl ether, and methyl formate. Reactions involving excited radicals (recently produced by other surface or bulk reactions) appear to be influential in producing COMs, although their influence is likely limited to reactions involving the highly-abundant CO molecule on the grain surfaces, and/or the excited radical CH$_3$, whose production through the addition of H to CH$_2$ is especially exothermic. In the three-body excited formation model, the efficiency of the 3-BEF mechanism for methyl formate production must indeed be optimized, as this mechanism is otherwise more efficient than is required to agree with observations. Although the gas-phase dimethyl ether abundance in particular remains difficult to reproduce, the models presented here tested only three plausible excited-formation mechanisms; alternative surface-formation processes for this molecule could be active.

Further application of the new mechanisms and formulations presented here into astrochemical models of the later stages of star formation, such as the hot core/corino stage, is currently underway.

The main conclusions of our study are enumerated below:

\begin{enumerate}
\item Non-diffusive reactions between newly-formed radicals and nearby species on dust grains, which we label three-body reactions, appear to influence strongly the production of complex organics and other species on interstellar dust grains and in their ice mantles, producing abundances similar to those detected in the gas phase toward hot star-forming sources.

\item We propose a new surface/bulk mechanism in which the energy of formation of a newly-formed radical allows it to overcome an activation energy barrier against reaction with a nearby, stable species (the three-body excited-formation process, 3-BEF). For key molecules/processes, especially reactions between excited radicals and abundant grain-surface CO, this mechanism appears strongly to influence production, beyond the effects of the regular three-body mechanism. Except for these key processes, the 3-BEF process is expected to be inefficient in most cases.

\item Grain-surface molecule production is enhanced by the three-body excited-formation mechanism sufficiently to explain observed {\em gas-phase} abundances of methyl formate and acetaldehyde in prestellar core \object{L1544}. Chemical desorption allows these grain surface-formed molecules to enter the gas phase.

\item Dimethyl ether is still under-produced in the model, when compared with observations of L1544. Other plausible grain-surface production mechanisms, such as reactions between CH$_2$ and methanol, remain to be tested in the models.

\item Repetitive H-abstraction by H atoms from COMs on grain surfaces, followed by recombination with another H atom and the possible desorption of the product into the gas phase, gives chemical desorption a greater influence than its basic efficiency of around 1\% would otherwise suggest. This cyclic amplification effect brings the required surface-formed COMs into the gas phase effectively enough to reproduce abundances as described above. The effect should be especially important in regions where gas-phase H abundances remain relatively high.

\item Specific to the L1544 models, the position of the methanol peak is located further outward than observation, but it is still associated with the region where CO starts to freeze out significantly. The off-center peaks in COM column densities toward L1544 are most likely related to the interplay between rising COM fractional abundances at larger radii and rising gas density at smaller radii.

\item The surface-diffusion rate of atomic H is important to radical lifetimes, which affects the efficiency of non-diffusive mechanisms that rely on reactive radicals being available on the grain surfaces. Thus the choice of diffusion barrier for H has a strong effect on the production of COMs in chemical models that consider non-diffusive chemistry.

\item Photodissociation-induced non-diffusive chemistry within the bulk ices produces abundances of O$_2$ and related species on the order of 1\% of water. This suggests that interstellar (and perhaps later) UV-processing of grain-surface ices may be sufficient to reproduce observed cometary values, regardless of the precise temperature of the dust grains.

\item The broader inclusion of various non-diffusive grain-surface/ice chemical reactions in interstellar chemical models now seems imperative.
\end{enumerate}

\acknowledgments

We thank the anonymous referees for helpful comments and suggestions. We are grateful to E. Herbst and P. Caselli for useful discussions. This work was partially funded by a grant from the National Science Foundation (Grant number AST 19-06489).

\section{Appendix}
 
\begin{deluxetable}{llll}
\tablewidth{0pt}
\tabletypesize{\footnotesize}
\tablecolumns{4}
\tablecaption{Gas-phase reactions newly included in chemical network \label{gas_rxn}}
\scriptsize
\tablehead{
\colhead{Reaction} & \colhead{Rate coefficient} & \colhead{Reference }& \colhead{Type }}
\startdata
$\ce{CH3OCH2} + \ce{H3+}  \rightarrow \ce{CH3OCH3} + \ce{H2}            $  & $3.21 \times 10^{ -9}~(T/300~\textrm{K})^{-0.5}~\textrm{cm}^{3}~\textrm{s}^{-1}$  & a & ion-neutral\\
$\ce{CH3OCH2} + \ce{H3O+}  \rightarrow \ce{CH3OCH3} + \ce{H2O}           $  & $1.48 \times 10^{ -9}~(T/300~\textrm{K})^{-0.5}~\textrm{cm}^{3}~\textrm{s}^{-1}$  & a & ion-neutral\\
$\ce{CH3OCH2} + \ce{HCO+}  \rightarrow \ce{CH3OCH3} + \ce{CO}            $  & $1.29 \times 10^{ -9}~(T/300~\textrm{K})^{-0.5}~\textrm{cm}^{3}~\textrm{s}^{-1}$  & a & ion-neutral\\
$\ce{CH3OCH2} + \ce{C+}    \rightarrow \ce{CH3OCH3} + \ce{C}    + \ce{He}$  & $1.76 \times 10^{ -9}~(T/300~\textrm{K})^{-0.5}~\textrm{cm}^{3}~\textrm{s}^{-1}$  & a & ion-neutral \\
$\ce{CH3OCH2} + \ce{He+}   \rightarrow \ce{CH2+}    + \ce{CH3O} + \ce{He}$  & $7.06 \times 10^{-10}~(T/300~\textrm{K})^{-0.5}~\textrm{cm}^{3}~\textrm{s}^{-1}$  & a & ion-neutral \\
$\ce{CH3OCH2} + \ce{He+}   \rightarrow \ce{CH3O+}   + \ce{CH2}  + \ce{He}$  & $7.06 \times 10^{-10}~(T/300~\textrm{K})^{-0.5}~\textrm{cm}^{3}~\textrm{s}^{-1}$  & a & ion-neutral \\
$\ce{CH3OCH2} + \ce{He+}   \rightarrow \ce{CH3+}    + \ce{H2CO} + \ce{He}$  & $7.06 \times 10^{-10}~(T/300~\textrm{K})^{-0.5}~\textrm{cm}^{3}~\textrm{s}^{-1}$  & a & ion-neutral \\
$\ce{CH3OCH2} + \ce{He+}   \rightarrow \ce{H2CO+}   + \ce{CH3}  + \ce{He}$  & $7.06 \times 10^{-10}~(T/300~\textrm{K})^{-0.5}~\textrm{cm}^{3}~\textrm{s}^{-1}$  & a & ion-neutral \\
$\ce{CH3OCH2} + \ce{CH2OH} \rightarrow \ce{CH3OCH3} + \ce{H2CO}          $  & $1.00 \times 10^{-11}~\textrm{cm}^{3}~\textrm{s}^{-1}$  & b & neutral-neutral \\
$\ce{CH3OCH2} + \ce{CH3O}  \rightarrow \ce{CH3OCH3} + \ce{H2CO}          $  & $1.00 \times 10^{-11}~\textrm{cm}^{3}~\textrm{s}^{-1}$  & b & neutral-neutral \\
$\ce{CH3OCH2} + \ce{HCO}   \rightarrow \ce{CH3OCH3} + \ce{CO}            $  & $1.00 \times 10^{-11}~\textrm{cm}^{3}~\textrm{s}^{-1}$  & b & neutral-neutral \\
$\ce{CH3OCH2} + \ce{COOH}  \rightarrow \ce{CH3OCH3} + \ce{CO2}           $  & $1.00 \times 10^{-11}~\textrm{cm}^{3}~\textrm{s}^{-1}$  & b & neutral-neutral \\
$\ce{CH3OCH2 ->T[\textrm{CRUV}]CH3} + \ce{H2CO}                          $  & $5.00 \times 10^{2} \times \zeta_{0}~\textrm{s}^{-1}$  & c, e  & cosmic ray-induced photodissociation\\
$\ce{CH3OCH2 ->T[\textrm{CRUV}]CH2} + \ce{CH3O}                          $  & $5.00 \times 10^{2} \times \zeta_{0}~\textrm{s}^{-1}$  & c, e  & cosmic ray-induced photodissociation \\
$\ce{CH3OCH2 ->T[\textrm{UV}]CH3} + \ce{H2CO}                            $  & $5.00 \times 10^{-10} \, \textrm{exp}(-1.7~A_\textrm{V})~\textrm{s}^{-1}$  & d, e & external UV photodissociation\\
$\ce{CH3OCH2 ->T[\textrm{UV}]CH2} + \ce{CH3O}                            $  & $5.00 \times 10^{-10} \, \textrm{exp}(-1.7~A_\textrm{V})~\textrm{s}^{-1}$  & d, e & external UV photodissociation \\
\enddata
\tablenotetext{a}{Ion-molecule rates calculated using the method of \citet{herbstleung86}.}
\tablenotetext{b}{Generic rate coefficients are assumed as per \citet{garrod13}.}
\tablenotetext{c}{The cosmic ray ionization rate, $\zeta_{0}$, is set to $1.3 \times 10^{-17} \textrm{s}^{-1}$; generic prefactors are assumed based on like processes.}
\tablenotetext{d}{Generic rate coefficients based on like processes.}
\tablenotetext{e}{Same processes are assumed for grain-surface/ice species with a factor 3 smaller rate.}

\end{deluxetable}

\begin{deluxetable}{lccc}
\tablewidth{0pt}
\tabletypesize{\footnotesize}
\tablecolumns{4}
\tablecaption{Important and/or newly-included grain-surface and ice-mantle reactions \label{solid_rxn}}
\scriptsize
\tablehead{
\colhead{Reaction} & \colhead{$E_\textrm{A}$ (K)} & \colhead{Width (\AA)} & \colhead{Notes}}
\startdata
$\ce{CH3}+\ce{CO} \rightarrow \ce{CH3CO}$ & 2,870 & 1.0 & \tablenotemark{a} \tablenotemark{b}  \\
$\ce{CH3}+\ce{H2CO} \rightarrow \ce{CH3OCH2}$ & 2,870  & 1.0 & \tablenotemark{a} \tablenotemark{c}  \\
$\ce{CH3O}+\ce{CO} \rightarrow \ce{CH3OCO}$ & 3,967  & 1.0 & \tablenotemark{a} \tablenotemark{d}  \\
$\ce{CH2}+\ce{CH3O} \rightarrow \ce{CH3OCH2}$ & 0  & 1.0 & \\
$\ce{H}+\ce{CH3OCH2} \rightarrow \ce{CH3OCH3}$ & 0  & 1.0 & \\
$\ce{H}+\ce{CH3OCH3} \rightarrow \ce{CH3OCH2}+\ce{H2}$ & 4,450  & 1.0 & \tablenotemark{e}\\
%$\ce{CH2}+\ce{H2CO} \rightarrow \ce{CH3CHO}$ & 817  & 1.0 & \tablenotemark{c} \\
%$\ce{CH2}+\ce{H2CO} \rightarrow \ce{CH3}+\ce{HCO}$ & 817  & 1.0 & \tablenotemark{c} \\
$\ce{H}+\ce{CH2CO} \rightarrow \ce{CH2CHO}$ & 3050  & 1.0 & \tablenotemark{f} \\
$\ce{H}+\ce{CH2CO} \rightarrow \ce{CH3CO}$& 1320 & 1.0 & \tablenotemark{f} \\
$\ce{H}+\ce{CH3CO} \rightarrow \ce{CH3CHO}$ & 0 & 1.0 &\\
$\ce{O}+\ce{CH4} \rightarrow \ce{CH3}+\ce{OH}$ & 4380 & 1.0 & \tablenotemark{g} \\
\enddata
\tablecomments{The full set of the bulk-ice reactions is available in the machine readable format. This version only shows important and/or newly-included reactions.}
\tablenotetext{a}{Also treated using the 3-BEF scheme, see \S~2.5}
\tablenotetext{b}{Estimate using Evans-Polanyi relation.}
\tablenotetext{c}{Estimate based on $\ce{CH3}+\ce{CO}$ reaction.}
\tablenotetext{d}{\citet{huynh08}}
\tablenotetext{e}{\citet{takahashi07}}
\tablenotetext{f}{\citet{senosiain16}}
\tablenotetext{g}{\citet{westenberg67}}
\end{deluxetable}

\end{document}